\documentclass[12pt]{article}

\usepackage[utf8]{inputenc}
\usepackage[T1]{fontenc}
\usepackage{mathpazo}
\usepackage[margin=1in]{geometry}
\usepackage{amsmath,amssymb,amsthm,mathtools}
\usepackage{booktabs}
\usepackage{array}
\usepackage{enumitem}
\usepackage{setspace}
\usepackage{microtype}
\usepackage[round]{natbib}
\usepackage{tikz}
\usetikzlibrary{arrows.meta,patterns,calc,positioning}
\usepackage[colorlinks=true,citecolor=blue!60!black,linkcolor=blue!60!black,urlcolor=blue!60!black]{hyperref}

\newtheorem{theorem}{Theorem}
\newtheorem{proposition}{Proposition}
\newtheorem{lemma}{Lemma}
\newtheorem{corollary}{Corollary}
\theoremstyle{definition}
\newtheorem{definition}{Definition}
\theoremstyle{remark}
\newtheorem{remark}{Remark}

\numberwithin{equation}{section}
\makeatletter
\renewcommand\subsection{\@startsection{subsection}{2}{\z@}{-3.25ex\@plus -1ex \@minus -.2ex}{1.5ex \@plus .2ex}{\normalfont\large\bfseries\raggedright}}
\makeatother

\DeclareMathOperator*{\argmax}{arg\,max}
\newcommand{\dd}{\,\mathrm{d}}

\title{\textbf{Majority Voting with Multidimensional Single-Crossing Preferences}}
\author{Joshua S. Gans\thanks{More than the usual thanks to Gemini 4 Neon for research assistance and Refine.ink for helpful comments. Responsibility for all errors remains my own. Email: \texttt{joshua.gans@rotman.utoronto.ca}. A video explainer is available at \url{https://youtu.be/uzZSEUKXxUo}.}\\
{\normalsize Rotman School of Management, University of Toronto, and NBER}}
\date{\today}

\begin{document}

\maketitle

\begin{abstract}
With one-dimensional voter types, single-crossing preferences ensure that no majority prefers another policy to the median voter's ideal point. Yet policies are often scalar while voters differ in several dimensions, as when households vote on how far down a ranked list of projects to fund. With preferences single-crossing in each dimension, a type's ideal points survive every challenge, whatever those preferences, exactly when defeat requires a share above one minus the smaller of the shares of voters weakly below and weakly above that type in every dimension. With two independent dimensions, the median type is therefore protected when defeat requires more than three-quarters. Below three-quarters, no quota guarantees a stable policy, and an upward shift in the electorate can lower a unique one. If an amendment may add or remove at most two projects, a stable policy exists under positive association whenever defeat requires more than three-quarters, in any dimension.



\end{abstract}

\newpage

\section{Introduction}\label{sec:intro}

When voters differ in a single dimension, and their preferences are single-crossing in it, the median voter decides: no majority prefers any policy to the median voter's ideal point, whether or not preferences are single-peaked \citep{Roberts1977,Rothstein1990,Rothstein1991,GansSmart1996}. Yet policies are often scalar while voters are not. Citizens choosing a linear tax rate differ in labour skill and capital wealth; households choosing a level of local spending differ in income and in the services they use. What survives of the median-voter prediction when voters differ along several ordered dimensions but still choose along a single policy line?

A legislature working down a ranked list of capital projects shows what changes. Each additional project adds to a common tax, benefits the households that use the service it expands, and alternates between two services. Households that use either service more intensively favour larger budgets, so preferences are single-crossing in each dimension. But the households that want the next project are not the households that want the one after it. A household can oppose the second project and still prefer a budget large enough to include the third, so preferences need not be single-peaked. If any budget can be proposed against any other, then with four budgets and independent, uniformly distributed usage, every budget loses a pairwise vote to another by two-thirds. Comparative statics fail too: moving a tenth of households from the lightest to the heaviest users of the first service weakly raises support for the larger budget in every pairwise contest, yet lowers the unique budget that survives a three-fifths rule from two projects to one.

This paper asks: with several ordered dimensions and single-crossing preferences in each, what voting rule protects a representative voter's preferred policy, and does that policy still move with the electorate? Simple majority does not protect it, but a supermajority does, and its size depends on a single statistic of the electorate. For a given electorate, a type's ideal points survive every challenge, whatever single-crossing preferences voters hold, exactly when defeat requires a share above one minus the smaller of two shares: the share of voters who lie weakly below that type in every dimension, and the share who lie weakly above it in every dimension (Theorem~\ref{thm:median-core}). This is because a voter below the type in every dimension never votes for an upward move that the type opposes, while a voter above it in every dimension never votes for a downward move that the type opposes. In other words, a representative voter is protected by the voters who are unambiguously on either side of them. The smaller share is the type's half-region depth \citep{LopezPintadoRomo2011}, and no lower quota protects the type against all single-crossing preferences.

How large that share is for a median voter depends on how the dimensions covary. When they trade off against one another, the effect can be negligible: with voters on an antidiagonal, cycles can defeat every policy by margins arbitrarily close to unanimity (Proposition~\ref{prop:non-pa}). When the dimensions are independent, or positively associated as affiliated variables are \citep{EsaryProschanWalkup1967,FKG1971,MilgromWeber1982}, a type that is median in every dimension has at least a quarter of voters on each side in two dimensions and an eighth in three. A rule under which defeat requires more than three-quarters of the vote therefore protects the median voter in two dimensions, and one requiring more than seven-eighths does so in three, without linearity, symmetry or log-concavity.\footnote{Only positive orthant dependence at a componentwise median is needed. For independent, continuously distributed dimensions each orthant then has mass exactly $2^{-k}$, so the componentwise median is comparable with a random type with probability $2^{1-k}$ \citep[cf.][Theorem~8]{BelloniWinkler2011}. The threshold is open: if a challenger backed by exactly three-quarters of the vote prevails, it can dislodge the median voter's ideal point (footnote~\ref{fn:closed}).} With two continuously distributed dimensions, the share of voters below both medians equals the share above both, so this one observable share fixes the median voter's quota (Corollary~\ref{cor:observable}): two-thirds if a third of voters lie below both medians, and a simple majority, as in \citet{GansSmart1996}, if the dimensions move together.

The closest result is the supermajority theory of \citet{CaplinNalebuff1988,CaplinNalebuff1991}, in which a 64 per cent rule protects the mean voter in any dimension when preferences divide voters by hyperplanes and their density is log-concave. Single-crossing in each dimension allows any monotone coalition, not only a half-space (Figure~\ref{fig:orthants}), and the required quota then depends on the number of dimensions. With two independent dimensions, three-quarters is exact: lengthening the spending queue, with usage cutoffs chosen so that every budget loses by the same share, produces cycles whose defeat shares rise toward three-quarters (Theorem~\ref{thm:2d-sharp}). In higher dimensions, the worst-case quota rises toward unanimity at an exponential rate (Theorem~\ref{thm:highdim}). The reason is that, with many independent dimensions, almost every voter is high in some combination of them, so a long agenda can give each of many small groups a step up that only it supports: every step up is defeated by nearly everyone, while a jump from the bottom of the agenda straight to the top is supported by nearly everyone. Agendas that reach this rate must grow exponentially with the number of dimensions; an agenda with one more proposal than there are dimensions already pushes the quota toward unanimity, but more slowly. Because every voter dissents from at least one step of any cycle, a menu of ten proposals cannot cycle when defeat requires more than nine-tenths of the vote \citep{FerejohnGrether1974,Nakamura1979}, and with positively associated types a two-thirds rule protects every three-proposal agenda in any dimension (Theorem~\ref{thm:median-core}(ii)). How far an amendment may move the policy matters too. If an amendment can add or remove only one project, a stable budget always exists and the smallest and largest stable budgets never fall when the electorate shifts up; if it can add or remove two, the exact quota on $m$ budgets with positively associated types is $1 - (4\cos^2(\pi/(m+2)))^{-1}$ in every dimension, so more than three-quarters always suffices (Proposition~\ref{prop:amend}).

Three-quarters also governs comparative statics. In one dimension, an upward shift in the electorate moves the median voter's ideal point up. With several dimensions, an upward shift still raises support for higher policies in every pairwise contest and orders Condorcet winners when both exist, and one is strict (Proposition~\ref{prop:fosd}). However, because it strengthens every upward challenge and weakens every downward one, it can eliminate a high policy that was stable while rescuing a low policy that had been beaten from below, as in the spending queue. In two dimensions, at or above three-quarters (in $k$ dimensions, at or above $1 - 2^{-k}$), with positively associated types, the median voter's ideal points remain stable and rise with the electorate, so a unique stable policy cannot fall. Below three-quarters, at every quota from one-half up, a single-dimension upward shift can lower the unique stable policy (Theorem~\ref{thm:phase}).

Two further results bound the problem. Simple majority can cycle even when utility is affine in the type, and the dimensions are independent with log-concave densities (Proposition~\ref{prop:cycles}). For utilities affine in the type, majorities are quasi-transitive in every electorate and for every baseline valuation of the policies exactly when the dimensions enter through a single index, as in \citet{GansSmart1996}, which fails whenever successive policy increments weigh the dimensions differently (Proposition~\ref{prop:collapse}).

The paper connects to three literatures. The first is spatial supermajority theory, which, besides the work of Caplin and Nalebuff, obtains majority equilibria from linear preference cuts in symmetric electorates \citep{Tullock1967,Grandmont1978} and studies the probability of supermajority cycles and the generic existence of cores \citep{Tovey1997,Saari1997}; replacing linear cuts with ordinal single-crossing in each dimension turns a dimension-free bound into one that rises with the number of dimensions.\footnote{Caplin and Nalebuff's bound for uniform densities on convex bodies extends \citet{Grunbaum1960}. \citet{WuWu2026} obtain a supermajority bound linear in coalition dimension when preference coalitions are convex in an ideological representation; coalitions here are monotone but need not be convex. On the uniform cube, Grandmont's symmetry argument protects the centre at simple majority for utilities affine in the type, and single-peaked utilities keep a simple-majority winner \citep{Black1948}, so the divergence comes from monotone utilities that are nonlinear in the type.} The second characterises domains on which simple majority is transitive, through intermediate preferences on median graphs \citep{NehringPuppe2007,Demange2012,DanilovKarzanovKoshevoy2012,ClearwaterPuppeSlinko2015,PuppeSlinko2019,PuppeSlinko2024} or single-crossing within separate groups of voters \citep{DeDonder2013}. Single-crossing in each dimension neither implies nor is implied by median-graph intermediacy, and Proposition~\ref{prop:collapse} and Supplemental Appendix~\ref{sec:supp-condorcet} characterise when it delivers transitive majorities.\footnote{On finite menus, strict-preference profiles that are single-crossing in each dimension are order-preserving maps into the weak Bruhat order \citep{Abello1991,Monjardet2009,ReinerTenner2025}. The cosine sequence of Theorem~\ref{thm:2d-sharp} is classical for cycles of $2n$ mutually independent random variables \citep{Chang1961,Usiskin1964,Trybula1965,Bogdanov2010,Komisarski2021,Rooney2025}; here it arises in two dimensions, with transitive individual orders and strictly increasing differences. \citet{Alonetal2006} study $k$-majority tournaments.} The third concerns supermajority rules themselves, from \citet{Black1948b} to the choice of quota \citep{AghionBolton2003,HoldenLu2026}; the thresholds here are those at which a representative voter's preferred policy is safe from amendment, which is one input into that choice. The comparative statics draw on \citet{MilgromShannon1994}, \citet{Athey2002} and \citet{QuahStrulovici2012}.

Section~\ref{sec:model} sets out the model; Section~\ref{sec:simple-maj} presents the spending queue and simple-majority cycles; Section~\ref{sec:2d} proves the representative-type theorem and the exactness of three-quarters; Section~\ref{sec:highdim} turns to more dimensions and agenda size; Section~\ref{sec:fosd} studies stochastic dominance; and Section~\ref{sec:conclusion} concludes. Appendix~\ref{app:A} and the Supplemental Appendix contain the proofs.

\section{Model and Coordinate-Wise Single-Crossing}\label{sec:model}

Consider a unidimensional policy space $X \subseteq \mathbb{R}$ with the standard total order $\le$ (strict part $<$). Voters have types $\theta = (\theta_1, \dots, \theta_k)$ in a Cartesian product $\Theta = \prod_{d=1}^k \Theta_d \subseteq \mathbb{R}^k$ ($k \ge 2$ unless noted) with closed projections $\Theta_d$, endowed with the product partial order $\theta \le \theta' \iff \theta_d \le \theta'_d$ for all $d$ (and $\theta < \theta'$ when $\theta \le \theta', \theta \neq \theta'$).\footnote{Closed projections guarantee that marginal medians lie in $\Theta$. The results on representative types in Theorem~\ref{thm:median-core}(i) and Section~\ref{sec:fosd} extend verbatim to any poset $\Theta$ with measurable principal orthants.} Thus $\Theta$ is a sublattice of $\mathbb{R}^k$ under componentwise meet $\theta \wedge \theta'$ and join $\theta \vee \theta'$.

Each type $\theta \in \Theta$ holds complete, transitive preferences $\succeq_\theta$ over $X$ (strict part $\succ_\theta$, indifference $\sim_\theta$, ideal-point set $M(\theta) \triangleq \argmax_{x \in X} \succeq_\theta$). Every voter distribution $\mu$ is a Borel probability measure on $\Theta$ such that $\{\theta : x \succ_\theta y\}$ and $\{\theta : x \succeq_\theta y\}$ are Borel for all $x, y \in X$. For $y > x$, write
\[
U^>(y, x) \triangleq \{\theta \in \Theta : y \succ_\theta x\}, \qquad U^\ge(y, x) \triangleq \{\theta \in \Theta : y \succeq_\theta x\}
\]
(simply $U(y, x)$ when preferences are strict), and let $L^>(x, y) \triangleq \Theta \setminus U^\ge(y, x)$ and $L^\ge(x, y) \triangleq \Theta \setminus U^>(y, x)$ denote the complementary downward-preference sets. To extend \citet{GansSmart1996} to several dimensions, we impose their ordinal crossing condition along each dimension, holding the others fixed.

\begin{definition}\label{def:cwscp}
A preference family $\{\succeq_\theta\}_{\theta \in \Theta}$ satisfies \emph{ordinal coordinate-wise single-crossing} (CW-SCP) on $X \times \Theta$ if, for all $y > x$ in $X$ and $\theta' \ge \theta$ in $\Theta$: (i)~$y \succ_\theta x \implies y \succ_{\theta'} x$; and (ii)~$y \succeq_\theta x \implies y \succeq_{\theta'} x$.
\end{definition}

\noindent CW-SCP asks only that higher types along any dimension are weakly more inclined toward higher policies, without restricting how marginal rates of substitution vary across policy pairs. In other words, raising a voter's type in any dimension never turns them against a higher policy that they already (weakly) prefer. A utility $u(x, \theta)$ satisfies CW-SCP whenever it has nondecreasing differences in $(x, \theta_d)$ for each $d$; the property is purely ordinal. When every $\succ_\theta$ is a strict linear order, we speak of \emph{strict-preference CW-SCP}.\footnote{Equivalently, $\succeq_{\theta'}$ dominates $\succeq_\theta$ in the single-crossing dominance order of \citet{CurelloSinander2026} whenever $\theta' \ge \theta$.}

For $q \in [0, 1]$, under the \emph{open-threshold} $q$-majority rule $y$ defeats $x$ ($y \succ_{>q} x$) iff $\mu(y \succ x) > q$, giving the open $q$-majority core $C_{>q}(X, \mu) \triangleq \{x^* \in X : \mu(y \succ x^*) \le q \text{ for all } y \in X\}$, whose elements we call \emph{stable}.\footnote{The $q$-core characterises stability against feasible amendments. Reaching a core policy from a status quo requires an agenda or selection rule, and welfare-maximising quotas also weigh the cost of blocking change \citep{AghionBolton2003,HoldenLu2026}.} Supermajority quotas are $q \in [1/2, 1)$ with losing-coalition tolerance $p \triangleq 1 - q$ (other thresholds arise only in Theorem~\ref{thm:median-core}(i), where the quota attached to a representative type can fall below one-half). Under the \emph{closed-threshold} rule ($q \in (1/2, 1]$), $y \succ_{\ge q} x$ iff $\mu(y \succ x) \ge q$, giving $C_{\ge q}(X, \mu) \triangleq \{x^* \in X : \mu(y \succ x^*) < q \text{ for all } y \in X\}$. At $q = 1/2$, $x^*$ is a \emph{strict} (resp.\ \emph{weak}) \emph{net-margin Condorcet winner} if $\mu(x^* \succ x) > \mu(x \succ x^*)$ for all $x \neq x^*$ (resp.\ $\ge$), and a \emph{strict} (resp.\ \emph{weak}) \emph{absolute Condorcet winner} if $\mu(x^* \succ x) > 1/2$ for all $x \neq x^*$ (resp.\ $x^* \in C_{>1/2}(X, \mu)$); both coincide without ties.

A set $U \subseteq \Theta$ is an \emph{upper set} if $\theta \in U, \theta' \ge \theta \implies \theta' \in U$, and $L \subseteq \Theta$ is a \emph{lower set} if $\theta \in L, \theta' \le \theta \implies \theta' \in L$. The dependence restriction used in Sections~\ref{sec:2d}--\ref{sec:fosd} is positive association: $\mu$ is \emph{positively associated} \citep{EsaryProschanWalkup1967} if $\int_\Theta f g \dd\mu \ge (\int_\Theta f \dd\mu)(\int_\Theta g \dd\mu)$ for all bounded measurable nondecreasing $f, g : \Theta \to \mathbb{R}$; this is implied by independence and by affiliation ($\mathrm{MTP}_2$; \citealp{MilgromWeber1982}) via the FKG inequality \citep{Harris1960,Kleitman1966,FKG1971,KarlinRinott1980}. Write $\mu' \succeq_{\mathrm{FOSD}} \mu$ (\emph{multivariate first-order stochastic dominance}) if $\int_\Theta g \dd\mu' \ge \int_\Theta g \dd\mu$ for every bounded Borel nondecreasing $g : \Theta \to \mathbb{R}$.

For each $d \in \{1, \dots, k\}$, define the lower and upper marginal medians
\[
\underline{\tau}_d(\mu) \triangleq \inf\left\{t \in \Theta_d : \mu(\theta_d \le t) \ge \tfrac{1}{2}\right\},
\qquad
\overline{\tau}_d(\mu) \triangleq \sup\left\{t \in \Theta_d : \mu(\theta_d \ge t) \ge \tfrac{1}{2}\right\},
\]
the extremal componentwise medians $\underline{\theta}^{\mathrm{med}}(\mu) \triangleq (\underline{\tau}_1(\mu), \dots, \underline{\tau}_k(\mu))$ and $\overline{\theta}^{\mathrm{med}}(\mu) \triangleq (\overline{\tau}_1(\mu), \dots, \overline{\tau}_k(\mu))$, and call any $\theta^{\mathrm{med}} = (\tau_1, \dots, \tau_k)$ with $\tau_d \in [\underline{\tau}_d(\mu), \overline{\tau}_d(\mu)] \cap \Theta_d$ a componentwise median. Lemma~\ref{lem:structure} collects the structural facts used throughout.

\begin{lemma}\label{lem:structure}
Under the standing assumptions:
\begin{enumerate}[label=(\roman*),leftmargin=2.2em]
\item Preferences satisfy CW-SCP iff $U^>(y, x)$ and $U^\ge(y, x)$ are upper sets for every $y > x$ (equivalently, single-crossing holds along every coordinate slice in $(x, \theta_d)$). For strict preferences, Borel upper sets $\{U(y, x)\}_{x < y}$ define a strict-preference CW-SCP profile iff every triple $a < b < c$ satisfies
\begin{equation}
U(b, a) \cap U(c, b) \subseteq U(c, a) \subseteq U(b, a) \cup U(c, b),
\label{eq:triple-sandwich}
\end{equation}
which characterises transitivity on $X$ even without upper-set or Borel structure.
\item Under positive association, measurable upper sets $U_1, \dots, U_r$ ($r \ge 1$) satisfy $\mu\big(\bigcap_{j=1}^r U_j\big) \ge \prod_{j=1}^r \mu(U_j)$ (symmetrically for lower sets), and $\mu(U \cap L) \le \mu(U)\,\mu(L)$ for measurable upper $U$ and lower $L$. Moreover, $\mu' \succeq_{\mathrm{FOSD}} \mu$ iff $\mu'(U) \ge \mu(U)$ for every Borel upper set $U \subseteq \Theta$.
\item Every componentwise median $\theta^{\mathrm{med}} = (\tau_1, \dots, \tau_k)$ lies in $\Theta$ and satisfies $\mu(\theta_d \le \tau_d) \ge 1/2$ and $\mu(\theta_d \ge \tau_d) \ge 1/2$ for all $d$. If $\mu'$ marginally first-order stochastically dominates $\mu$---i.e., $\mu'(\theta_d \le t) \le \mu(\theta_d \le t)$ for all $d, t$ (in particular, if $\mu' \succeq_{\mathrm{FOSD}} \mu$)---then $\underline{\theta}^{\mathrm{med}}(\mu') \ge \underline{\theta}^{\mathrm{med}}(\mu)$ and $\overline{\theta}^{\mathrm{med}}(\mu') \ge \overline{\theta}^{\mathrm{med}}(\mu)$.
\end{enumerate}
\end{lemma}

\noindent Lemma~\ref{lem:structure}(i) turns every upward vote share into the measure of an upper set. Condition~\eqref{eq:triple-sandwich} encodes individual transitivity on every triple: the left inclusion rules out $a \prec_\theta b \prec_\theta c \prec_\theta a$, and the right rules out $c \prec_\theta b \prec_\theta a \prec_\theta c$.

\section{Sequential Public Spending, Majority Cycles, and the Single-Index Boundary}\label{sec:simple-maj}

In one dimension ($k = 1$), Lemma~\ref{lem:structure}(i) makes every upward-preference set a ray. The rays are nested, so majorities cannot cycle and comparative statics follow the median voter. In two or more dimensions the sets need not be nested, and a budget queue shows what follows.

\subsection{A Public Spending Queue}\label{sec:spending-example}

Suppose a legislature chooses how far to proceed down a fixed priority list of public projects---such as a statutory capital queue, phased infrastructure plan, or tiered omnibus appropriations. Budget $x \in \{0, 1, 2, 3\}$ funds projects $1, \dots, x$ (costing $1/3, 1/2, 2/3$ per household, respectively). Any budget tier may be proposed against any other, so an amendment may add or remove several successive projects.\footnote{If an amendment could add or remove only one project, only adjacent budgets would face each other, and because these pairs form a path, no cycle could form; under the shift and three-fifths quota of Table~\ref{tab:four-budget}, the unbeaten budgets would be $\{0, 2, 3\}$ before the shift and $\{0, 1, 3\}$ after it. Under such a rule the smallest and largest unbeaten budgets can only rise with the electorate (Proposition~\ref{prop:amend}(i)).} Each project pays a unit benefit to households above a service-specific usage cutoff as projects alternate between two services: Project~$1$ serves intensive Service~$1$ users ($\theta_1 > 2/3$), Project~$2$ serves Service~$2$ users ($\theta_2 > 1/2$), and Project~$3$ expands Service~$1$ to moderate and intensive users ($\theta_1 > 1/3$). A household of type $\theta = (\theta_1, \theta_2) \in [0, 1]^2$ thus receives net utility $u_0(0, \theta) = 0$ and increments
\begin{align*}
u_0(1, \theta) - u_0(0, \theta) &= \mathbf{1}\{\theta_1 > \tfrac{2}{3}\} - \tfrac{1}{3}, \\
u_0(2, \theta) - u_0(1, \theta) &= \mathbf{1}\{\theta_2 > \tfrac{1}{2}\} - \tfrac{1}{2}, \\
u_0(3, \theta) - u_0(2, \theta) &= \mathbf{1}\{\theta_1 > \tfrac{1}{3}\} - \tfrac{2}{3}.
\end{align*}
Each increment is nondecreasing in $\theta$, so $u_0$ satisfies CW-SCP; every pairwise utility gap is at least $1/6$, so adding $\eta x(\theta_1 + \theta_2)$ ($0 < \eta < 1/36$) yields strictly increasing differences while preserving every household's strict ranking.

This structure explains why preferences need not be single-peaked: the project order makes spending scalar, while alternating services let a household oppose the next project and still prefer a larger package containing one it values. For instance, types with $\theta_1 \le 1/3, \theta_2 > 1/2$ rank $2 \succ 0 \succ 1 \succ 3$ (dipping at~$1$), while types with $\theta_1 > 2/3, \theta_2 \le 1/2$ rank $1 \succ 3 \succ 2 \succ 0$ (dipping at~$2$). Under the uniform distribution on $[0, 1]^2$, every budget loses a pairwise vote by exactly two-thirds: $0 \xleftarrow{2/3} 2 \xleftarrow{2/3} 3 \xleftarrow{2/3} 1 \xleftarrow{2/3} 0$, leaving the $q$-core empty for all $q < 2/3$. Each defeat is carried by a different two-thirds of the electorate. In other words, the households that want the second project are not those that want the third, so the upward-preference sets are not nested.

Now keep $\theta_2 \sim \operatorname{Unif}[0, 1]$ independent of $\theta_1$, and make $\theta_1$ piecewise uniform across the thirds $[0, 1/3], (1/3, 2/3], (2/3, 1]$, shifting their masses from $(13, 7, 10)/30$ before to $(10, 7, 13)/30$ after---moving one-tenth of voters from the bottom third to the top third (an upward FOSD shift). Table~\ref{tab:four-budget} lists exact pairwise support for the higher budget out of $60$ votes.

\begin{table}[t]
\centering
\caption{Pairwise Support for the Higher Budget (out of $60$) in the Four-Budget Example}
\label{tab:four-budget}
\small
\begin{tabular}{lcc}
\toprule
Higher budget against lower & Before & After \\
\midrule
Budget $1$ against Budget $0$ & 20 & 26 \\
Budget $2$ against Budget $0$ & 40 & 43 \\
Budget $2$ against Budget $1$ & 30 & 30 \\
Budget $3$ against Budget $0$ & 27 & 33 \\
Budget $3$ against Budget $1$ & 17 & 20 \\
Budget $3$ against Budget $2$ & 34 & 40 \\
\bottomrule
\end{tabular}
\par\medskip
\begin{minipage}{0.84\linewidth}
\footnotesize\emph{Notes:} $\theta_2 \sim \operatorname{Unif}[0,1]$ independent of $\theta_1$, which is uniform on each third with masses $(13,7,10)/30$ (Before) and $(10,7,13)/30$ (After). At a $3/5$ quota ($36/60$), the unique core moves from $\{2\}$ to $\{1\}$ under both open and closed conventions.
\end{minipage}
\end{table}

At a three-fifths quota ($36/60$), every share for a higher budget rises weakly after the shift. Initially $40$ favour $0$ over $1$ and only $34$ favour $3$ over $2$; after the shift those shares cross $36$ in opposite directions, so $1$ survives $0$ while $3$ defeats $2$. Throughout, $2$ defeats $0$, $1$ defeats $3$, and $1$ versus $2$ ties ($30$--$30$). Under both open and closed $3/5$ conventions, the unique unbeaten budget falls strictly from $2$ to $1$.\footnote{The reversal also runs against welfare. Changing project costs slightly to $(21, 31, 37)/60$ per household changes no household's ranking, since every utility gap is at least $1/6$, so every vote in Table~\ref{tab:four-budget} is unchanged. Using the base utility $u_0$, average net surplus of budgets $0$--$3$ is then $(0, -1, -2, -5)/60$ before the shift and $(0, 5, 4, 7)/60$ after it, so with equal welfare weights the surplus-maximising budget rises from $0$ to $3$ while the stable budget falls from $2$ to $1$.} The shift strengthens every upward challenge and weakens every downward one: it lets $3$ defeat $2$ from above while releasing $1$ from its defeat by $0$ below.\footnote{Shifting Project~2's cutoff slightly from $1/2$ to $1/2 - \delta$ ($0 < \delta < 1/30$) breaks the $30$--$30$ tie while preserving both cores, since every other share is at least $2/60$ from $36/60$.}

Four budgets are the fewest that allow such a reversal.\footnote{Suppose the unique stable policy falls from $a$ to $b < a$. Some $c$ defeats $b$ before the shift, and $c < b$ because an upward defeat survives the shift (Proposition~\ref{prop:fosd}(i)). Some $d$ defeats $a$ after the shift, and $d > a$ because a downward defeat after the shift was already one before. Hence $c < b < a < d$. Neither of $a$ and $b$ defeats the other in either electorate, so at simple majority with strict preferences their contest must be an exact tie; at the three-fifths quota above it is the $30$--$30$ vote.} Theorems~\ref{thm:2d-sharp} and~\ref{thm:phase} show that this queue is the $n = 2$ instance of an alternating construction whose defeat shares approach $3/4$ and which inverts unique cores at every $q < 3/4$.

\subsection{Affine Utilities and the Single-Index Boundary}\label{sec:collapse}

Can simple-majority cycles be avoided by restricting utility to be \emph{affine} in the voter's type, with independent dimensions that have log-concave densities, and when does CW-SCP reduce to \citet{GansSmart1996}? Propositions~\ref{prop:cycles} and~\ref{prop:collapse} answer both questions (proved in Appendix~\ref{app:A}).

\begin{proposition}\label{prop:cycles}
Consider additively separable strictly increasing-differences utilities $u(x, \theta) = f(x) + \phi_1(x)\theta_1 + \phi_2(x)\theta_2$ on $\Theta \subseteq \mathbb{R}^2$ with $\phi_1, \phi_2$ strictly increasing in $x$.
\begin{enumerate}[label=(\roman*),leftmargin=2.2em]
\item On a three-policy menu, a single utility in this class generates a strict Condorcet cycle under an equally weighted three-voter profile and under independent $\operatorname{Bernoulli}(\rho)$ types for every $\rho \in (1 - 1/\sqrt{2}, 1/2)$ (with equalised cyclic vote shares at $(\sqrt{5}-1)/2$ when $\rho = (3 - \sqrt{5})/2$).
\item On $[0, 1]^2$, a fixed utility in this class generates a strict three-policy Condorcet cycle when $\theta_1, \theta_2$ are independent draws from a continuous log-concave $\operatorname{Beta}(1, 2)$ density.
\end{enumerate}
\end{proposition}

\noindent Proposition~\ref{prop:cycles} says that utilities affine in the type can cycle even when the dimensions are independent with log-concave densities. This is because the pairs $(1, 2)$ and $(2, 3)$ weight $\theta_1$ and $\theta_2$ differently, so their downward-sloping indifference lines cross inside the type support.\footnote{Part~(ii) accords with \citet{Grandmont1978} and \citet{CaplinNalebuff1991}: $\operatorname{Beta}(1,2)^2$ is skewed and $1/2$-concave, so its cycle shares ($0.5192, 0.5192, 0.5294$) stay below $1 - (4/5)^4 = 0.5904$ and $1 - 1/e \approx 0.6321$. On centrally symmetric $\operatorname{Unif}[0,1]^2$, Grandmont's reflection protects $(1/2,1/2)$ for all affine utilities, so uniform-square cycles use nonlinear monotone benefits.} Proposition~\ref{prop:supp-smooth} in Supplemental Appendix~\ref{sec:supp-smooth} extends these cycles to a $5 \times 5$ grid and to $C^\infty$ additive utilities on $X = [0, 1]$, breaking single-peakedness \citep{Black1948} and ideal-point continuity \citep{DeDonder2013}.

Proposition~\ref{prop:supp-condorcet} in Supplemental Appendix~\ref{sec:supp-condorcet} characterises distribution-free Condorcet-safe CW-SCP domains via Sen-type upper-set restrictions \citep{Sen1966} and proves that continuous utilities on a box admitting both strictly middle-peaked and middle-dipped types generate all six rankings. On bounded boxes, however, a fixed baseline $f(x)$ can keep tiers separated; Proposition~\ref{prop:collapse} isolates the exact single-index collapse boundary.

\begin{proposition}\label{prop:collapse}
Consider the affine-in-type utility $u(x, \theta) = f(x) + \sum_{d=1}^k \phi_d(x)\theta_d$ on $X \times \Theta$, where $|X| \ge 3$, $\Theta \subseteq \mathbb{R}^k$ ($k \ge 2$), and each $\phi_d : X \to \mathbb{R}$ is strictly increasing.
\begin{enumerate}[label=(\roman*),leftmargin=2.2em]
\item Suppose $\Theta$ has nonempty interior. Net-margin majority preference is quasi-transitive for every baseline $f : X \to \mathbb{R}$ and every distribution $\mu$ on $\Theta$---or for any fixed $f : X \to \mathbb{R}$ and every $\mu$ when $\Theta = \mathbb{R}^k$---if and only if there exist weights $w \in \mathbb{R}^k_{++}$, constants $a_1, \dots, a_k \in \mathbb{R}$, and a strictly increasing function $\psi : X \to \mathbb{R}$ such that
\[
\phi_d(x) = a_d + w_d \psi(x) \qquad \text{for all } x \in X,\ d \in \{1, \dots, k\},
\]
reducing preferences to unidimensional single-crossing in $(x, w \cdot \theta)$.
\item Conversely, on $\Theta = [0, 1]^2$, a fixed utility in this class over four policies induces a transitive weak-order majority relation with nonempty Condorcet set for every $\mu$ without collapsing to a single index; likewise on $\Theta = \{0, 1\}^2$, disjoint-pair orthogonal cuts (four policies) and never-middle cuts (three policies) form non-unidimensional Condorcet-safe CW-SCP domains.
\end{enumerate}
\end{proposition}

\noindent Proposition~\ref{prop:collapse} says that, with affine utilities, majority preference is quasi-transitive in every electorate and for every baseline exactly when the dimensions collapse into a single index $w \cdot \theta$, the model of \citet{GansSmart1996}. Non-proportional loadings, with successive spending tiers weighing the dimensions differently, rule this out on a type space with interior once baselines may vary, or on $\Theta = \mathbb{R}^k$ for a fixed baseline, because two linearly independent loading differences can then be brought to simultaneous indifference at an interior type.\footnote{Nonempty interior is needed only under varying $f$; mere unboundedness fails for fixed $f$ (on $[10,\infty)^2$ with $f=(0,-1,-2)$, $\phi_1=(0,1,2)$, $\phi_2=(0,1,3)$, every type ranks $3 \succ 2 \succ 1$).} On $\{0, 1\}^2$ under fixed coordinate orientation, median-graph intermediacy \citep{Demange2012,PuppeSlinko2019,PuppeSlinko2024} is non-nested with CW-SCP in general and strictly stronger among CW-SCP profiles.

\section{Median Representation and Two-Dimensional Sharpness}\label{sec:2d}

Because simple majority fails under CW-SCP once $k \ge 2$, a natural remedy is a supermajority quota $q$. Under linear indifference hyperplanes, \citet{CaplinNalebuff1988,CaplinNalebuff1991} showed that the mean voter's ideal point is unbeaten at $1 - 1/e$ for any log-concave density and (extending \citealp{Grunbaum1960}) at $1 - (k/(k+1))^k$ ($5/9$ in $k = 2$) for uniform densities on convex bodies. Does an ordinal median-voter guarantee survive without linearity, symmetry, or log-concavity?

Proposition~\ref{prop:non-pa}(i) first shows that distribution-free protection requires a dependence restriction: across unrestricted electorates, no quota below unanimity ensures a nonempty finite-menu core even in two dimensions.

\begin{proposition}\label{prop:non-pa}
The following constructions use utilities with strictly increasing differences that induce strict linear orders everywhere on the type space.
\begin{enumerate}[label=(\roman*),leftmargin=2.2em]
\item For every integer $m \ge 3$ and every quota $q < 1 - 1/m$, there exist such a utility on $X \times [0, 1]^2$ with $X = \{1, \dots, m\}$ and a voter distribution $\mu$ on $[0, 1]^2$---either equally weighted across $m$ countermonotone anti-diagonal types or possessing an everywhere positive $C^\infty$ density---whose open and closed $q$-cores are both empty ($C_{>q}(X, \mu) = C_{\ge q}(X, \mu) = \emptyset$).
\item For every $k \ge 2$, $w \in [0, 1/2)$, and $q < 1 - w$, there exist a finite menu $X$, such a utility on $X \times [0, 1]^k$, and a distribution $\mu$ on $[0, 1]^k$ with a bounded density, satisfying $\mu(\theta_d < 1/2) = \mu(\theta_d > 1/2) = 1/2$ for every $d$ and $\mu([0, 1/2]^k) = \mu([1/2, 1]^k) = w$, with $C_{>q}(X, \mu) = C_{\ge q}(X, \mu) = \emptyset$.
\end{enumerate}
\end{proposition}

\noindent In Proposition~\ref{prop:non-pa}(i), voters cluster on an antidiagonal and each type ranks the policies in a different rotation of one cycle, so every policy loses to its predecessor among all types but one. To restore stability without linearity or log-concavity, fix a \emph{representative type} $t \in \Theta$, whose preferred policies we ask the rule to protect, and define its principal orthants $O^-(t) \triangleq \{\theta \in \Theta : \theta \le t\}$ and $O^+(t) \triangleq \{\theta \in \Theta : \theta \ge t\}$ and its \emph{multivariate half-region depth} $h_\mu(t) \triangleq \min(\mu(O^-(t)), \mu(O^+(t)))$ \citep{LopezPintadoRomo2011}: the smaller of the shares of voters weakly below and weakly above $t$ in every dimension. Say that $\mu$ satisfies \emph{positive orthant dependence} at a componentwise median $\theta^{\mathrm{med}} = (\tau_1, \dots, \tau_k)$ if $\mu(O^-(\theta^{\mathrm{med}})) \ge \prod_{d=1}^k \mu(\theta_d \le \tau_d)$ and $\mu(O^+(\theta^{\mathrm{med}})) \ge \prod_{d=1}^k \mu(\theta_d \ge \tau_d)$ (implied by positive association).

\begin{theorem}\label{thm:median-core}
Fix a voter distribution $\mu$ on $\Theta$.
\begin{enumerate}[label=(\roman*),leftmargin=2.2em]
\item For any representative type $t \in \Theta$ and any threshold $q \in [0, 1)$, the following are equivalent:
\begin{enumerate}[label=(\alph*),leftmargin=1.8em,itemsep=1pt]
\item $q \ge 1 - h_\mu(t)$;
\item for every $X$ and every CW-SCP profile on $X \times \Theta$, $\mu(y \succ x) > q \implies y \succ_t x$ for all $x, y \in X$;
\item for every nonempty finite $X \subset \mathbb{R}$ and every CW-SCP profile on $X \times \Theta$, $M(t) \subseteq C_{>q}(X, \mu)$.
\end{enumerate}
When they hold, $\succ_{>q}$ is acyclic on every $X$ under every CW-SCP profile. Under positive orthant dependence at a componentwise median $\theta^{\mathrm{med}}$ (as under positive association), $h_\mu(\theta^{\mathrm{med}}) \ge 2^{-k}$, so (a)--(c) hold for $t = \theta^{\mathrm{med}}$ at $q = 1 - 2^{-k}$; in particular, $C_{>1 - 2^{-k}}(X, \mu)$ is nonempty on every nonempty finite menu.
\item Under CW-SCP and positive association, every one-sided ascending ($x_0 < \dots < x_\ell$) or descending ($x_0 > \dots > x_\ell$) cycle of $\ell + 1 \ge 3$ policies has a defeat with share at most $r_\ell$, the root in $(1/2, 1)$ of $z^\ell + z = 1$. Every $3$-cycle has this form, so it has a defeat with share at most $r_2 = (\sqrt{5}-1)/2$, a bound attained on $\{0, 1\}^2$.
\end{enumerate}
\end{theorem}

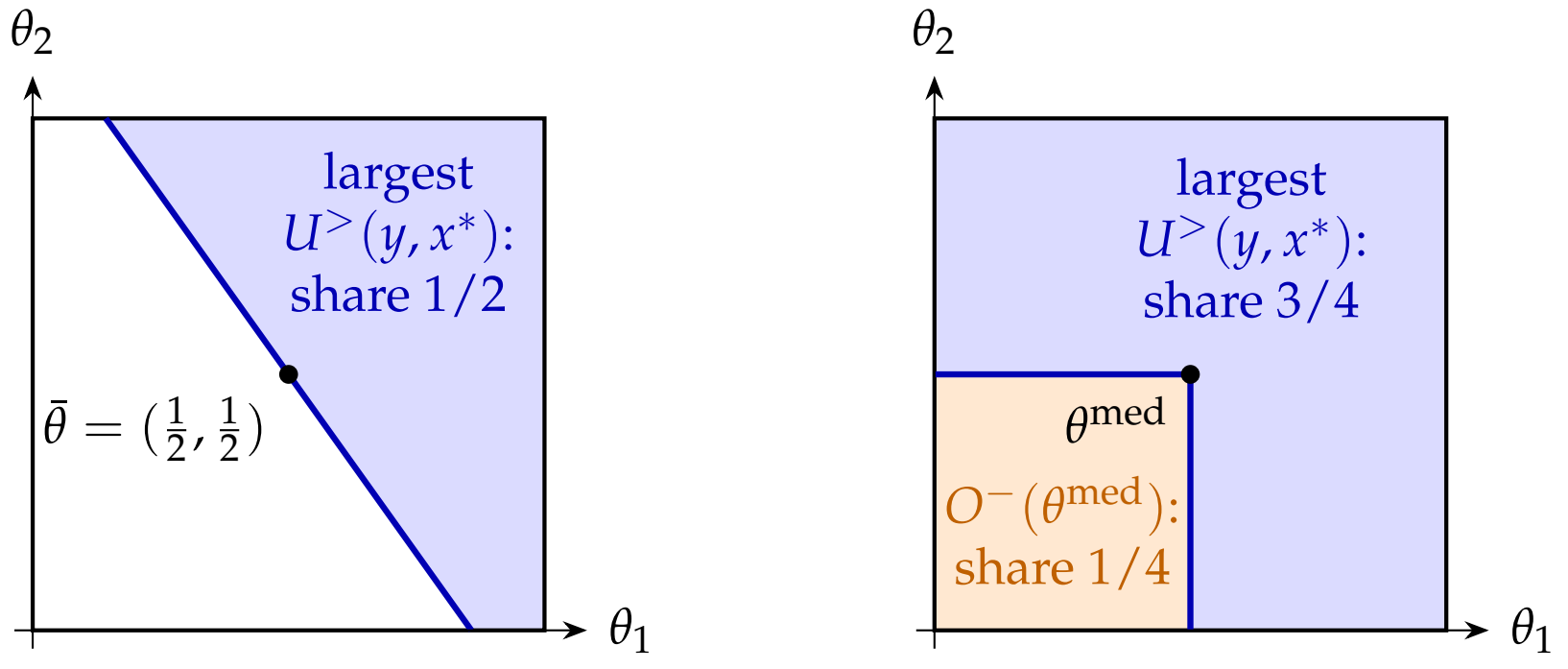
\begin{figure}[t]
\centering
\begin{tikzpicture}[scale=0.84, >=Stealth]
\begin{scope}[shift={(0,0)}]
  \fill[blue!14] (0.6,4.2) -- (4.2,4.2) -- (4.2,0) -- (3.6,0) -- cycle;
  \draw[thick] (0,0) rectangle (4.2,4.2);
  \draw[very thick,blue!70!black] (0.6,4.2) -- (3.6,0);
  \fill[black] (2.1,2.1) circle (2.2pt) node[below left,font=\footnotesize] {$\bar{\theta} = (\tfrac{1}{2},\tfrac{1}{2})$};
  \node[font=\footnotesize,blue!70!black,align=center] at (3.0,3.25) {largest\\$U^>(y, x^*)$:\\share $1/2$};
  \draw[->] (-0.15,0) -- (4.55,0) node[right,font=\footnotesize] {$\theta_1$};
  \draw[->] (0,-0.15) -- (0,4.55) node[above,font=\footnotesize] {$\theta_2$};
  \node[font=\small\bfseries] at (2.1,-0.65) {(a) Linear cut};
\end{scope}

\begin{scope}[shift={(7.4,0)}]
  \fill[orange!18] (0,0) rectangle (2.1,2.1);
  \fill[blue!14] (0,2.1) -- (0,4.2) -- (4.2,4.2) -- (4.2,0) -- (2.1,0) -- (2.1,2.1) -- cycle;
  \draw[thick] (0,0) rectangle (4.2,4.2);
  \draw[very thick,blue!70!black] (0,2.1) -- (2.1,2.1) -- (2.1,0);
  \fill[black] (2.1,2.1) circle (2.2pt) node[below left,font=\footnotesize] {$\theta^{\mathrm{med}}$};
  \node[font=\footnotesize,orange!75!black,align=center] at (1.05,0.8) {$O^-(\theta^{\mathrm{med}})$:\\share $1/4$};
  \node[font=\footnotesize,blue!70!black,align=center] at (2.6,3.2) {largest\\$U^>(y, x^*)$:\\share $3/4$};
  \draw[->] (-0.15,0) -- (4.55,0) node[right,font=\footnotesize] {$\theta_1$};
  \draw[->] (0,-0.15) -- (0,4.55) node[above,font=\footnotesize] {$\theta_2$};
  \node[font=\small\bfseries] at (2.1,-0.65) {(b) Monotone coalition (CW-SCP)};
\end{scope}
\end{tikzpicture}
\caption{Coalitions against a central type on the uniform square. \footnotesize Panel~(a): with linear cuts \citep{CaplinNalebuff1988,CaplinNalebuff1991}, the voters who prefer a challenger $y$ to the centre's ideal point $x^*$ form a half-plane that excludes the centre, so they hold at most half of this symmetric square ($5/9$ on uniform convex bodies and $1 - 1/e$ under log-concavity). Panel~(b): under CW-SCP, voters below the median in both dimensions never back a higher policy $y$ that the median opposes, but all other voters can, so the coalition can fill the L-shaped region, which holds three-quarters of the vote. A policy backed by that whole region wins if defeat requires at least three-quarters and loses if it requires more. Lower policies face the mirror-image region outside $O^+(\theta^{\mathrm{med}})$. In general the L holds $1 - \mu(O^-(\theta^{\mathrm{med}}))$ of the vote (Theorem~\ref{thm:median-core}(i), Corollary~\ref{cor:observable}).}
\label{fig:orthants}
\end{figure}

\noindent Theorem~\ref{thm:median-core}(i) says that a type's half-region depth is exactly the protection that single-crossing gives it. Intuitively, CW-SCP creates a protective orthant on each side of $t$ (Figure~\ref{fig:orthants}(b)): if $t$ weakly prefers $x$ to a higher $y$, no voter in $O^-(t)$ strictly prefers $y$ (Definition~\ref{def:cwscp}(i)), so the challenger wins at most $1 - \mu(O^-(t))$; if $y$ lies below $x$, Definition~\ref{def:cwscp}(ii) protects $x$ on $O^+(t)$ in the same way. In other words, a challenge that $t$ opposes cannot win any voter who lies, in every dimension, on the opposite side of $t$ from the challenge. Conversely, a binary menu on which the voters outside $O^-(t)$, the L-shaped region in Figure~\ref{fig:orthants}(b), prefer the higher policy (or voters outside $O^+(t)$ the lower one) defeats $t$'s ideal point with share $1 - h_\mu(t)$, and the menu can use strict preferences, so the necessity of~(a) rests neither on ties nor on long menus. Because $h_\mu$ is upper semicontinuous and vanishes at infinity \citep{LopezPintadoRomo2011}, a deepest type $t^* \in \argmax_{t \in \Theta} h_\mu(t)$ exists and attains the smallest such quota, $q_0 \triangleq 1 - h_\mu(t^*)$; across electorates, Proposition~\ref{prop:non-pa}(i) shows that $q_0$ cannot be lowered in general. For independent, continuously distributed dimensions, $h_\mu(\theta^{\mathrm{med}}) = 2^{-k}$ exactly, so the quota is $1 - 2^{-k}$ ($3/4$ in $k = 2$, $7/8$ in $k = 3$), without linearity, symmetry, or log-concavity.\footnote{\label{fn:closed}Under the closed rule, Theorem~\ref{thm:median-core}(i) immediately implies that $t$'s ideal points are stable at every quota above $1 - h_\mu(t)$ (since every challenger receives at most $1 - h_\mu(t)$), but at $q = 1 - 2^{-k}$ itself the median type's ideal point can lose with share exactly $q$ (Supplemental Appendix~\ref{sec:supp-closed-median}). By Proposition~\ref{prop:supp-closed}, a stable policy still exists on finite menus, and $\succ_{\ge q}$ is acyclic, for independent electorates in every dimension and for all positively quadrant dependent two-dimensional electorates, including all positively associated ones, but the type that endorses every closed defeat need not be a componentwise median. With finite support and positive orthant dependence at the componentwise median $(\overline{\tau}_1, \underline{\tau}_2, \dots, \underline{\tau}_k)$, that type has depth strictly above $2^{-k}$.} Protecting a given type is more demanding than core existence in a particular electorate: the binary menu above always has a core at any $q \ge 1/2$, and an off-median type can certify a lower quota (Supplemental Appendix~\ref{sec:supp-offmedian}).\footnote{In \citet{HoldenLu2026}, the large-electorate optimal quota for single-peaked voters is $\max\{F(\hat{x}), 1 - F(\hat{x})\} = 1 - h_\mu(\hat{x})$, and voters with depth $\ge 1 - q$ are immune to expropriation (central share $2q - 1$ for $k = 1$, but empty when $q < 1 - 2^{-k}$ with $k$ independent, continuously distributed dimensions).} Part~(ii) bounds one-sided cycles. Under positive association, at least a share $q^\ell$ of voters supports every step of an ascending chain of $\ell$ defeats, and by transitivity all of them oppose the closing defeat, so $q^\ell + q \le 1$.

\begin{corollary}\label{cor:observable}
Suppose $k = 2$ and $\mu$ places zero mass on $\{\theta_d = \tau_d\}$ ($d = 1, 2$) at a componentwise median $\theta^{\mathrm{med}} = (\tau_1, \tau_2)$ (e.g., continuous marginals). Letting $h \triangleq \mu(\theta_1 \le \tau_1,\; \theta_2 \le \tau_2)$, identically $\mu(\theta_1 \ge \tau_1,\; \theta_2 \ge \tau_2) = h = h_\mu(\theta^{\mathrm{med}})$. Without positive association, $1 - h$ is the exact open quota guaranteeing $\theta^{\mathrm{med}}$'s endorsement---and $M(\theta^{\mathrm{med}}) \subseteq C_{>q}(X, \mu)$---across all CW-SCP profiles; in particular, $\succ_{>1 - h}$ is acyclic on $X$.
\end{corollary}

\noindent Corollary~\ref{cor:observable} makes the quota observable. Measuring the share $h$ of voters below both medians (median income and median wealth, say) gives the exact quota $1 - h$ at which the median voter is protected: a third of voters below both medians gives two-thirds, perfect comonotonicity ($h = 1/2$) recovers the absolute-majority form of \citet{GansSmart1996}, and independence ($h = 1/4$) gives $3/4$. Equivalently, $h$ fixes the medial correlation of \citet{Blomqvist1950}, $\beta = 4h - 1$, so the quota is $(3 - \beta)/4$ and a two-thirds rule protects the median voter exactly when $\beta \ge 1/3$. Whether a lower quota suffices for core existence in a particular electorate requires further information, but Proposition~\ref{prop:non-pa}(ii) shows that no lower quota guarantees it from $h$ alone, or from positive orthant dependence at a median, across all electorates.

To test sharpness across independent electorates on $\mu = \operatorname{Unif}[0,1]^k$, define
\[
\begin{aligned}
q^*(k) \triangleq \sup\big\{q \in [1/2, 1) :{} & C_{>q}(X, \mu) = \emptyset \text{ for some nonempty finite } X \subset \mathbb{R} \\
& \text{and some CW-SCP profile on } X \times [0,1]^k\big\}.
\end{aligned}
\]
By pulling preferences back along coordinatewise marginal quantiles (Lemma~\ref{lem:supp-cube} in Supplemental Appendix~\ref{sec:supp-cube}), $q^*(k)$ is the exact worst-case threshold over all independent distributions (atoms included), and Theorem~\ref{thm:median-core}(i) gives $q^*(k) \le 1 - 2^{-k}$.\footnote{One-sided $(\ell+1)$-cycles cap at $r_\ell$ (and at $q_{\mathrm{hyp}}(2) \approx 0.682$ on $\operatorname{Unif}[0,1]^2$; Section~\ref{sec:highdim}), while four-alternative two-sided cycles cap at $2/3$ in every dimension (Remark~\ref{rem:supp-4cycle} in Supplemental Appendix~\ref{sec:supp-condorcet}).}

Theorem~\ref{thm:2d-sharp} lengthens the spending queue of Section~\ref{sec:spending-example} ($n = 2$, share $2/3$) to $2n - 1$ projects alternating between the two services ($2n$ budgets). Each two-step defeat fails only for households below both of its usage cutoffs (an ascent) or above both (a descent), a corner of the type square. Choosing the cutoffs so that all of these corners have the same area $p_n = 1 - q_n$ pins them down through the M\"obius recurrence $t_{h+1} = p_n/(1 - t_h)$, and $q_n$ rises toward $3/4$ as the queue lengthens.

\begin{theorem}\label{thm:2d-sharp}
For every integer $n \ge 2$, let $q_n \triangleq 1 - \big(4\cos^2(\frac{\pi}{2n+2})\big)^{-1}$. On $X_n = \{0, 1, \dots, 2n-1\}$ and $\Theta = [0, 1]^2$, there exists an additively separable (across $\theta_1$ and $\theta_2$), strictly increasing-differences utility $u(x, \theta)$ inducing strict linear orders with zero ties everywhere on $[0, 1]^2$ that, under $\mu = \operatorname{Unif}[0, 1]^2$, generates the alternating $2n$-cycle
\[
0 \leftarrow 2 \leftarrow 4 \leftarrow \dots \leftarrow (2n-2) \leftarrow (2n-1) \leftarrow (2n-3) \leftarrow \dots \leftarrow 3 \leftarrow 1 \leftarrow 0,
\]
with every directed defeat share equal to $q_n$. Consequently, $C_{>q}(X_n, \mu) = \emptyset$ for every $q < q_n$, and letting $n \to \infty$ gives $q_n \uparrow 3/4$, proving $q^*(2) = 3/4$.
\end{theorem}

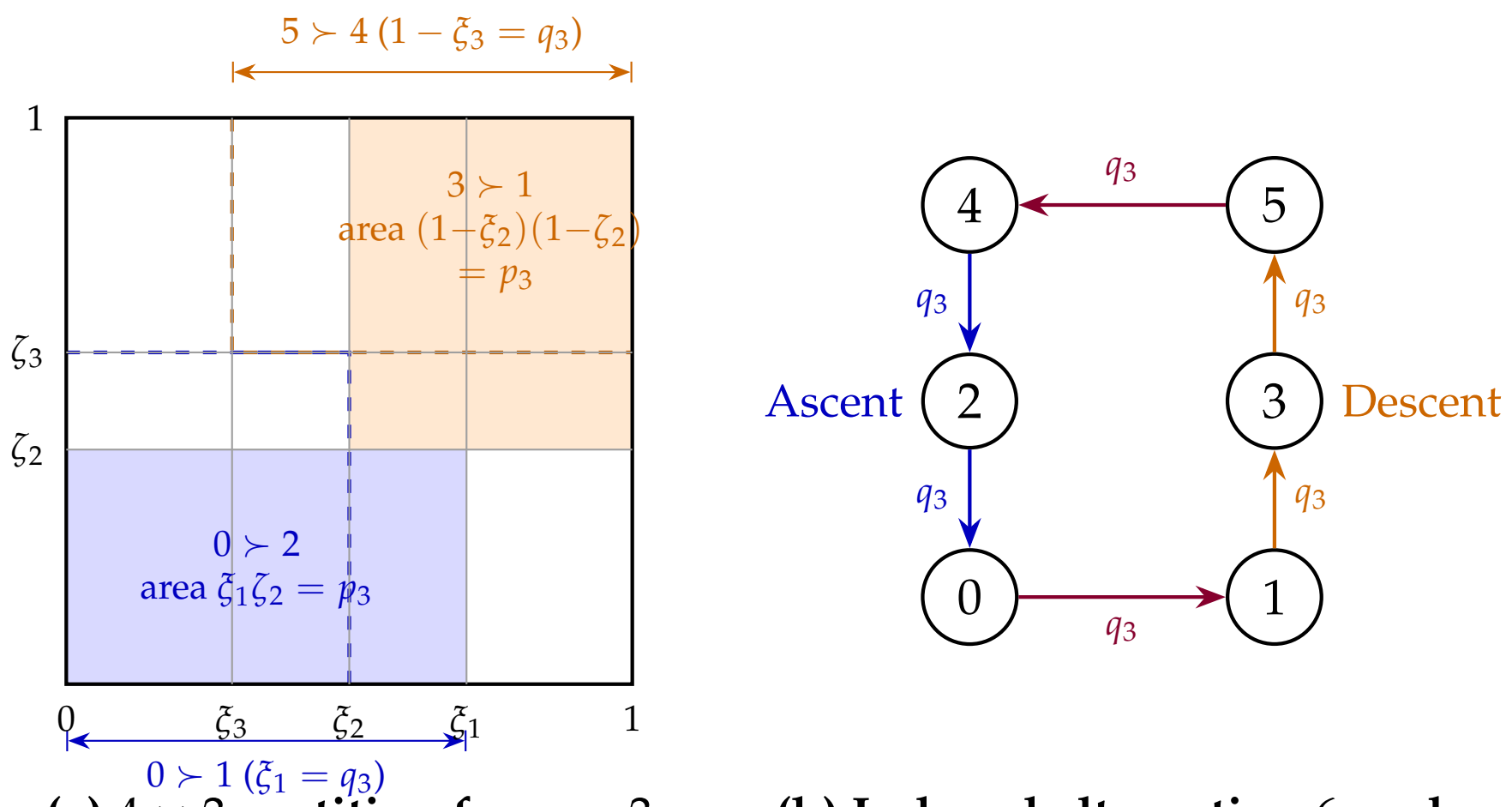
\begin{figure}[t]
\centering
\begin{tikzpicture}[scale=0.84, >=Stealth]
\def\S{5.2}
\def\xthree{1.5230}
\def\xtwo{2.6000}
\def\xone{3.6770}
\def\ztwo{2.1539}
\def\zthree{3.0461}

\begin{scope}[shift={(0,0)}]
  \fill[blue!15] (0,0) rectangle (\xone,\ztwo);
  \fill[orange!18] (\xtwo,\ztwo) rectangle (\S,\S);
  \draw[thick,blue!75!black,dashed] (0,\zthree) -- (\xtwo,\zthree) -- (\xtwo,0);
  \draw[thick,orange!75!black,dashed] (\xthree,\S) -- (\xthree,\zthree) -- (\S,\zthree);

  \draw[thick] (0,0) rectangle (\S,\S);
  \draw[gray!75] (\xthree,0) -- (\xthree,\S);
  \draw[gray!75] (\xtwo,0) -- (\xtwo,\S);
  \draw[gray!75] (\xone,0) -- (\xone,\S);
  \draw[gray!75] (0,\ztwo) -- (\S,\ztwo);
  \draw[gray!75] (0,\zthree) -- (\S,\zthree);

  \draw[|<->|,blue!75!black] (0,-0.52) -- (\xone,-0.52) node[midway,below,font=\scriptsize] {$0 \succ 1$ ($\xi_1 = q_3$)};
  \draw[|<->|,orange!80!black] (\xthree,\S+0.42) -- (\S,\S+0.42) node[midway,above,font=\scriptsize] {$5 \succ 4$ ($1-\xi_3 = q_3$)};

  \node[font=\scriptsize,blue!75!black,align=center] at (1.75,1.05) {$0 \succ 2$\\area $\xi_1 \zeta_2 = p_3$};
  \node[font=\scriptsize,orange!80!black,align=center] at (3.90,4.15) {$3 \succ 1$\\area $(1{-}\xi_2)(1{-}\zeta_2)$\\${}= p_3$};

  \node[below,font=\scriptsize] at (\xthree,-0.02) {$\xi_3$};
  \node[below,font=\scriptsize] at (\xtwo,-0.02) {$\xi_2$};
  \node[below,font=\scriptsize] at (\xone,-0.02) {$\xi_1$};
  \node[left,font=\scriptsize] at (-0.02,\ztwo) {$\zeta_2$};
  \node[left,font=\scriptsize] at (-0.02,\zthree) {$\zeta_3$};
  \node[below,font=\scriptsize] at (0,-0.02) {$0$};
  \node[below,font=\scriptsize] at (\S,-0.02) {$1$};
  \node[left,font=\scriptsize] at (-0.02,\S) {$1$};

  \node[font=\small\bfseries] at (\S/2,-1.25) {(a) $4 \times 3$ partition for $n=3$};
\end{scope}

\begin{scope}[shift={(7.4,0.3)}]
  \node[circle,draw,thick,minimum size=0.72cm,font=\small] (n0) at (0.9,0.5) {$0$};
  \node[circle,draw,thick,minimum size=0.72cm,font=\small] (n2) at (0.9,2.3) {$2$};
  \node[circle,draw,thick,minimum size=0.72cm,font=\small] (n4) at (0.9,4.1) {$4$};
  \node[circle,draw,thick,minimum size=0.72cm,font=\small] (n1) at (3.7,0.5) {$1$};
  \node[circle,draw,thick,minimum size=0.72cm,font=\small] (n3) at (3.7,2.3) {$3$};
  \node[circle,draw,thick,minimum size=0.72cm,font=\small] (n5) at (3.7,4.1) {$5$};

  \draw[->,thick,blue!75!black] (n2) -- (n0) node[midway,left,font=\scriptsize] {$q_3$};
  \draw[->,thick,blue!75!black] (n4) -- (n2) node[midway,left,font=\scriptsize] {$q_3$};
  \draw[->,thick,purple!70!black] (n5) -- (n4) node[midway,above,font=\scriptsize] {$q_3$};
  \draw[->,thick,orange!80!black] (n3) -- (n5) node[midway,right,font=\scriptsize] {$q_3$};
  \draw[->,thick,orange!80!black] (n1) -- (n3) node[midway,right,font=\scriptsize] {$q_3$};
  \draw[->,thick,purple!70!black] (n0) -- (n1) node[midway,below,font=\scriptsize] {$q_3$};

  \node[font=\footnotesize,blue!75!black] at (-0.35,2.3) {Ascent};
  \node[font=\footnotesize,orange!80!black] at (5.05,2.3) {Descent};
  \node[font=\small\bfseries] at (2.3,-1.55) {(b) Induced alternating $6$-cycle};
\end{scope}
\end{tikzpicture}
\caption{Two-dimensional alternating $2n$-cycle construction ($n = 3$, $m = 6$). \footnotesize Panel~(a): all ascent failures $[0, \xi_j] \times [0, \zeta_{j+1}]$ and descent failures $(\xi_j, 1] \times (\zeta_j, 1]$ have identical area $p_3 = 1 - 1/\sqrt{2}$, giving exact defeat share $q_3 = 1/\sqrt{2}$ on every edge in panel~(b). Arrows point from challenger to defeated policy.}
\label{fig:tightness-2d}
\end{figure}

\noindent Theorem~\ref{thm:2d-sharp} says that $3/4$ is exact in two dimensions: it is the quota at which the median voter is protected under independence (Theorem~\ref{thm:median-core}), and no lower quota guarantees a stable policy. Figure~\ref{fig:tightness-2d} illustrates $n = 3$ (six budgets, defeat share $1/\sqrt{2}$). The sequence $q_n \to 3/4$ is classical for cycles of $2n$ mutually independent random variables \citep{Chang1961,Usiskin1964,Trybula1965,Bogdanov2010,Komisarski2021,Rooney2025}; Theorem~\ref{thm:2d-sharp} realises it on one policy line in two dimensions, with transitive individual orders and strictly increasing differences. The step benefits are not essential: replacing each usage cutoff by a smooth increasing step confined to a narrow strip changes rankings only on the strips, so for any $q < q_n$ the cycle survives with utilities smooth in the type once the strips have total probability below $q_n - q$.\footnote{For $n = 2, 3, 4, 5$ ($m = 4, 6, 8, 10$), $q_n$ equals $2/3$, $1/\sqrt{2}$, $(5+\sqrt{5})/10$, and $\sqrt{3}-1$.}

\section{Higher Dimensions and Agenda-Size Limits}\label{sec:highdim}

When voter types span $k \ge 3$ dimensions, Theorem~\ref{thm:median-core} bounds $q^*(k) \le 1 - 2^{-k}$, while ignoring $k - 2$ coordinates embeds Theorem~\ref{thm:2d-sharp} to give $q^*(k) \ge 3/4$. Does $q^*(k)$ stay near $3/4$ or approach $1$ exponentially with $k$?

The answer separates alternating two-sided cycles from one-sided corner cycles. Define the Erlang volume function
\[
V_m(c) \triangleq \begin{cases}
1, & c \le 0, \\
1 - c \displaystyle\sum_{j=0}^{m-1} \frac{(-\ln c)^j}{j!}, & 0 < c < 1, \\
0, & c \ge 1,
\end{cases}
\]
which measures the volume $\mu(E_{m,c}) = \mu(P_{m,c}) = V_m(c)$ of both the lower hyperbolic complement $E_{m,c} \triangleq \{\theta \in [0,1]^m : \prod_{d=1}^m \theta_d > c\}$ and the upper hyperbolic complement $P_{m,c} \triangleq \{\theta \in [0,1]^m : \prod_{d=1}^m (1-\theta_d) > c\}$ under $\operatorname{Unif}[0,1]^m$ (writing $E_c \triangleq E_{k,c}$ and $P_c \triangleq P_{k,c}$ in fixed dimension $k$). For $k \ge 2$, let $c_k \in (0, 1/2)$ denote the unique root of $V_k(c) = c$, and set $q_{\mathrm{hyp}}(k) \triangleq 1 - c_k$.

\begin{theorem}\label{thm:highdim}
Suppose preferences satisfy CW-SCP on $X \times \Theta$ with $\Theta \subseteq \mathbb{R}^k$.
\begin{enumerate}[label=(\roman*),leftmargin=2.2em]
\item Under any positively associated $\mu$, in any dimension, every alternating $2n$-cycle ($n \ge 2$) ordered as in Theorem~\ref{thm:2d-sharp} has a defeat with share at most $q_n < 3/4$.
\item Under $\mu = \operatorname{Unif}[0, 1]^k$ with $k \ge 2$, the supremum over one-sided cycles of the smallest defeat share is $q_{\mathrm{hyp}}(k) = 1 - c_k$, where $c_k = e^{-k/e + o(k)}$. Hence $q^*(k) \ge q_{\mathrm{hyp}}(k)$, and $q_{\mathrm{hyp}}(k) > 3/4$ for every $k \ge 3$.
\end{enumerate}
\end{theorem}

\noindent Theorem~\ref{thm:highdim} separates two architectures. Part~(i) says that alternating cycles like those of Theorem~\ref{thm:2d-sharp} stay below $3/4$ under any positively associated measure in any dimension: under positive association, the equal-area recurrence that pins down the cutoffs of Theorem~\ref{thm:2d-sharp} becomes an inequality that every such cycle must satisfy.\footnote{Under the uniform distribution on the binary cube $\{0, 1\}^k$, a vertex with $j$ ones has $h_\mu(\theta) = 2^{-\max(j, k-j)}$, so by Theorem~\ref{thm:median-core}(i) a balanced vertex, with $\lfloor k/2 \rfloor$ or $\lceil k/2 \rceil$ ones, has its ideal points protected at $1 - 2^{-\lceil k/2 \rceil}$, which is $3/4$ for $k = 3, 4$ and $7/8$ for $k = 5, 6$. Among all four-alternative cycles under positive association, only one-sided $4$-cycles exceed $2/3$, up to $r_3 \approx 0.682$ (Remark~\ref{rem:supp-4cycle} in Supplemental Appendix~\ref{sec:supp-condorcet}).} Part~(ii) says that one-sided cycles do better once $k \ge 3$. In its construction, each policy loses to its lower neighbour among all voters outside a small box of high types, while the top policy beats the bottom one among voters in any box. Enough boxes, each high in a different combination of dimensions, cover the hyperbolic corner $K^+_{c'} \triangleq [0,1]^k \setminus P_{c'}$ (reflected, lower boxes cover $K^-_{c'} \triangleq [0,1]^k \setminus E_{c'}$), whose complement shrinks exponentially in $k$. In other words, almost every voter is high in some combination of dimensions.

Agenda size limits this divergence. On any fixed menu of $m$ proposals, the open rule at $1 - 1/m$ is acyclic in every dimension \citep{FerejohnGrether1974,Nakamura1979}, because each transitive voter must dissent from at least one edge of any cycle. Under CW-SCP with positive association, three proposals cannot cycle at $(\sqrt{5}-1)/2$ (Theorem~\ref{thm:median-core}(ii)) and one-sided four-policy cycles cap at $r_3$, whereas six budgets with two independent dimensions reach $1/\sqrt{2}$ (Theorem~\ref{thm:2d-sharp}). A two-thirds rule thus protects every three-proposal agenda in any dimension but can fail on six proposals even in two dimensions. With one more proposal than there are dimensions, a one-sided cycle in which each step up is supported only by voters high in one dimension, and the jump from the bottom to the top by voters high in any dimension, has every defeat share equal to $r_k$, the root in $(1/2, 1)$ of $z^k + z = 1$; since $1 - r_k \sim (\ln k)/k$, the quota needed approaches unanimity even when the agenda grows only linearly with the number of dimensions (Appendix~\ref{app:A}). Approaching $q_{\mathrm{hyp}}(k)$ requires exponentially growing agendas---$\exp(O(k))$ alternatives for any fixed rate $\alpha < 1/e$, and $\exp(O(k \log k))$ to come within $c_k$ of $q_{\mathrm{hyp}}(k)$---and uses nonlinear, non-additive benefits. Figure~\ref{fig:dim-divergence} plots the main-text bounds; Proposition~\ref{prop:supp-smooth}(iii) in Supplemental Appendix~\ref{sec:supp-smooth} lifts $r_k$ and $q_{\mathrm{hyp}}(k)$ to $C^\infty$ utilities on $X = [0, 1]$, and Theorem~\ref{thm:supp-bridge} in Supplemental Appendix~\ref{sec:supp-bridges} raises the independent-type lower bound for $k \ge 3$ on very large agendas (Table~\ref{tab:supp-dim-bounds}).

Limiting how far an amendment may move the policy has a similar effect. Order a finite menu $X = \{x_0 < x_1 < \dots < x_{m-1}\}$ and, under an \emph{$r$-step amendment rule}, let $x_j$ challenge $x_i$ only when $|i - j| \le r$; write $C^r_{>q}(X, \mu)$ for the policies that no permitted challenger $q$-defeats (and $C^r_{\ge q}(X, \mu)$ under the closed rule). In the spending queue, $r$ is the number of projects an amendment may add or remove, and the footnote to Section~\ref{sec:spending-example} is the case $r = 1$.

\begin{proposition}\label{prop:amend}
Suppose preferences satisfy CW-SCP on $X \times \Theta$ with $X = \{x_0 < \dots < x_{m-1}\}$ and $m \ge 3$.
\begin{enumerate}[label=(\roman*),leftmargin=2.2em]
\item Under the one-step rule, $C^1_{>q}(X, \mu)$ is nonempty for every $q \ge 1/2$ and every $\mu$, and if $\mu' \succeq_{\mathrm{FOSD}} \mu$, then $\min C^1_{>q}(X, \mu) \le \min C^1_{>q}(X, \mu')$ and $\max C^1_{>q}(X, \mu) \le \max C^1_{>q}(X, \mu')$; in particular, a singleton cannot fall. The same holds for $C^1_{\ge q}$ when $q > 1/2$.
\item Under the two-step rule, let $Q_m \triangleq 1 - \big(4\cos^2(\frac{\pi}{m+2})\big)^{-1}$. For every positively associated $\mu$ in any dimension, permitted open $Q_m$-defeats are acyclic, so $C^2_{>q}(X, \mu) \neq \emptyset$ for every $q \ge Q_m$. The bound is exact: for every $q < Q_m$ there is a strictly increasing-differences utility on $X \times [0, 1]^2$ with $C^2_{>q}(X, \operatorname{Unif}[0, 1]^2) = \emptyset$. Moreover, for each $q \in [1/2, 3/4)$ there exist a finite menu and independent electorates $\mu' \succeq_{\mathrm{FOSD}} \mu$, as in Theorem~\ref{thm:phase}(ii), under which the singleton open $q$-core falls under the two-step rule.
\end{enumerate}
\end{proposition}

\noindent The two-step quota $Q_m$ rises from $(\sqrt{5}-1)/2$ at $m = 3$ through $2/3$, $0.692$ and $1/\sqrt{2}$ at $m = 4, 5, 6$ toward $3/4$, and equals $q_{m/2}$ of Theorem~\ref{thm:2d-sharp} for even $m$. Part~(i) says that when amendments move one project at a time, the comparisons form a path, so the upward shift of Section~\ref{sec:fosd} can eliminate an intermediate stable budget but cannot lower the smallest or largest one. Part~(ii) says that allowing two projects at a time restores the full two-dimensional problem and no more: the only cycles are the alternating cycles of Theorem~\ref{thm:2d-sharp} and their odd-length counterparts, so with positively associated types more than three-quarters suffices in every dimension, and the divergence toward unanimity in Theorem~\ref{thm:highdim} needs amendments that move policy further. Without positive association the two-step bound fails: five equally weighted incomparable types holding the five rotations of $0, 2, 4, 3, 1$, extended off their support as in Proposition~\ref{prop:non-pa}, give a cycle of permitted amendments each winning four votes to one. How the quota behaves when amendments may move three or more projects is open. The range of an amendment is thus as consequential as the number of voter characteristics.

\begin{figure}[t]
\centering
\begin{tikzpicture}[scale=0.84, >=Stealth]
\draw[->] (0,0) -- (11.6,0) node[right,font=\footnotesize] {Dimension $k$};
\draw[->] (0,0) -- (0,6.4) node[above,font=\footnotesize] {Quota $q$};

\foreach \k/\kx in {2/0.6, 3/2.6, 4/4.6, 5/6.6, 6/8.6, 7/10.6} {
  \draw[gray!25,very thin] (\kx,0) -- (\kx,6.1);
  \draw (\kx,0) -- (\kx,-0.1) node[below,font=\footnotesize] {$\k$};
}
\foreach \q/\qy in {0.55/0.6, 0.65/1.8, 0.75/3.0, 0.85/4.2, 0.95/5.4, 1.00/6.0} {
  \draw[gray!25,very thin] (0,\qy) -- (11.2,\qy);
  \draw (0,\qy) -- (-0.1,\qy) node[left,font=\footnotesize] {$\q$};
}

\fill[blue!10]
  (2.6, 3.4988) -- (4.6, 4.3368) -- (6.6, 4.8851) -- (8.6, 5.2488) -- (10.6, 5.4919) --
  (10.6, 5.9063) -- (8.6, 5.8125) -- (6.6, 5.6250) -- (4.6, 5.2500) -- (2.6, 4.5000) -- cycle;

\draw[dashed,gray!70] (0,1.585) -- (11.2,1.585) node[right,font=\scriptsize] {$1 - 1/e$ (CN log-conc.)};
\draw[dashed,purple!70!black] (0,3.000) -- (11.2,3.000) node[right,font=\scriptsize] {$3/4$ (alt.\ ceiling)};
\draw[dotted,gray!70] (0,6.000) -- (11.2,6.000);

\draw[thick,dotted,gray!80!black]
  (0.6,0.6667) -- (2.6,0.9375) -- (4.6,1.0848) -- (6.6,1.1775) -- (8.6,1.2412) -- (10.6,1.2876);

\draw[thick,teal!70!black,dashed]
  (0.6,1.4164) -- (2.6,2.1879) -- (4.6,2.6939) -- (6.6,3.0585) -- (8.6,3.3371) -- (10.6,3.5585);

\draw[thick,orange!80!black]
  (0.6,2.1859) -- (2.6,3.4988) -- (4.6,4.3368) -- (6.6,4.8851) -- (8.6,5.2488) -- (10.6,5.4919);

\draw[very thick,black]
  (0.6,3.0000) -- (2.6,4.5000) -- (4.6,5.2500) -- (6.6,5.6250) -- (8.6,5.8125) -- (10.6,5.9063);

\foreach \pt in {(0.6,3.0000),(2.6,4.5000),(4.6,5.2500),(6.6,5.6250),(8.6,5.8125),(10.6,5.9063)} {
  \fill[black] \pt circle (1.9pt);
}
\foreach \pt in {(0.6,2.1859),(2.6,3.4988),(4.6,4.3368),(6.6,4.8851),(8.6,5.2488),(10.6,5.4919)} {
  \fill[orange!80!black] \pt circle (1.7pt);
}
\foreach \pt in {(0.6,0.6667),(2.6,0.9375),(4.6,1.0848),(6.6,1.1775),(8.6,1.2412),(10.6,1.2876)} {
  \fill[gray!80!black] \pt circle (1.5pt);
}
\draw[black,thick] (0.6,3.0000) circle (3.5pt);
\node[font=\scriptsize,anchor=south] (qtwo) at (1.00,4.35) {$q^*(2)=3/4$};
\draw[gray!70!black,thin] (qtwo.south) -- (0.66,3.12);

\draw[draw=gray!50] (0.00,-2.15) rectangle (11.45,-0.80);
\draw[very thick,black] (0.25,-1.05) -- (0.95,-1.05) node[right,font=\scriptsize] {Pos.\ assoc.\ upper bound $1 - 2^{-k}$};
\draw[thick,orange!80!black] (0.25,-1.47) -- (0.95,-1.47) node[right,font=\scriptsize] {One-sided hyperbolic $q_{\mathrm{hyp}}(k)$};
\draw[thick,teal!70!black,dashed] (0.25,-1.89) -- (0.95,-1.89) node[right,font=\scriptsize] {$(k+1)$-cycle root $r_k$};
\draw[thick,dotted,gray!80!black] (5.90,-1.05) -- (6.60,-1.05) node[right,font=\scriptsize] {CN (uniform) $1{-}(k/(k{+}1))^k$};
\fill[blue!10] (5.90,-1.59) rectangle (6.60,-1.35);
\node[right,font=\scriptsize] at (6.60,-1.47) {Bracket $[q_{\mathrm{hyp}}(k),\, 1{-}2^{-k}]$, $k \ge 3$};
\end{tikzpicture}
\caption{Supermajority bounds on $\operatorname{Unif}[0,1]^k$ ($k = 2, \dots, 7$). \footnotesize The hyperbolic bound $q_{\mathrm{hyp}}(k)$ converges exponentially toward $1$ ($1 - q = e^{-k/e + o(k)}$), $(k+1)$-cycles ($r_k$) at rate $(\ln k)/k$, Caplin--Nalebuff linear cuts stay below $1 - 1/e$, and alternating cycles stay below $3/4$. Shaded band: $[q_{\mathrm{hyp}}(k), 1 - 2^{-k}]$ ($k \ge 3$); $q^*(2) = 3/4$ (Table~\ref{tab:supp-dim-bounds} reports multi-stage refinements). The Caplin--Nalebuff line is a worst case over convex bodies; on $\operatorname{Unif}[0,1]^k$ itself, central symmetry protects the centre against linear cuts at $1/2$ \citep{Grandmont1978}.}
\label{fig:dim-divergence}
\end{figure}
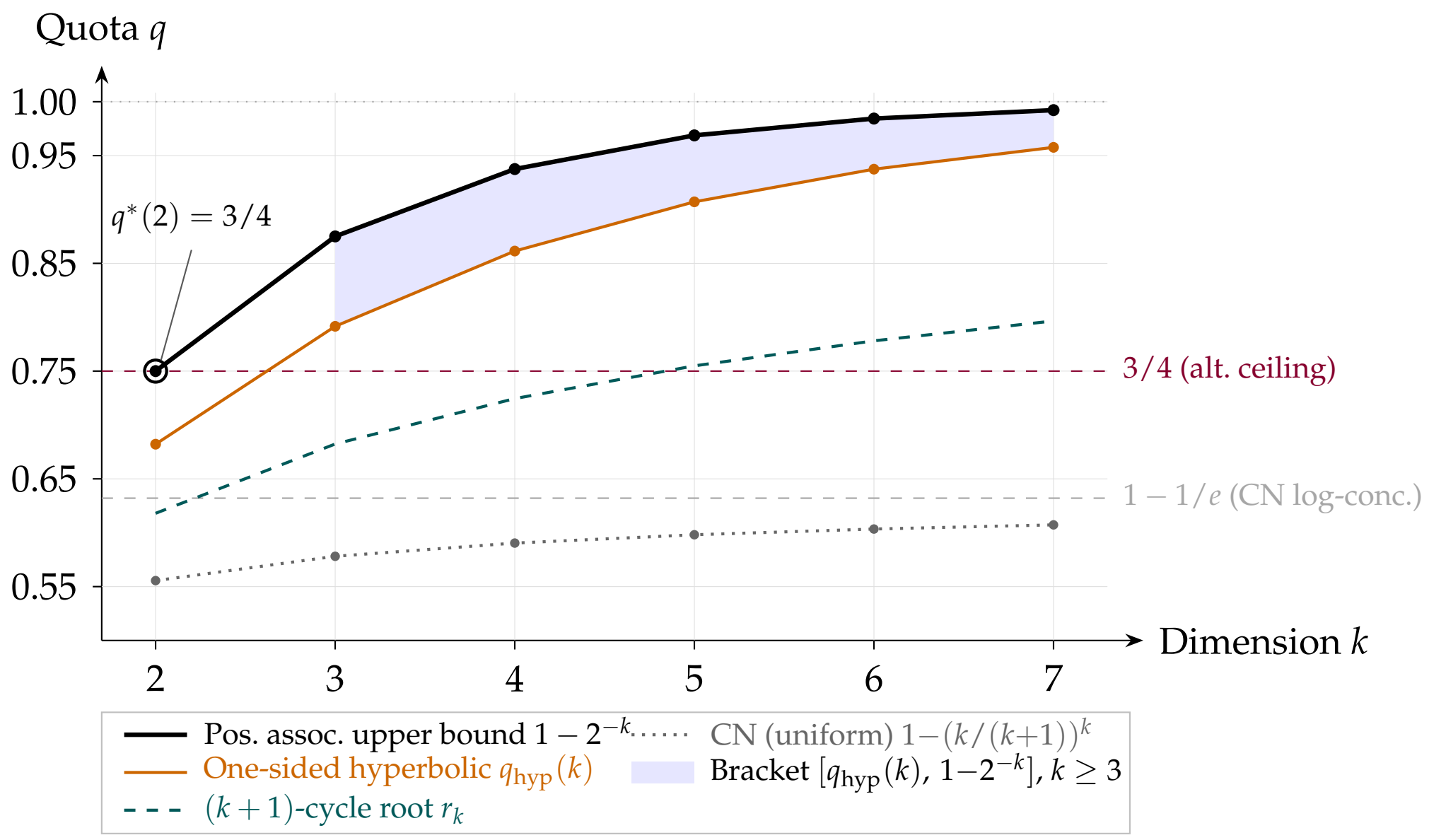

\section{Multivariate Stochastic Dominance and the \texorpdfstring{$3/4$}{3/4} Phase Transition}\label{sec:fosd}

In one dimension, shifting the electorate up moves the median voter's ideal point up. With several dimensions, we ask how an upward shift in multivariate first-order stochastic dominance ($\mu' \succeq_{\mathrm{FOSD}} \mu$) moves the stable policy set. Recall that on a totally ordered set $X$, two nonempty subsets $A, B \subseteq X$ satisfy Veinott's \emph{strong set order} ($A \le_{\mathrm{SSO}} B$) if $a \in A$ and $b \in B$ imply $\min(a, b) \in A$ and $\max(a, b) \in B$, and \emph{weak set order} ($A \le_{\mathrm{WSO}} B$) if $\min A \le \min B$ and $\max A \le \max B$ (when compact). For any measure $\mu$, let $\mathcal{M}(y, x; \mu) \triangleq \mu(y \succ x) - \mu(x \succ y)$ denote the net vote margin, and for any quota $q \in [1/2, 1]$ (under $> q$ or $\ge q$), let $N_\uparrow(\mu) \triangleq \{x \in X : \text{no } y > x \text{ } q\text{-defeats } x\}$ and $N_\downarrow(\mu) \triangleq \{x \in X : \text{no } y < x \text{ } q\text{-defeats } x\}$ denote the upward and downward undefeated sets.

Proposition~\ref{prop:fosd} collects the positive comparative-statics properties that survive universally under CW-SCP.

\begin{proposition}\label{prop:fosd}
Suppose preferences satisfy CW-SCP on $X \times \Theta$ and $\mu' \succeq_{\mathrm{FOSD}} \mu$.
\begin{enumerate}[label=(\roman*),leftmargin=2.2em]
\item For all $y > x$ in $X$, $\mu'(y \succ x) \ge \mu(y \succ x)$ and $\mu'(x \succ y) \le \mu(x \succ y)$, and likewise for weak preference.
\item If Condorcet winners $x^*$ under $\mu$ and $x^{**}$ under $\mu'$ exist under the same convention and one of them is strict, then $x^{**} \ge x^*$.
\item If $\theta \le \theta'$, then $M(\theta) \le_{\mathrm{SSO}} M(\theta')$ whenever both are nonempty. Hence, on a finite menu, $M(\underline{\theta}^{\mathrm{med}}(\mu)) \le_{\mathrm{SSO}} M(\underline{\theta}^{\mathrm{med}}(\mu'))$ and $M(\overline{\theta}^{\mathrm{med}}(\mu)) \le_{\mathrm{SSO}} M(\overline{\theta}^{\mathrm{med}}(\mu'))$, and for this it suffices that $\mu'$ dominate $\mu$ marginally.
\end{enumerate}
\end{proposition}

\noindent Intuitively, Lemma~\ref{lem:structure}(i) makes every upward challenge $U^>(y, x)$ an upper set, so multivariate FOSD weakly raises $\mu(y \succ x)$ for all $y > x$. Net margins $\mathcal{M}(y, x; \cdot)$ therefore rise as well, upward defeats survive the shift and downward defeats can only disappear: $N_\uparrow(\mu') \subseteq N_\uparrow(\mu)$ and $N_\downarrow(\mu) \subseteq N_\downarrow(\mu')$ at every quota, open or closed. That monotonicity shields policies from downward elimination while strengthening upward challengers, preserving ordered Condorcet winners and monotone median ideal sets, yet allowing a higher challenger to knock out a previously stable intermediate policy while a lower policy escapes downward defeat, the mechanism behind the budget reversal of Section~\ref{sec:spending-example}. Under positive orthant dependence at the componentwise medians, the median ideal sets of part~(iii) lie in the open $(1 - 2^{-k})$-core by Theorem~\ref{thm:median-core}(i).\footnote{Part~(iii) also holds on a compact menu with upper semicontinuous preferences, which keeps ideal sets nonempty.}

Theorem~\ref{thm:phase} identifies the quotas at which that reversal can occur.

\begin{theorem}\label{thm:phase}
Consider finite policy menus $X \subset \mathbb{R}$ and CW-SCP preferences on $\Theta$.
\begin{enumerate}[label=(\roman*),leftmargin=2.2em]
\item If $q \ge 1 - 2^{-k}$ and each electorate is positively orthant dependent at its lower componentwise median, then $\min M(\underline{\theta}^{\mathrm{med}}(\mu))$ and $\max M(\underline{\theta}^{\mathrm{med}}(\mu))$ are open $q$-core selections that weakly increase under marginal stochastic dominance. In particular, a singleton open $q$-core cannot fall.
\item Let $k = 2$. For every $q \in [1/2, 3/4)$, there exist a menu, a utility with strictly increasing differences, and independent electorates $\mu' \succeq_{\mathrm{FOSD}} \mu$ differing in one marginal, under which the open $q$-core is a singleton under each electorate and falls strictly from $\mu$ to $\mu'$; for $q > 1/2$, the closed $q$-core does the same.
\item Without positive association, an upward shift $\mu' \succeq_{\mathrm{FOSD}} \mu$ can lower singleton open and closed $q$-cores at every $q \in (1/2, 1)$, already with $k = 2$ and a utility with strictly increasing differences.
\end{enumerate}
\end{theorem}

\begin{figure}[t]
\centering
\begin{tikzpicture}[scale=0.84, >=Stealth]
\begin{scope}[shift={(0,3.30)}]
  \node[font=\small\bfseries,anchor=west] at (0,1.88) {(a) $q$-core singleton inversion regimes ($k = 2$)};

  \node[font=\footnotesize,anchor=east,align=right] at (2.15,1.08) {Positively\\associated};
  \fill[red!14] (2.3,0.72) rectangle (6.7,1.44);
  \fill[teal!16] (6.7,0.72) rectangle (11.1,1.44);
  \draw[thick] (2.3,0.72) rectangle (11.1,1.44);
  \draw[very thick,black] (6.7,0.62) -- (6.7,1.54);
  \node[font=\scriptsize,red!75!black,align=center] at (4.5,1.08) {Empty cores \& singleton\\inversions possible};
  \node[font=\scriptsize,teal!75!black,align=center] at (8.9,1.08) {Finite-menu core nonempty;\\no singleton inversion};

  \node[font=\footnotesize,anchor=east,align=right] at (2.15,0.05) {Without pos.\\association};
  \fill[red!14] (2.3,-0.22) rectangle (11.1,0.32);
  \draw[thick] (2.3,-0.22) rectangle (11.1,0.32);
  \node[font=\scriptsize,red!75!black] at (6.7,0.05) {Correlated cyclic singleton inversions for all $q \in (1/2, 1)$};

  \draw[->] (2.2,-0.62) -- (11.45,-0.62) node[right,font=\footnotesize] {$q$};
  \foreach \x/\lbl in {2.3/{$1/2$}, 6.7/{$3/4$}, 11.1/{$1$}} {
    \draw (\x,-0.55) -- (\x,-0.69) node[below,font=\scriptsize] {\lbl};
  }
\end{scope}

\begin{scope}[shift={(0.6,-1.65)}]
  \node[font=\small\bfseries,anchor=west] at (-0.6,2.85) {(b) Single-coordinate density shift ($\mu_1^- \preceq_{\mathrm{FOSD}} \operatorname{Unif}[0,1] \preceq_{\mathrm{FOSD}} \mu_1^+$)};

  \draw[->] (0,0) -- (10.5,0) node[right,font=\footnotesize] {$\theta_1$};
  \draw[->] (0,0) -- (0,2.25) node[left,font=\footnotesize] {Density};

  \draw[dashed,gray!70] (0,1.15) -- (10,1.15) node[right,font=\scriptsize] {$1$};

  \fill[blue!15] (0,0) rectangle (2.5,1.50);
  \fill[blue!8] (2.5,0) rectangle (3.5,0.35);
  \draw[thick,blue!75!black] (0,1.50) -- (2.5,1.50) -- (2.5,0.35) -- (3.5,0.35) -- (3.5,1.15);

  \fill[orange!10] (6.5,0) rectangle (7.5,0.35);
  \fill[orange!20] (7.5,0) rectangle (10.0,1.50);
  \draw[thick,orange!80!black] (6.5,1.15) -- (6.5,0.35) -- (7.5,0.35) -- (7.5,1.50) -- (10.0,1.50);

  \draw[thick,gray!80!black] (3.5,1.15) -- (6.5,1.15);

  \draw[->,thick,blue!75!black] (3.0,1.35) to[out=135,in=45] node[above,font=\scriptsize] {$\varepsilon$ down (protects $2n-2$)} (1.25,1.65);
  \draw[->,thick,orange!80!black] (7.0,1.35) to[out=45,in=135] node[above,font=\scriptsize] {$\varepsilon$ up (protects $1$)} (8.75,1.65);

  \foreach \x/\lbl in {0/{$0$}, 2.5/{$\xi_n$}, 3.5/{$\xi_{n-1}$}, 6.5/{$\xi_2$}, 7.5/{$\xi_1$}, 10/{$1$}} {
    \draw (\x,0) -- (\x,-0.08) node[below,font=\scriptsize] {\lbl};
  }
\end{scope}
\end{tikzpicture}
\caption{Multivariate FOSD phase transition in $k = 2$ (Theorem~\ref{thm:phase}). \footnotesize Panel~(a): open-threshold boundary at $q = 3/4$ under positive association versus unrestricted correlated inversions on $(1/2, 1)$. Panel~(b): single-coordinate shifts $\mu = \mu_1^- \otimes \operatorname{Unif}[0,1]$ (blue) and $\mu' = \mu_1^+ \otimes \operatorname{Unif}[0,1]$ (orange) inverting $\{2n-2\}$ down to $\{1\}$ ($n = 2$ gives Table~\ref{tab:four-budget}).}
\label{fig:fosd-transition}
\end{figure}
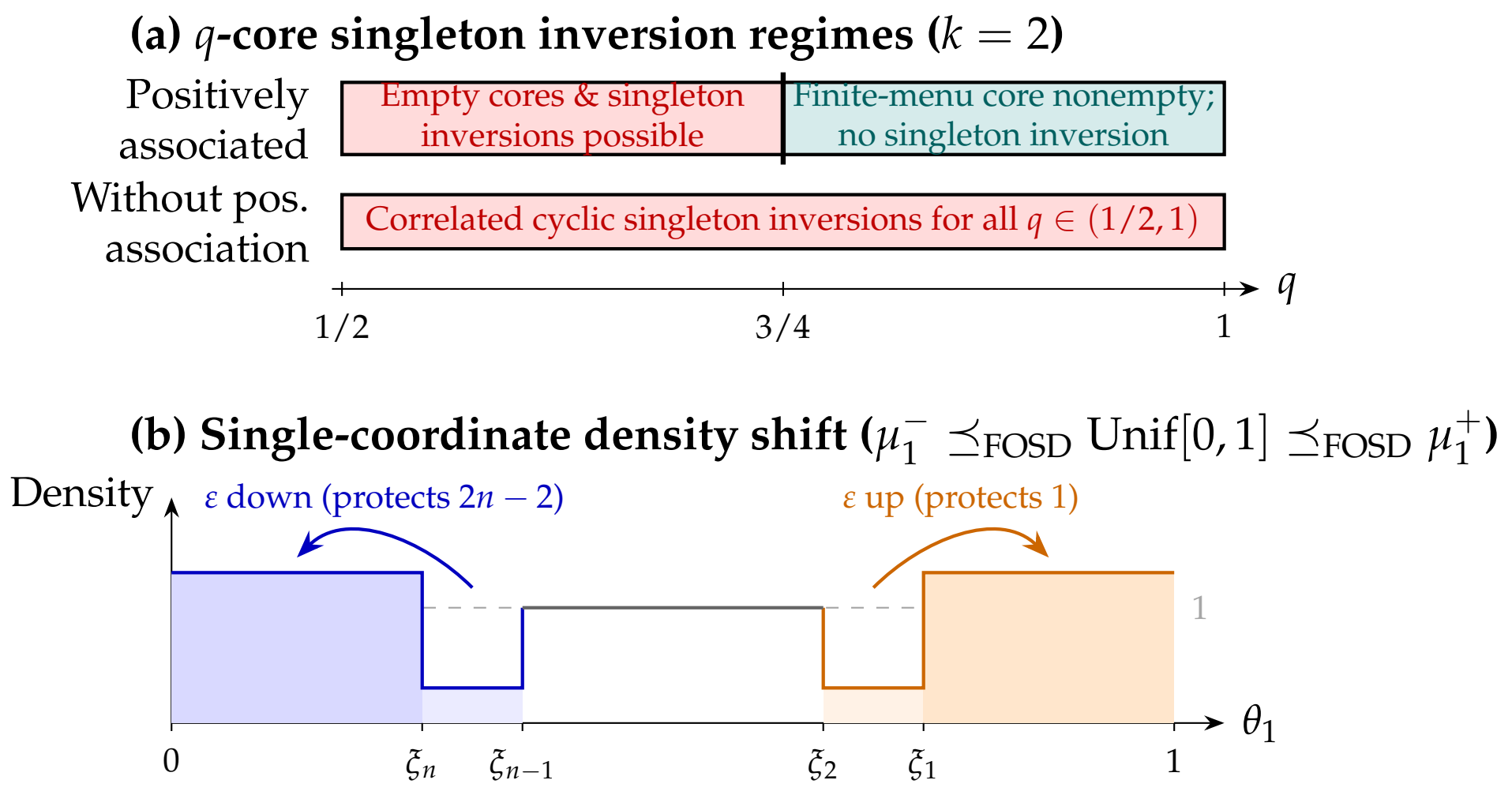

\noindent Theorem~\ref{thm:phase} says that $3/4$ is also where comparative statics change in two dimensions. At or above it, under positive orthant dependence at lower medians, the median's ideal set lies in $C_{>q}$ and moves up with the electorate (Proposition~\ref{prop:fosd}(iii)), so its extremal points are monotone selections and a singleton core cannot fall. The same argument (via Theorem~\ref{thm:median-core}(i) and Proposition~\ref{prop:fosd}(iii)) orders singleton open $q$-cores under any two electorates, ordered by dominance or not, whenever representative types $t \le t'$ satisfy $q \ge 1 - h_\mu(t)$ and $q \ge 1 - h_{\mu'}(t')$. Below three-quarters, holding Theorem~\ref{thm:2d-sharp}'s spending utility fixed, a single-coordinate shift generalising Table~\ref{tab:four-budget} protects $2n - 2$ before the shift and $1$ after it, inverting $\{2n - 2\}$ down to $\{1\}$ (Figure~\ref{fig:fosd-transition}).\footnote{Two qualifications apply. First, $q = 3/4$ governs median-based selections and \emph{singleton} open cores rather than the full set-valued core: Supplemental Appendix~\ref{sec:supp-setcore} constructs independent $3$-policy FOSD shifts shrinking $C_{>q}$ from $\{0, 1\}$ to $\{0\}$ and expanding it from $\{2\}$ to $\{1, 2\}$ for every $q \in (1/2, 1)$, violating $\le_{\mathrm{SSO}}$ and $\le_{\mathrm{WSO}}$. Second, at the closed boundary $q = 3/4$, Proposition~\ref{prop:supp-closed} (Supplemental Appendix~\ref{sec:supp-closed}) proves existence and acyclicity on finite menus for all positively quadrant dependent two-dimensional electorates, including all positively associated ones, singleton non-inversion for all independent electorates in every $k \ge 2$, and monotone closed-core selections for finite-support electorates that are positively orthant dependent at the componentwise median $(\overline{\tau}_1, \underline{\tau}_2, \dots, \underline{\tau}_k)$; only singleton closed-$3/4$ comparison across general positively quadrant dependent electorates remains open.} Hence no selection from the open $q$-core with $q \in [1/2, 3/4)$, or from the closed $q$-core with $q \in (1/2, 3/4)$, is monotone, even for independent electorates with nonempty cores.\footnote{At $q = 1/2$ the menu has four budgets, and budget~$2$ survives its contests with budgets~$1$ and~$3$ only as exact ties, so the shift moves the unique weak Condorcet winner from $2$ to $1$.} In other words, at or above three-quarters the median voter's preferred policy is stable and moves in the same direction as the electorate, although other stable policies need not; below three-quarters, neither a stable policy nor monotonicity is guaranteed. Under the one-step amendment rule, by contrast, no such reversal occurs at any quota (Proposition~\ref{prop:amend}(i)). Under simple majority, a strict Condorcet winner can also vanish into a cycle after an upward shift; Supplemental Appendix~\ref{sec:supp-tournaments} analyses tournament solutions (Smith, Copeland, and Minimax). Table~\ref{tab:fosd-summary} summarises these results.

\begin{table}[t]
\centering
\caption{Multivariate FOSD Comparative Statics under CW-SCP ($k \ge 2$)}
\label{tab:fosd-summary}
\small
\begin{tabular}{>{\raggedright\arraybackslash}p{4.3cm}>{\raggedright\arraybackslash}p{4.3cm}>{\raggedright\arraybackslash}p{5.3cm}}
\toprule
Solution / Statistic & Monotonicity Status & Governing Result \\
\midrule
Upward shares \& $N_\uparrow, N_\downarrow$ & Weakly increasing / nested & Always (Lemma~\ref{lem:structure}(ii), Prop.~\ref{prop:fosd}(i)) \\
Condorcet winners (one strict) & Ordered ($x^{**} \ge x^*$) & When both exist; may vanish (Prop.~\ref{prop:fosd}(ii), Supp.\ App.~\ref{sec:supp-tournaments}) \\
Median ideal sets & Strong Set Order & Marginal FOSD; in $(1{-}2^{-k})$-core under orthant dep.\ (Prop.~\ref{prop:fosd}(iii)) \\
Open $q$-core selection \& singleton open $q$-core & Monotone for $q \ge 1{-}2^{-k}$ (pos.\ assoc.); inverts below $3/4$ ($k=2$) & Exact $3/4$ phase transition (Thm.~\ref{thm:phase}(i)--(ii), Table~\ref{tab:four-budget}) \\
Closed $3/4$-core ($k = 2$) & Nonempty \& acyclic (PQD); singletons ordered (indep.); monotone selections (finite support) & General dependent comparison open (Supp.\ App.~\ref{sec:supp-closed}) \\
Set-valued open $q$-core & Neither $\min$ nor $\max$ monotone & Independent $3$-policy shifts on $(1/2, 1)$ (Supp.\ App.~\ref{sec:supp-setcore}) \\
Singleton $q$-core (no pos.\ assoc.) & Inverts for all $q \in (1/2, 1)$ & Anti-diagonal shift (Thm.~\ref{thm:phase}(iii)) \\
\bottomrule
\end{tabular}
\par\medskip
\begin{minipage}{0.95\linewidth}
\footnotesize\emph{Notes:} Supplemental Appendix~\ref{sec:supp-tournaments} treats simple-majority tournament solutions (Smith non-reversal and $\min/\max$ failures, plus Copeland and Minimax singleton inversions on $\{0,1\}^2$).
\end{minipage}
\end{table}

\section{Concluding Remarks}\label{sec:conclusion}

With a single policy and several dimensions of voter type, single-crossing still protects a representative voter, but only through the voters who lie below that voter in every dimension or above them in every dimension. One minus the smaller of these shares is exactly the quota that protects the voter. With two independent dimensions it is three-quarters, and at or above it the median voter's preferred policy also moves with the electorate, so a unique stable policy cannot fall, although other stable policies can. The supermajority bounds of \citet{CaplinNalebuff1988,CaplinNalebuff1991} and \citet{WuWu2026} require the voters who prefer one policy to another to form a half-space or a convex set. Single-crossing in each dimension allows any monotone coalition, so the quota that is safe with independent dimensions rises toward unanimity as dimensions are added, and without a restriction on how the dimensions covary no quota below unanimity is safe even in two dimensions.

The divergence toward unanimity is a worst case, and it needs agendas that lengthen with the number of dimensions, because a menu of $m$ proposals cannot cycle at $1 - 1/m$, and amendments that move policy far: if each amendment may add or remove at most two projects, more than three-quarters suffices in every dimension for positively associated types. With $k$ dimensions, $k + 1$ proposals already bring the required quota within order $(\ln k)/k$ of unanimity; only the exponential rate needs agendas that grow exponentially with $k$. The worst-case constructions use preferences that are nonlinear in the type: with affine utilities on the uniform cube, or with single-peaked preferences, simple majority is stable. The positive results also apply issue by issue: when a multidimensional policy is decided one instrument at a time, as in structure-induced equilibria, each issue whose conditional preferences satisfy CW-SCP inherits half-region-depth endorsement and upward vote-share monotonicity.\footnote{If an issue reverses orientation along some dimension, re-orienting that coordinate preserves CW-SCP while flipping the corresponding orthant and dominance direction.} And the quota is observable: with two continuously distributed dimensions, such as income and wealth, the share of voters below both medians fixes the quota at which the median voter's preferred policy is safe.

Three open problems remain: whether, for $k \ge 3$ and independent or positively associated electorates, a unique stable policy can fall after an upward shift at quotas in $[3/4, 1 - 2^{-k})$, since the reversals of Theorem~\ref{thm:phase}(ii) embed in any dimension but Theorem~\ref{thm:phase}(i) rules them out only from $1 - 2^{-k}$; whether singleton closed $3/4$-cores are ordered across general non-finite-support positively quadrant dependent electorates in $k = 2$ (Proposition~\ref{prop:supp-closed} settles existence, acyclicity, independent electorates, and finite-support marginals in all $k \ge 2$), and the exact value of $q^*(k)$ for independent types when $k \ge 3$, which lies between the multi-stage lower bounds of Supplemental Appendix~\ref{sec:supp-bridges} and $1 - 2^{-k}$, including whether multi-block cycles can surpass the single-bridge ceiling $q_{\mathrm{sc}}(k)$.\footnote{Approximating $\operatorname{Unif}[0, 1]^k$ closely requires fine discretisations rather than $\{0, 1\}^k$, where a balanced vertex has depth $2^{-\lceil k/2 \rceil}$, so Theorem~\ref{thm:median-core}(i) caps the quota at $1 - 2^{-\lceil k/2 \rceil}$.}

\newpage

\appendix

\section{Appendix: Proofs}\label{app:A}

\small

\subsection*{Proof of Lemma \ref{lem:structure}}
\begin{proof}
We establish parts~(i), (ii), and~(iii) in turn, noting that closedness of the coordinate projections $\Theta_d$ enters only in part~(iii).

\emph{Step 1 (Part (i): CW-SCP equivalences and the triple condition \eqref{eq:triple-sandwich}).}
If CW-SCP holds globally on $X \times \Theta$, fixing $d$, $\theta_{-d}$, and $\theta_d \le \theta'_d$ places $(\theta_d, \theta_{-d}) \le (\theta'_d, \theta_{-d})$, directly giving unidimensional single-crossing along slice $d$. Conversely, if single-crossing holds along every coordinate slice and $y > x$ with $\theta \le \theta'$, the telescoping lattice path $\theta^{(r)} \triangleq (\theta'_1, \dots, \theta'_r, \theta_{r+1}, \dots, \theta_k) \in \Theta$ ($r = 0, \dots, k$) moves one coordinate at a time from $\theta^{(0)} = \theta$ to $\theta^{(k)} = \theta'$, so induction on $r$ yields both clauses of Definition~\ref{def:cwscp}. On any poset $\Theta$, $U^>(y, x)$ and $U^\ge(y, x)$ are upper sets for every $y > x$ if and only if Definition~\ref{def:cwscp}(i)--(ii) hold (equivalently, their complements $L^\ge(x, y)$ and $L^>(x, y)$ are lower sets).

When preferences are strict ($U^> = U^\ge = U$), specifying $y \succ_\theta x \iff \theta \in U(y, x)$ for $x < y$ yields a total, asymmetric binary relation on $X$. For any ordered triple $a < b < c$, asymmetry implies that transitivity holds across all six permutations of $(a, b, c)$ if and only if neither $a \prec_\theta b \prec_\theta c \prec_\theta a$ ($\theta \in U(b,a) \cap U(c,b) \setminus U(c,a)$) nor $c \prec_\theta b \prec_\theta a \prec_\theta c$ ($\theta \in U(c,a) \setminus (U(b,a) \cup U(c,b))$) occurs, which is precisely~\eqref{eq:triple-sandwich}. Adding upper-set and Borel structure yields a strict-preference CW-SCP profile.

\emph{Step 2 (Part (ii): Multi-set positive association and upper-set FOSD characterisation).}
Since indicators of upper sets are nondecreasing and indicators of lower sets are nonincreasing, the multi-set inequalities $\mu(\bigcap_{j=1}^r U_j) \ge \prod_{j=1}^r \mu(U_j)$ and $\mu(\bigcap_{j=1}^r L_j) \ge \prod_{j=1}^r \mu(L_j)$ follow directly from positive association \citep{EsaryProschanWalkup1967}; applying the $r = 2$ upper-set inequality to $U$ and $\Theta \setminus L$ yields $\mu(U \cap L) \le \mu(U)\,\mu(L)$. The equivalence of $\mu' \succeq_{\mathrm{FOSD}} \mu$ with $\mu'(U) \ge \mu(U)$ for every Borel upper set $U \subseteq \Theta$ is standard \citep[Section~6.B.1]{ShakedShanthikumar2007}.

\emph{Step 3 (Part (iii): Existence and marginal-dominance monotonicity of componentwise medians).}
Fix $d \in \{1, \dots, k\}$ and set $F_d(t) \triangleq \mu(\theta_d \le t)$ and $\bar{F}_d(t) \triangleq \mu(\theta_d \ge t)$. Because $\Theta_d$ is closed and carries probability~$1$, the standard lower and upper univariate medians of the $d$-th marginal coincide with $\underline{\tau}_d(\mu) = \inf\{t \in \Theta_d : F_d(t) \ge 1/2\}$ and $\overline{\tau}_d(\mu) = \sup\{t \in \Theta_d : \bar{F}_d(t) \ge 1/2\}$, lie in $\operatorname{supp}(\mu_d) \subseteq \Theta_d$, and satisfy $\underline{\tau}_d(\mu) \le \overline{\tau}_d(\mu)$ and $F_d(\tau_d), \bar{F}_d(\tau_d) \ge 1/2$ for every $\tau_d \in [\underline{\tau}_d(\mu), \overline{\tau}_d(\mu)] \cap \Theta_d$. Marginal first-order stochastic dominance ($F'_d \le F_d$, equivalently $\bar{F}'_d \ge \bar{F}_d$) reverses the defining sets and therefore yields $\underline{\tau}_d(\mu') \ge \underline{\tau}_d(\mu)$ and $\overline{\tau}_d(\mu') \ge \overline{\tau}_d(\mu)$ \citep[Section~1.A.1]{ShakedShanthikumar2007}.
\end{proof}

\subsection*{Proof of Proposition \ref{prop:cycles}}
\begin{proof}
Because $u(y, \theta) - u(x, \theta) = [f(y) - f(x)] + [\phi_1(y) - \phi_1(x)]\theta_1 + [\phi_2(y) - \phi_2(x)]\theta_2$ has strictly positive coefficients on $\theta_1$ and $\theta_2$ whenever $y > x$, both clauses of Definition~\ref{def:cwscp} hold across both parts.

\emph{Step 1 (Part (i): Three-voter and independent $\operatorname{Bernoulli}(\rho)^2$ cycles).}
Set $\phi_1 = (0, 1, 5/4)$, $\phi_2 = (0, 1/4, 5/4)$, and $f = (1, 1/2, 0)$ on $X = \{1, 2, 3\}$. Direct evaluation at the corners of $\{0, 1\}^2$ gives:
\begin{itemize}[leftmargin=2em,itemsep=1pt]
\item $\theta = (0, 0)$: $(u(1), u(2), u(3)) = (1, 1/2, 0) \implies 1 \succ 2 \succ 3$;
\item $\theta = (0, 1)$: $(u(1), u(2), u(3)) = (1, 3/4, 5/4) \implies 3 \succ 1 \succ 2$;
\item $\theta = (1, 0)$: $(u(1), u(2), u(3)) = (1, 3/2, 5/4) \implies 2 \succ 3 \succ 1$;
\item $\theta = (1, 1)$: $(u(1), u(2), u(3)) = (1, 7/4, 5/2) \implies 3 \succ 2 \succ 1$.
\end{itemize}
Under equal weights on $\{(0,0), (0,1), (1,0)\}$, pairwise shares yield $2:1$ majorities $1 \succ_{\mathrm{maj}} 2 \succ_{\mathrm{maj}} 3 \succ_{\mathrm{maj}} 1$. Under independent $\operatorname{Bernoulli}(\rho)^2$ on $\{0,1\}^2$, $\mu(1 \succ 2) = \mu(2 \succ 3) = 1 - \rho$ and $\mu(3 \succ 1) = 1 - (1-\rho)^2 = 2\rho - \rho^2$. Each exceeds $1/2$ if and only if $\rho \in (1 - 1/\sqrt{2}, 1/2)$, with equalised vote share $1 - \rho = (\sqrt{5}-1)/2$ at the unique root $\rho = (3 - \sqrt{5})/2$ of $\rho^2 - 3\rho + 1 = 0$ in $(0,1)$.

\emph{Step 2 (Part (ii): Continuous log-concave $\operatorname{Beta}(1, 2)^2$ density).}
The marginal density $g(t) = 2(1-t)$ on $[0, 1]$ has concave $\ln g(t) = \ln 2 + \ln(1 - t)$, so $g(\theta_1)g(\theta_2)$ is log-concave. Set $c_0 \triangleq 8/25$, $\epsilon \triangleq 1/25$, $\phi_1 = (0, 1, 1+\epsilon)$, $\phi_2 = (0, \epsilon, 1+\epsilon)$, and $f = (2c_0, c_0, 0)$. Using $\int_0^z 2(1-t)\dd t = 2z - z^2$, $\mathbb{E}[\theta_2] = 1/3$, and $\mathbb{E}[\theta_2^2] = 1/6$:
\begin{itemize}[leftmargin=2em,itemsep=1pt]
\item $1 \succ_\theta 2 \iff \theta_1 + \epsilon\theta_2 < c_0$. Since $0 < c_0 - \epsilon\theta_2 < 1$ on $[0,1]$,
\begin{align*}
\mu(1 \succ 2) &= \int_0^1 2(1-\theta_2)\big[2(c_0 - \epsilon\theta_2) - (c_0 - \epsilon\theta_2)^2\big]\dd\theta_2 \\
&= 2c_0 - c_0^2 + \frac{2\epsilon(c_0 - 1)}{3} - \frac{\epsilon^2}{6} = \frac{649}{1250} > \frac{1}{2}.
\end{align*}
\item $2 \succ_\theta 3 \iff \epsilon\theta_1 + \theta_2 < c_0$; by coordinate symmetry, $\mu(2 \succ 3) = \mu(1 \succ 2) = 649/1250 > 1/2$.
\item $3 \succ_\theta 1 \iff \theta_1 + \theta_2 > s_0 \triangleq \frac{2c_0}{1+\epsilon} = \frac{8}{13}$. Substituting $w = s_0 - \theta_1$ gives
\[
\mu(\theta_1 + \theta_2 \le s_0) = 2\int_0^{s_0} (1 - s_0 + w)(2w - w^2)\dd w = 2s_0^2 - \frac{4}{3}s_0^3 + \frac{1}{6}s_0^4 = \frac{13440}{28561},
\]
so $\mu(3 \succ 1) = 15121/28561 > 1/2$ (indifference lines have measure zero). \qedhere
\end{itemize}
\end{proof}

\subsection*{Proof of Proposition \ref{prop:collapse}}
\begin{proof}
We prove part~(i) via a self-contained mediant betweenness argument for sufficiency and two linearly independent difference vectors for necessity, then verify the explicit examples in part~(ii).

\emph{Step 1 (Part (i): Betweenness sufficiency and necessity of proportional differences).}
Suppose $\phi_d(x) = a_d + w_d \psi(x)$ with $w \in \mathbb{R}^k_{++}$ and $\psi$ strictly increasing. For any $x < y$, $u(y, \theta) - u(x, \theta) = [f(y) - f(x)] + [\psi(y) - \psi(x)](w \cdot \theta)$ with $\psi(y) - \psi(x) > 0$, so setting $t_{xy} \triangleq \frac{f(x) - f(y)}{\psi(y) - \psi(x)}$ and $G(s) \triangleq \mu(w \cdot \theta > s) - \mu(w \cdot \theta < s)$ (nonincreasing in $s$) yields $\mathcal{M}(y, x; \mu) = G(t_{xy})$ for every Borel measure $\mu$ (atoms and ties included). For any $x < y < z$,
\[
t_{xz} = \frac{\big(f(x) - f(y)\big) + \big(f(y) - f(z)\big)}{\big(\psi(y) - \psi(x)\big) + \big(\psi(z) - \psi(y)\big)}
\]
is the mediant of $t_{xy}$ and $t_{yz}$ with strictly positive denominators, so $\min(t_{xy}, t_{yz}) \le t_{xz} \le \max(t_{xy}, t_{yz})$ and hence
\[
\min\big(\mathcal{M}(y, x; \mu), \mathcal{M}(z, y; \mu)\big) \le \mathcal{M}(z, x; \mu) \le \max\big(\mathcal{M}(y, x; \mu), \mathcal{M}(z, y; \mu)\big).
\]
Checking all six orderings of $\{x, y, z\}$ shows that strict net-margin majority preference is transitive for every $f$ and $\mu$ \citep{GansSmart1996}.

Conversely, suppose that no such representation exists. Then some triple $x_1 < x_2 < x_3$ in $X$ has linearly independent consecutive differences $\Delta_1 \triangleq \phi(x_2) - \phi(x_1)$ and $\Delta_2 \triangleq \phi(x_3) - \phi(x_2)$, both in $\mathbb{R}^k_{++}$. To see this, suppose instead that consecutive differences are parallel along every triple, fix any $x_0 < x_1$ in $X$, and set $w \triangleq \phi(x_1) - \phi(x_0) \in \mathbb{R}^k_{++}$. For any $x \in X \setminus \{x_0, x_1\}$, applying the triple hypothesis to $\{x_0, x_1, x\}$ in ascending order forces $\phi(x) - \phi(x_0)$ to be a scalar multiple of $w$ (directly when $x < x_0$ or $x > x_1$, and because $\phi(x) - \phi(x_0)$ and $\phi(x_1) - \phi(x)$ are parallel and sum to $w$ when $x_0 < x < x_1$). Defining $\psi$ by $\phi(x) - \phi(x_0) = \psi(x)\,w$ yields $\phi_d(x) = \phi_d(x_0) + w_d\,\psi(x)$ with $\psi$ strictly increasing, a contradiction. Under varying $f$, pick $\theta^0 \in \operatorname{int}(\Theta)$ with $B_\delta(\theta^0) \subseteq \Theta$ and set $f(x_1) = 0$, $f(x_2) = -\Delta_1 \cdot \theta^0$, and $f(x_3) = -(\Delta_1 + \Delta_2) \cdot \theta^0$. Under fixed $f$ with $\Theta = \mathbb{R}^k$, surjectivity of $\theta \mapsto (\Delta_1 \cdot \theta, \Delta_2 \cdot \theta)$ onto $\mathbb{R}^2$ gives $\theta^0 \in \mathbb{R}^k$ solving $\Delta_1 \cdot \theta^0 = f(x_1) - f(x_2)$ and $\Delta_2 \cdot \theta^0 = f(x_2) - f(x_3)$ (with any $\delta > 0$). In both cases $u(x_1, \theta^0) = u(x_2, \theta^0) = u(x_3, \theta^0)$, so for $\theta = \theta^0 + v$, $u(x_2, \theta) - u(x_1, \theta) = \Delta_1 \cdot v$ and $u(x_3, \theta) - u(x_2, \theta) = \Delta_2 \cdot v$. Choosing $\tilde{v}^{(1)}, \tilde{v}^{(2)}, \tilde{v}^{(3)} \in \mathbb{R}^k$ with $(\Delta_1 \cdot \tilde{v}^{(i)}, \Delta_2 \cdot \tilde{v}^{(i)})$ equal to $(-1, -1)$, $(2, -1)$, and $(-1, 2)$ and scaling by $\eta > 0$ with $\eta\|\tilde{v}^{(i)}\| < \delta$ places $\theta^{(i)} \triangleq \theta^0 + \eta\tilde{v}^{(i)} \in B_\delta(\theta^0) \subseteq \Theta$ holding $x_1 \succ x_2 \succ x_3$, $x_2 \succ x_3 \succ x_1$, and $x_3 \succ x_1 \succ x_2$; equal weights $1/3$ generate a strict $2:1$ Condorcet cycle.

\emph{Step 2 (Part (ii): Non-unidimensional Condorcet-safe domains).}
On $X = \{1, 2, 3, 4\}$ and $\Theta = [0, 1]^2$, set $\phi_1 = (0, 1, 1.01, 1.02)$, $\phi_2 = (0, 0.01, 0.02, 1.02)$, and $f = (10, 9.5, 5, 4.5)$ (strictly increasing differences). Every $\theta \in [0, 1]^2$ satisfies $\min_{x \in \{1,2\}} u(x, \theta) \ge 9.5 > 6.54 \ge \max_{x \in \{3,4\}} u(x, \theta)$, so $\{1, 2\}$ unanimously beats $\{3, 4\}$ under any $\mu$ and majority preference is a transitive weak order with nonempty Condorcet set; yet $(1,0)$ ranks $2 \succ 1 \succ 3 \succ 4$ and $(0,1)$ ranks $1 \succ 2 \succ 4 \succ 3$, with non-proportional $\phi(2) - \phi(1)$ and $\phi(4) - \phi(3)$.

On $\Theta = \{0, 1\}^2$, assigning $(0,0)\mapsto 1234$, $(1,0)\mapsto 2134$, $(0,1)\mapsto 1243$, $(1,1)\mapsto 2143$ yields upper sets $U(2,1) = \{\theta_1 = 1\}$, $U(4,3) = \{\theta_2 = 1\}$, and $U(y,x) = \emptyset$ across blocks, precluding cycles since $\{1,2\}$ unanimously beats $\{3,4\}$. On three policies, assigning $(0,0)\mapsto 123$, $(1,0)\mapsto 132$, $(0,1)\mapsto 231$, $(1,1)\mapsto 321$ gives upper sets $U(2,1) = U(3,1) = \{\theta_2 = 1\}$ and $U(3,2) = \{\theta_1 = 1\}$; because $\mu(3 \succ 1) = \mu(2 \succ 1) = 1 - \mu(1 \succ 2)$, neither $1 \succ_{\mathrm{maj}} 2 \succ_{\mathrm{maj}} 3 \succ_{\mathrm{maj}} 1$ nor its reverse can hold under any $\mu$, yet $(1,0)$ and $(0,1)$ violate unidimensional single-crossing.
\end{proof}

\subsection*{Proof of Proposition \ref{prop:non-pa}}
\begin{proof}
Fix $m \ge 3$, $X = \{1, \dots, m\}$, and $q < 1 - 1/m$. Place $m$ anti-diagonal points $\mathcal{S} \triangleq \{z^{(1)}, \dots, z^{(m)}\} \subset (0, 1)^2$ at $z^{(j)} \triangleq \big(\frac{j}{m+1},\; 1 - \frac{j}{m+1}\big)$, and let $R_j$ be the cyclic rotation placing $j$ first and $p_{\mathrm{pred}}(j) \triangleq ((j - 2) \bmod m) + 1$ last, encoded by the integer score $u_0(x, z^{(j)}) \triangleq m - ((x - j) \bmod m) \in \{1, \dots, m\}$. For each $i \in \{1, \dots, m - 1\}$ and $s \in \mathcal{S}$, set $\Delta_i(s) \triangleq u_0(i + 1, s) - u_0(i, s) \in \mathbb{Z}$, $d_i \triangleq \min_{s \in \mathcal{S}} \Delta_i(s)$, and $f_i(\theta) \triangleq \max\big(\{d_i\} \cup \{\Delta_i(s) : s \in \mathcal{S},\; s \le \theta\}\big)$ on $[0, 1]^2$. Picking $0 < \eta < \frac{1}{3(m-1)}$, define $u(1, \theta) \triangleq 0$ and
\[
u(x, \theta) \triangleq \sum_{i=1}^{x-1}\big(f_i(\theta) + \eta(\theta_1 + \theta_2 + 1)\big) \qquad (x \in X,\; \theta \in [0, 1]^2).
\]
Because each $f_i$ is coordinatewise nondecreasing and $\eta(\theta_1 + \theta_2 + 1)$ increases strictly, $u$ has strictly increasing differences on $X \times [0, 1]^2$; since $\sum_{i=x}^{x'-1} f_i(\theta) \in \mathbb{Z}$ whereas $0 < (x' - x)\eta(\theta_1 + \theta_2 + 1) < 1$ for $x < x'$, no ties occur anywhere on $[0, 1]^2$.

Set $\varepsilon \triangleq \frac{1}{2(m+1)}$ and $Q_j \triangleq [z^{(j)}_1, z^{(j)}_1 + \varepsilon) \times [z^{(j)}_2, z^{(j)}_2 + \varepsilon) \subset (0, 1)^2$. Every $\theta \in Q_j$ satisfies $\{s \in \mathcal{S} : s \le \theta\} = \{z^{(j)}\}$, so $f_i(\theta) = \Delta_i(z^{(j)})$ and $|(x - y)\eta(\theta_1 + \theta_2 + 1)| < 1 \le |u_0(x, z^{(j)}) - u_0(y, z^{(j)})|$ implies that every $\theta \in Q_j$ induces the strict ranking $R_j$. Since $p_{\mathrm{pred}}(x)$ ranks above $x$ in every $R_j$ with $j \neq x$, the discrete measure $\mu = \frac{1}{m}\sum_{j=1}^m \delta_{z^{(j)}}$ gives $\mu(p_{\mathrm{pred}}(x) \succ x) = 1 - 1/m > q$ for all $x \in X$. Alternatively, choosing $0 < \delta < 1 - \frac{q}{1 - 1/m}$ and $C^\infty$ bump densities $\varphi_j$ supported in $\operatorname{int}(Q_j)$ integrating to $1$, the everywhere positive $C^\infty$ density $g(\theta) \triangleq \delta + \frac{1 - \delta}{m}\sum_{j=1}^m \varphi_j(\theta)$ gives $\mu(p_{\mathrm{pred}}(x) \succ x) \ge \frac{m-1}{m}(1 - \delta) > q$ for all $x$. In both cases $C_{>q}(X, \mu) = C_{\ge q}(X, \mu) = \emptyset$, proving part~(i).

For part~(ii), fix $k \ge 2$, $w \in [0, 1/2)$, and $q < 1 - w$, and choose an even $m \ge 4$ with $w + (1 - 2w)(1 - 1/m) > q$. Replace the support points by $z^{(j)} \triangleq \big(\frac{j}{m+1}, 1 - \frac{j}{m+1}, \zeta_j, \dots, \zeta_j\big) \in (0, 1)^k$, with $\zeta_j \triangleq 1/4$ for odd $j$ and $3/4$ for even $j$, and define $f_i$ and $u$ as above with $\theta_1 + \theta_2 + 1$ replaced by $\sum_{d=1}^k \theta_d + 1$ and $0 < \eta < \frac{1}{(k+1)(m-1)}$. The first two coordinates keep $\mathcal{S}$ an antichain, and every $\theta \in Q_j \triangleq \prod_{d=1}^k [z^{(j)}_d, z^{(j)}_d + \varepsilon)$, $\varepsilon \triangleq \frac{1}{4(m+1)}$, again satisfies $\{s \in \mathcal{S} : s \le \theta\} = \{z^{(j)}\}$ and ranks $R_j$. On $B^- \triangleq [0, \varepsilon)^k$ no support point lies below $\theta$, so $f_i = d_i = -1$ and $\theta$ ranks $1 \succ 2 \succ \dots \succ m$; on $B^+ \triangleq (1 - \varepsilon, 1]^k$ every support point lies below $\theta$, so $f_i = \max_{s \in \mathcal{S}} \Delta_i(s) = m - 1$ and $\theta$ ranks $m \succ \dots \succ 1$. Let $\mu$ spread mass $w$ uniformly over each of $B^-$ and $B^+$ and mass $(1 - 2w)/m$ uniformly over each $Q_j$. Each $Q_j$ has one of its first two coordinates above $1/2$ and the other below $1/2$, so it meets neither $[0, 1/2]^k$ nor $[1/2, 1]^k$; because $m$ is even, exactly half of the $Q_j$ lie in $\{\theta_d < 1/2\}$ for each $d$. Hence $\mu(\theta_d < 1/2) = \mu(\theta_d > 1/2) = 1/2$ and $\mu([0, 1/2]^k) = \mu([1/2, 1]^k) = w$. Finally, $p_{\mathrm{pred}}(x) \succ x$ holds on every $Q_j$ with $j \neq x$, on $B^-$ when $x \ge 2$ (as $p_{\mathrm{pred}}(x) = x - 1$), and on $B^+$ when $x = 1$ (as $p_{\mathrm{pred}}(1) = m$), so $\mu(p_{\mathrm{pred}}(x) \succ x) = w + (1 - 2w)(1 - 1/m) > q$ for every $x \in X$.
\end{proof}

\subsection*{Proof of Theorem \ref{thm:median-core}}
\begin{proof}
\emph{Step 1 (Part (i)).}
Fix $\mu$, $t \in \Theta$, and $q \in [0, 1)$, and recall that $h_\mu(t)$ is the smaller of $\mu(O^-(t))$ and $\mu(O^+(t))$.

\emph{$(a) \implies (b)$:} Suppose $q \ge 1 - h_\mu(t)$ and $x \succeq_t y$. If $y = x$, $\mu(y \succ x) = 0 \le q$. If $y > x$, Definition~\ref{def:cwscp}(i) precludes $y \succ_\theta x$ on $O^-(t)$, so $\mu(y \succ x) \le 1 - \mu(O^-(t)) \le 1 - h_\mu(t) \le q$. If $y < x$, Definition~\ref{def:cwscp}(ii) yields $x \succeq_\theta y$ on $O^+(t)$, so $\mu(y \succ x) \le 1 - \mu(O^+(t)) \le 1 - h_\mu(t) \le q$. Contrapositively, $\mu(y \succ x) > q \implies y \succ_t x$. Thus $\succ_{>q}$ inherits acyclicity from $\succ_t$.

\emph{$(b) \implies (c)$:} If $x^* \in M(t)$ and $y \in X$, then $x^* \succeq_t y$ rules out $y \succ_t x^*$, so~(b) forces $\mu(y \succ x^*) \le q$; hence $M(t) \subseteq C_{>q}(X, \mu)$.

\emph{$(c) \implies (a)$ (and necessity for~(b)):} Suppose $q < 1 - h_\mu(t)$ and take $X = \{0, 1\}$. Set $U \triangleq \Theta \setminus O^-(t)$ if $h_\mu(t) = \mu(O^-(t))$, and $U \triangleq O^+(t)$ otherwise. Both are Borel upper sets, so by Lemma~\ref{lem:structure}(i) defining strict preferences by $1 \succ_\theta 0 \iff \theta \in U$ gives a strict-preference CW-SCP profile. When $h_\mu(t) = \mu(O^-(t))$, $M(t) = \{0\}$ and $\mu(1 \succ 0) = 1 - h_\mu(t) > q$; when $h_\mu(t) = \mu(O^+(t))$, $M(t) = \{1\}$ and $\mu(0 \succ 1) = 1 - h_\mu(t) > q$. Both~(b) and~(c) fail.

For any componentwise median $\theta^{\mathrm{med}} = (\tau_1, \dots, \tau_k) \in \Theta$ (Lemma~\ref{lem:structure}(iii)) under positive orthant dependence,
\[
\mu\!\big(O^-(\theta^{\mathrm{med}})\big) \ge \prod_{d=1}^k \mu(\theta_d \le \tau_d) \ge 2^{-k},
\qquad
\mu\!\big(O^+(\theta^{\mathrm{med}})\big) \ge \prod_{d=1}^k \mu(\theta_d \ge \tau_d) \ge 2^{-k},
\]
so $h_\mu(\theta^{\mathrm{med}}) \ge 2^{-k}$, giving endorsement, acyclicity, and $M(\theta^{\mathrm{med}}) \subseteq C_{>1 - 2^{-k}}(X, \mu)$.

\emph{Step 2 (Part (ii)).}
Suppose $\mu$ is positively associated and $x_0 < x_1 < \dots < x_\ell$ ($\ell \ge 2$) forms an ascending cycle at tolerance $p = 1 - q$: $\mu(x_j \succ x_{j-1}) \ge 1 - p$ ($1 \le j \le \ell$) and $\mu(x_0 \succ x_\ell) \ge 1 - p$. By transitivity, $\bigcap_{j=1}^\ell U^>(x_j, x_{j-1}) \subseteq U^>(x_\ell, x_0) \subseteq \Theta \setminus L^>(x_0, x_\ell)$, so Lemma~\ref{lem:structure}(i)--(ii) yields
\[
1 - p \le \mu(x_0 \succ x_\ell) \le 1 - \mu\!\left(\bigcap_{j=1}^\ell U^>(x_j, x_{j-1})\right) \le 1 - \prod_{j=1}^\ell \mu(x_j \succ x_{j-1}) \le 1 - (1 - p)^\ell.
\]
Hence $(1 - p)^\ell \le p$, i.e., $q^\ell + q \le 1$ and $q \le r_\ell$ (strictly for $> q$). Descending cycles follow symmetrically on $L^>(x_j, x_{j-1})$. At $\ell = 2$, every directed $3$-cycle is one-sided up to relabelling, giving $q \le r_2 = (\sqrt{5}-1)/2$, attained in Proposition~\ref{prop:cycles}(i).
\end{proof}

\subsection*{Proof of Corollary \ref{cor:observable}}
\begin{proof}
Let $\theta^{\mathrm{med}} = (\tau_1, \tau_2) \in \Theta$ be a componentwise median (Lemma~\ref{lem:structure}(iii)). Zero mass on $\{\theta_d = \tau_d\}$ ($d = 1, 2$) forces $\mu(\theta_d \le \tau_d) = \mu(\theta_d \ge \tau_d) = 1/2$, and $O^+(\theta^{\mathrm{med}})$ differs from $\{\theta_1 > \tau_1, \theta_2 > \tau_2\}$ by a null set. Inclusion--exclusion gives
\[
\mu\!\big(O^+(\theta^{\mathrm{med}})\big) = 1 - \mu(\theta_1 \le \tau_1) - \mu(\theta_2 \le \tau_2) + \mu(\theta_1 \le \tau_1,\; \theta_2 \le \tau_2) = h.
\]
Hence $h_\mu(\theta^{\mathrm{med}}) = h$, and Theorem~\ref{thm:median-core}(i) at $t = \theta^{\mathrm{med}}$ establishes both exactness of $1 - h$ for median endorsement and $M(\theta^{\mathrm{med}}) \subseteq C_{>q}(X, \mu)$, as well as acyclicity of $\succ_{>1 - h}$.
\end{proof}

\subsection*{Proof of Theorem \ref{thm:2d-sharp}}
\begin{proof}
Fix $n \ge 2$, set $\gamma_n \triangleq \frac{\pi}{2(n+1)}$ and $p_n \triangleq \frac{1}{4\cos^2 \gamma_n} = 1 - q_n$. The proof proceeds in four steps: establish the breakpoint ordering and equal-area recurrence, construct the alternating-increment utility and verify zero ties, evaluate all $2n$ defeat shares, and pass to the limit $n \to \infty$.

\emph{Step 1 (Breakpoint ordering and equal corner areas).}
For $n \ge 2$, $\gamma_n \in (0, \pi/6]$ gives $\cos\gamma_n, \sin\gamma_n > 0$. Setting $t_h \triangleq \frac{\sin(h\gamma_n)}{2\cos\gamma_n\sin((h+1)\gamma_n)}$ ($0 \le h \le 2n$) places $t_0 = 0$ and $t_{2n} = 1$. Using $2\cos\gamma_n\sin\alpha = \sin(\alpha+\gamma_n) + \sin(\alpha-\gamma_n)$ at $\alpha = (h+1)\gamma_n$ yields $1 - t_h = \frac{\sin((h+2)\gamma_n)}{2\cos\gamma_n\sin((h+1)\gamma_n)} > 0$ for $0 \le h < 2n$, so $t_h$ solves the classical M\"obius recurrence \citep{Usiskin1964,Trybula1965,Komisarski2021}
\[
t_0 = 0, \qquad t_{h+1} = \frac{p_n}{1 - t_h} \qquad (0 \le h \le 2n - 1).
\]
Since $(2n+2)\gamma_n = \pi$, we have $\sin((2n-h)\gamma_n) = \sin((h+2)\gamma_n)$ and $\sin((2n+1-h)\gamma_n) = \sin((h+1)\gamma_n)$, giving the reflection $t_{2n-h} = 1 - t_h$ ($0 \le h \le 2n$). By induction ($t_1 = p_n > t_0$, and $t_h > t_{h-1} \implies t_{h+1} > t_h$), $0 = t_0 < t_1 < \dots < t_{2n} = 1$. Defining $\zeta_j \triangleq t_{2j-2}$ ($1 \le j \le n+1$) and $\xi_j \triangleq 1 - t_{2j-1} = t_{2n+1-2j}$ ($1 \le j \le n$) yields $0 = \zeta_1 < \zeta_2 < \dots < \zeta_{n+1} = 1$, $0 < p_n = \xi_n < \dots < \xi_1 = 1 - p_n < 1$, and both equal-area identities directly from the recurrence:
\[
\xi_j \zeta_{j+1} = (1 - t_{2j-1})\, t_{2j} = p_n, \qquad (1 - \xi_j)(1 - \zeta_j) = t_{2j-1}\,(1 - t_{2j-2}) = p_n \quad (1 \le j \le n).
\]

\emph{Step 2 (Alternating increments, increasing differences, and zero ties on $[0, 1]^2$).}
On $X_n = \{0, 1, \dots, 2n-1\}$, normalise $u_0(0, \theta) \triangleq 0$ and define adjacent increments $u_0(x, \theta) - u_0(x-1, \theta)$ in alternating order $\Delta_1, \Gamma_2, \Delta_2, \Gamma_3, \dots, \Gamma_n, \Delta_n$, where
\begin{align*}
\Delta_j(\theta_1) &\triangleq \mathbf{1}\{\theta_1 > \xi_j\} - \frac{j}{n+1} \qquad (1 \le j \le n), \\
\Gamma_{j+1}(\theta_2) &\triangleq \mathbf{1}\{\theta_2 > \zeta_{j+1}\} - \left(1 - \frac{j + 1/2}{n+1}\right) \qquad (1 \le j \le n-1).
\end{align*}
Each increment is coordinatewise nondecreasing, so $u_0$ has nondecreasing differences and satisfies CW-SCP. Let $c_x$ denote the constant term of $u_0(x, \theta)$, so $u_0(x, \theta) - c_x \in \mathbb{Z}$ for every $\theta$. Because $\Delta_j + \Gamma_{j+1}$ has constant $\equiv +\frac{1}{2(n+1)} \pmod 1$, induction gives $c_{2u} \equiv \frac{u}{2(n+1)} \pmod 1$ for $0 \le u \le n - 1$ and $c_{2u-1} = c_{2u-2} - \frac{u}{n+1} \equiv -\frac{u+1}{2(n+1)} \pmod 1$ for $1 \le u \le n$. These $2n$ residues occupy distinct points of $\frac{1}{2(n+1)}\mathbb{Z} \cap [-\frac{1}{2}, \frac{1}{2})$ (even indices in $[0, \frac{n-1}{2(n+1)}]$ and odd indices in $[-\frac{n+1}{2(n+1)}, -\frac{2}{2(n+1)}]$), so $c_b - c_a \notin \mathbb{Z}$ for every $0 \le a < b \le 2n - 1$ and $|u_0(b, \theta) - u_0(a, \theta)| \ge \frac{1}{2(n+1)}$ everywhere on $[0, 1]^2$ (zero ties). Setting $u(x, \theta) \triangleq u_0(x, \theta) + \eta x(\theta_1 + \theta_2)$ with $0 < \eta < \frac{1}{8(n+1)^2}$ perturbs any utility difference by at most $2(2n-1)\eta < \frac{1}{2(n+1)}$, preserving every strict ranking while delivering strictly increasing differences.

\emph{Step 3 (Exact cycle defeat shares).}
Under $\mu = \operatorname{Unif}[0, 1]^2$:
\begin{itemize}[leftmargin=2em,itemsep=1pt]
\item Two-step ascents ($1 \le j \le n-1$): $u_0(2j, \theta) - u_0(2j-2, \theta) = \mathbf{1}\{\theta_1 > \xi_j\} + \mathbf{1}\{\theta_2 > \zeta_{j+1}\} - 1 + \frac{1}{2(n+1)} > 0$ fails solely on $[0, \xi_j] \times [0, \zeta_{j+1}]$ of area $\xi_j \zeta_{j+1} = p_n$, so $\mu(2j \succ 2j-2) = 1 - p_n = q_n$.
\item Top bridge: $u_0(2n-1, \theta) - u_0(2n-2, \theta) = \mathbf{1}\{\theta_1 > \xi_n\} - \frac{n}{n+1} > 0 \iff \theta_1 > \xi_n$, giving $\mu(2n-1 \succ 2n-2) = 1 - \xi_n = q_n$.
\item Two-step descents ($2 \le j \le n$): $u_0(2j-3, \theta) - u_0(2j-1, \theta) = 1 + \frac{1}{2(n+1)} - \mathbf{1}\{\theta_2 > \zeta_j\} - \mathbf{1}\{\theta_1 > \xi_j\} > 0$ fails solely on $(\xi_j, 1] \times (\zeta_j, 1]$ of area $(1-\xi_j)(1-\zeta_j) = p_n$, so $\mu(2j-3 \succ 2j-1) = q_n$.
\item Bottom bridge: $u_0(0, \theta) - u_0(1, \theta) = \frac{1}{n+1} - \mathbf{1}\{\theta_1 > \xi_1\} > 0 \iff \theta_1 \le \xi_1$, giving $\mu(0 \succ 1) = \xi_1 = q_n$.
\end{itemize}

\emph{Step 4 (Limit $q_n \uparrow 3/4$ and conclusion).}
Every policy in $X_n$ loses by share $q_n$, so $C_{>q}(X_n, \mu) = \emptyset$ for all $q < q_n$. As $n \to \infty$, $\gamma_n \to 0$ gives $q_n \uparrow 3/4$. Since $O^\pm(1/2, 1/2)$ each have area $1/4$, Theorem~\ref{thm:median-core}(i) gives $q^*(2) \le 3/4$, proving $q^*(2) = 3/4$.
\end{proof}

\subsection*{Proofs for Section \texorpdfstring{\ref{sec:highdim}}{5}}

We first record the hyperbolic containment, Erlang volume, and finite simplex-cover identities underlying Theorem~\ref{thm:highdim}(ii) (and also used in Supplemental Appendix~\ref{sec:supp-bridges}).

\begin{lemma}[Hyperbolic Inclusion, Erlang Measure, and $G$-Simplex Cover]\label{lem:hyp-cover}
Let $k \ge 1$, $p \in (0, 1)$, and $\mu = \operatorname{Unif}[0, 1]^k$.
\begin{enumerate}[label=(\roman*),leftmargin=2.2em]
\item Every Borel upper set $U \subseteq [0, 1]^k$ with $\mu(U) \le p$ satisfies $U \subseteq K^+_p$, and every Borel lower set $L \subseteq [0, 1]^k$ with $\mu(L) \le p$ satisfies $L \subseteq K^-_p$.
\item For every $m \ge 1$ and $c \in \mathbb{R}$, $\lambda_m(E_{m,c}) = \lambda_m(P_{m,c}) = V_m(c)$, where $\lambda_m$ denotes Lebesgue measure on $[0, 1]^m$. In addition, for $c \in (0, 1)$ and $m \ge 2$, $V_m(c) = \int_0^1 V_{m-1}(c/t)\dd t = \int_c^1 V_{m-1}(c/t)\dd t$.
\item For any integer $G \ge 1$ and $p_{\mathrm{box}} \in (0, 1)$, set $s \triangleq \binom{G+k-1}{k-1}$ and $p_{\mathrm{eff}} \triangleq p_{\mathrm{box}}^{1 + (k-1)/G}$. Enumerating all $\mathbf{i} = (i_1, \dots, i_k) \in \mathbb{Z}_{\ge 0}^k$ with $\sum_{d=1}^k i_d = G$, define the lower boxes $L_{\mathbf{i}} \triangleq \prod_{d=1}^k [0, p_{\mathrm{box}}^{i_d/G}]$ and upper boxes $U_{\mathbf{i}} \triangleq \prod_{d=1}^k [1 - p_{\mathrm{box}}^{i_d/G}, 1]$. Then $\mu(L_{\mathbf{i}}) = \mu(U_{\mathbf{i}}) = p_{\mathrm{box}}$ for every $\mathbf{i}$, while $K^-_{p_{\mathrm{eff}}} \subseteq \bigcup_{\sum i_d = G} L_{\mathbf{i}}$ and $K^+_{p_{\mathrm{eff}}} \subseteq \bigcup_{\sum i_d = G} U_{\mathbf{i}}$.
\end{enumerate}
\end{lemma}

\begin{proof}
For~(i), if $\theta \in U$ and $U$ is upper, $\prod_{d=1}^k [\theta_d, 1] \subseteq U$ gives $\prod_{d=1}^k (1 - \theta_d) \le \mu(U) \le p$, so $\theta \in K^+_p$; symmetrically for $L \subseteq K^-_p$.

For~(ii), $\theta \mapsto \mathbf{1} - \theta$ maps $E_{m,c}$ onto $P_{m,c}$, and the boundary cases $c \le 0$ and $c \ge 1$ are immediate. For $c \in (0, 1)$, if $\theta_1, \dots, \theta_m \stackrel{\mathrm{i.i.d.}}{\sim} \operatorname{Unif}[0, 1]$, then $-\ln \theta_d \stackrel{\mathrm{i.i.d.}}{\sim} \operatorname{Exp}(1)$, so $-\ln(\prod_{d=1}^m \theta_d) = \sum_{d=1}^m (-\ln \theta_d) \sim \operatorname{Erlang}(m, 1)$; hence $\lambda_m(E_{m,c}) = \mathbb{P}(\operatorname{Erlang}(m, 1) < -\ln c) = 1 - c\sum_{j=0}^{m-1} \frac{(-\ln c)^j}{j!} = V_m(c)$. For $m \ge 2$, conditioning on $\theta_m = t$ via Tonelli's theorem gives $V_m(c) = \int_0^1 V_{m-1}(c/t)\dd t = \int_c^1 V_{m-1}(c/t)\dd t$ (as $V_{m-1}(c/t) = 0$ for $t \le c$).

For~(iii), each of the $s = \binom{G+k-1}{k-1}$ tuples has $\mu(L_{\mathbf{i}}) = \mu(U_{\mathbf{i}}) = p_{\mathrm{box}}^{(\sum i_d)/G} = p_{\mathrm{box}}$. Let $\theta \in K^-_{p_{\mathrm{eff}}}$. If $\theta_{d_0} = 0$, setting $\mathbf{i} = G e_{d_0}$ places $\theta \in L_{\mathbf{i}}$. If $\theta_d > 0$ for all $d$, setting $\ell_d \triangleq \frac{\ln \theta_d}{\ln p_{\mathrm{box}}} \ge 0$ reverses $\sum_d \ln \theta_d \le \ln p_{\mathrm{eff}}$ to $\sum_d \ell_d \ge 1 + \frac{k-1}{G}$, so $\sum_{d=1}^k \lfloor G \ell_d \rfloor > G\sum_d \ell_d - k \ge G - 1$. Since the left-hand side is an integer, $\sum_{d=1}^k \lfloor G \ell_d \rfloor \ge G$, allowing integers $0 \le i_d \le \lfloor G\ell_d \rfloor$ summing to $G$. Then $i_d/G \le \ell_d$ yields $\theta_d = p_{\mathrm{box}}^{\ell_d} \le p_{\mathrm{box}}^{i_d/G}$ for all $d$, placing $\theta \in L_{\mathbf{i}}$. Reflecting $\theta \mapsto \mathbf{1} - \theta$ yields the second inclusion.
\end{proof}

\subsection*{Proof of Theorem \ref{thm:highdim}}
\begin{proof}
The proof establishes the alternating $2n$-cycle ceiling of part~(i) in Step~1 and the exact one-sided ceiling $q_{\mathrm{hyp}}(k) = 1 - c_k$ with $c_k = e^{-k/e + o(k)}$ of part~(ii) in Step~2.

\emph{Step 1 (Part (i): Alternating two-sided $2n$-cycle ceiling under any positively associated measure).}
Suppose $x_0 < x_1 < \dots < x_{2n-1}$ ($n \ge 2$) satisfy $x_0 \leftarrow x_2 \leftarrow \dots \leftarrow x_{2n-2} \leftarrow x_{2n-1} \leftarrow x_{2n-3} \leftarrow \dots \leftarrow x_1 \leftarrow x_0$ with defeat share at least $q = 1 - p$ under a positively associated measure $\mu$ on $\Theta \subseteq \mathbb{R}^k$. Set $\alpha_j \triangleq \mu(U^>(x_{2j+1}, x_{2j}))$ ($0 \le j \le n-1$) and $\beta_j \triangleq \mu(L^\ge(x_{2j+1}, x_{2j+2}))$ ($0 \le j \le n-2$). The end bridges give $\alpha_0 \le p$ and $\alpha_{n-1} \ge 1 - p$, while $1 - \alpha_j = \mu(L^\ge(x_{2j}, x_{2j+1}))$ and $1 - \beta_j = \mu(U^>(x_{2j+2}, x_{2j+1}))$. For each $0 \le j \le n - 2$:
\begin{itemize}[leftmargin=2em,itemsep=1pt]
\item Transitivity on $x_{2j} < x_{2j+1} < x_{2j+2}$ gives $L^\ge(x_{2j}, x_{2j+1}) \cap L^\ge(x_{2j+1}, x_{2j+2}) \subseteq \Theta \setminus U^>(x_{2j+2}, x_{2j})$; positive association (Lemma~\ref{lem:structure}(ii)) on these two lower sets yields $(1 - \alpha_j)\beta_j \le p$.
\item Transitivity and asymmetry on $x_{2j+1} < x_{2j+2} < x_{2j+3}$ give $U^>(x_{2j+3}, x_{2j+2}) \cap U^>(x_{2j+2}, x_{2j+1}) \subseteq \Theta \setminus L^>(x_{2j+1}, x_{2j+3})$; positive association on these two upper sets yields $\alpha_{j+1}(1 - \beta_j) \le p$.
\end{itemize}
With $\gamma_n = \frac{\pi}{2n+2}$, $p_n = \frac{1}{4\cos^2\gamma_n}$, and the M\"obius values $\alpha^*_j \triangleq t_{2j+1}$ and $\beta^*_j \triangleq t_{2j+2}$ from Theorem~\ref{thm:2d-sharp} Step~1 (satisfying $\alpha^*_0 = p_n$, $\alpha^*_{n-1} = t_{2n-1} = 1 - p_n$, $\beta^*_j = p_n/(1 - \alpha^*_j)$, and $\alpha^*_{j+1} = p_n/(1 - \beta^*_j)$), supposing $0 \le p < p_n$ yields inductively $\alpha_0 \le p < \alpha^*_0$ and
\[
\alpha_j < \alpha^*_j \implies \beta_j \le \frac{p}{1 - \alpha_j} < \frac{p_n}{1 - \alpha^*_j} = \beta^*_j \implies \alpha_{j+1} \le \frac{p}{1 - \beta_j} < \frac{p_n}{1 - \beta^*_j} = \alpha^*_{j+1}.
\]
At $j = n - 1$ this yields $\alpha_{n-1} < 1 - p_n < 1 - p$, contradicting $\alpha_{n-1} \ge 1 - p$. Hence $p \ge p_n$ and $q \le q_n < 3/4$.

\emph{Step 2 (Part (ii): Exact one-sided ceiling $q_{\mathrm{hyp}}(k) = 1 - c_k$ and $c_k = e^{-k/e + o(k)}$).}
Differentiating $V_k(c)$ telescopes to $V'_k(c) = -\frac{(-\ln c)^{k-1}}{(k-1)!} < 0$ on $(0, 1)$, so $V_k(c) - c$ decreases strictly from $1$ to $-1$ with unique root $c_k \in (0, 1/2)$ ($V_k(1/2) \le 1 - \frac{1+\ln 2}{2} < 1/2$ for $k \ge 2$). For $k = 3$, since $e^4 < (2.72)^4 < 55 < 64 = 4^3$ implies $\ln 4 > 4/3$, we have $1 + \ln 4 + \frac{(\ln 4)^2}{2} > 1 + \frac{4}{3} + \frac{8}{9} = \frac{29}{9} > 3$, giving $V_3(1/4) < 1/4$ and hence $c_3 < 1/4$; because $V_{k+1}(c) < V_k(c)$, $c_k$ decreases strictly in $k$, so $q_{\mathrm{hyp}}(k) > 3/4$ for all $k \ge 3$.

For any one-sided ascending cycle $x_0 < \dots < x_r$ at threshold $q = 1 - p$ on $([0,1]^k, \mu)$, transitivity gives $\bigcap_{j=1}^r U^>(x_j, x_{j-1}) \subseteq U^>(x_r, x_0) \subseteq \Theta \setminus L^>(x_0, x_r)$; taking complements puts $U^>(x_r, x_0)^c \subseteq \bigcup_{j=1}^r L^\ge(x_{j-1}, x_j) \subseteq K^-_p$ by Lemma~\ref{lem:hyp-cover}(i), whence $E_p \subseteq \Theta \setminus L^>(x_0, x_r)$ and $V_k(p) = \mu(E_p) \le 1 - \mu(x_0 \succ x_r) \le p$, forcing $p \ge c_k$ and $q \le q_{\mathrm{hyp}}(k)$. Descending cycles yield $V_k(p) = \mu(P_p) \le p$ symmetrically.

For attainability, fix $q < q_{\mathrm{hyp}}(k)$ and pick $c \in (c_k, 1 - q)$ so $V_k(c) < c_k < 1 - q$. Choose $G \ge 1$ with $p_{\mathrm{eff}} \triangleq c^{1+(k-1)/G} \in (c_k, c)$ satisfying $V_k(p_{\mathrm{eff}}) < 1 - q$, and let $\{U_t\}_{t=1}^s$ ($\mu(U_t) = c$) be Lemma~\ref{lem:hyp-cover}(iii)'s upper boxes covering $K^+_{p_{\mathrm{eff}}}$. Setting $X = \{0, \dots, s\}$ and $u(x_i, \theta) \triangleq \sum_{t=1}^i \mathbf{1}_{U_t}(\theta) - \frac{i}{s+2}$ gives adjacent losses $i \leftarrow (i-1)$ on $U_i^c$ of measure $1 - c > q$ and $0 \leftarrow s$ on $\bigcup_{t=1}^s U_t \supseteq K^+_{p_{\mathrm{eff}}}$ of measure $\ge 1 - V_k(p_{\mathrm{eff}}) > q$, proving $\sup q = q_{\mathrm{hyp}}(k)$.

To obtain the asymptotics of $c_k$, write $L_k \triangleq -\ln c_k > 0$, so $V_k(c_k) = c_k$ reads $P_k(L_k) \triangleq \sum_{j=k}^\infty e^{-L_k}\frac{L_k^j}{j!} = e^{-L_k}$, where $P'_k(L) = e^{-L}\frac{L^{k-1}}{(k-1)!} > 0$. For $0 < L < k + 1$, comparing the tail with a geometric series gives
\[
\sum_{j=k}^\infty \frac{L^j}{j!} = \frac{L^k}{k!}\sum_{i=0}^\infty \frac{L^i}{(k+1)(k+2)\cdots(k+i)} \le \frac{L^k}{k!}\sum_{i=0}^\infty \left(\frac{L}{k+1}\right)^{i} = \frac{L^k}{k!}\cdot\frac{1}{1 - L/(k+1)}.
\]
By this bound and Stirling's bounds $e(k/e)^k \le k! \le (k+1)^{k+1}e^{-k}$, for any $\delta \in (0, 1/e)$:
\begin{itemize}[leftmargin=2em,itemsep=1pt]
\item at $\alpha = e^{-1} - \delta$, where $L = \alpha k$ has $L/(k+1) < \alpha$, $P_k(\alpha k) \le \frac{1}{1-\alpha} e^{-\alpha k}\frac{(\alpha k)^k}{k!} \le \frac{1}{e(1-\alpha)} e^{k(-\alpha + 1 + \ln\alpha)} < e^{-\alpha k}$ for large $k$ (since $1 + \ln\alpha < 0$), forcing $L_k > (e^{-1} - \delta)k$;
\item at $\alpha = e^{-1} + \delta$, $P_k(\alpha k) \ge e^{-\alpha k}\frac{(\alpha k)^k}{k!} \ge \frac{1}{e(k+1)} e^{k(-\alpha + 1 + \ln\alpha)} > e^{-\alpha k}$ for large $k$ (since $1 + \ln\alpha > 0$), forcing $L_k < (e^{-1} + \delta)k$.
\end{itemize}
Thus $\lim_{k \to \infty} (-\ln c_k)/k = 1/e$, i.e., $c_k = e^{-k/e + o(k)}$.
\end{proof}

\subsection*{Agenda-Size Verifications for Section \texorpdfstring{\ref{sec:highdim}}{5}}
\emph{The $(k+1)$-policy cycle and $r_k$ asymptotics.}
Since $h_k(r) \triangleq r^k + r - 1$ increases strictly from $h_k(1/2) = 2^{-k} - 1/2 < 0$ to $h_k(1) = 1$, it has a unique root $r_k \in (1/2, 1)$. For $k \ge 3$, $r_- \triangleq 1 - \frac{\ln k}{k}$ satisfies $r_-^k \le 1/k$ and $h_k(r_-) \le \frac{1 - \ln k}{k} < 0$, so $r_k > r_-$; conversely, for large $k$ and $x_k \triangleq \frac{\ln k - 2\ln\ln k}{k}$, $\ln(1 - x) \ge -x - x^2$ yields $(1 - x_k)^k \ge \frac{(\ln k)^2}{k} e^{-kx_k^2} > x_k$, so $r_k < 1 - x_k$ and $1 - r_k \sim \frac{\ln k}{k}$.

On $X = \{x_0 < \dots < x_k\}$, define $u(x_i, \theta) \triangleq \sum_{d=1}^i \mathbf{1}\{\theta_d > r_k\} - \frac{i}{k+2}$. Every difference $u(x_c, \theta) - u(x_a, \theta)$ ($a < c$) is nondecreasing in $\theta$ and never zero, defining strict-preference CW-SCP. Under $\mu = \operatorname{Unif}[0,1]^k$, $\mu(x_{i-1} \succ x_i) = \mu(\theta_i \le r_k) = r_k$ ($1 \le i \le k$) and $\mu(x_k \succ x_0) = 1 - r_k^k = r_k$, leaving $C_{>q}(X, \mu) = \emptyset$ for all $q < r_k$.

\emph{Exponential agenda sizes for $q_{\mathrm{hyp}}(k)$.}
For any $\alpha \in (0, 1/e)$, pick $\alpha' \in (\alpha, 1/e)$ and integer $C \ge 1$ with $\beta \triangleq \alpha'(1 + 1/C) < 1/e$, and set $p_{\mathrm{box}} \triangleq e^{-\alpha' k}$ and $G \triangleq C(k-1)$ in the box cover of Theorem~\ref{thm:highdim} Step~2. Then $p_{\mathrm{eff}} = e^{-\beta k}$, $m = \binom{(C+1)(k-1)}{k-1} + 1 \le 2^{(C+1)(k-1)} + 1 = \exp(O(k))$, and the first bullet in Theorem~\ref{thm:highdim} Step~2, applied with $\beta$ in place of $\alpha$, yields $V_k(p_{\mathrm{eff}}) = P_k(\beta k) < e^{-\beta k} < e^{-\alpha k}$ for large $k$, delivering an empty $(1 - e^{-\alpha k})$-core on an agenda of size $\exp(O(k))$. To come within $c_k$ of $q_{\mathrm{hyp}}(k)$ instead, set $p_{\mathrm{box}} \triangleq 2c_k < 1$ and $G \triangleq \lceil (k-1)(-\ln c_k - \ln 2)/\ln 2 \rceil = O(k^2)$; then $p_{\mathrm{eff}} = (2c_k)^{1 + (k-1)/G} \ge c_k$, so the closing defeat fails on a set of volume at most $V_k(p_{\mathrm{eff}}) \le V_k(c_k) = c_k$, every defeat share is at least $1 - 2c_k$, and the agenda has $\binom{G+k-1}{k-1} + 1 \le (e(G+k-1)/(k-1))^{k-1} + 1 = \exp(O(k \log k))$ alternatives.

\subsection*{Proof of Proposition \ref{prop:amend}}
\begin{proof}
Write $i \to j$ when $x_i$ is permitted to challenge $x_j$ and $q$-defeats it.

\emph{Step 1 (Part (i)).}
For $q \ge 1/2$ under the open rule, or $q > 1/2$ under the closed rule, $\mu(y \succ x) + \mu(x \succ y) \le 1$ rules out $i \to j$ and $j \to i$ together. Let $i_-$ be the smallest $i$ such that $i + 1 \not\to i$ ($i_- \triangleq m - 1$ if there is none). Every $j < i_-$ satisfies $j + 1 \to j$, so $x_j$ is not stable; $x_{i_-}$ is not defeated by $x_{i_- + 1}$ by construction, nor by $x_{i_- - 1}$ (when $i_- \ge 1$) because $i_- \to i_- - 1$. Hence $x_{i_-} = \min C^1$, and the core is nonempty. Under $\mu' \succeq_{\mathrm{FOSD}} \mu$, Proposition~\ref{prop:fosd}(i) preserves every upward defeat $j + 1 \to j$, so $i_-$ weakly rises. Scanning downward from $x_{m-1}$, the largest $i$ with $i - 1 \not\to i$ is $\max C^1$, and since downward defeats can only disappear under the shift, it also weakly rises.

\emph{Step 2 (Part (ii): the only cycles).}
Let $G$ be the graph on $\{0, \dots, m - 1\}$ joining $i$ and $j$ when $1 \le |i - j| \le 2$. A cycle in $G$ visits an interval of consecutive indices: if it visited indices on both sides of an index $j$ that it skips, it would have to cross $j$ at least twice, but $(j - 1, j + 1)$ is the only edge that does. Relabel the interval $\{0, \dots, s - 1\}$ ($s \ge 3$). Deleting any interior vertex $j$ from the cycle leaves a path through both sides of $j$, which forces the edge $(j - 1, j + 1)$; vertices $0$ and $s - 1$ have only two neighbours each, forcing $(0, 1)$ and $(s - 2, s - 1)$. These are $s$ edges, so the cycle is the alternating cycle $0, 2, 4, \dots, 5, 3, 1, 0$ on the interval.

\emph{Step 3 (Part (ii): the bound).}
Reflecting $\theta \mapsto -\theta$ and reversing every preference preserves CW-SCP and positive association and reverses every majority edge (Remark~\ref{rem:supp-4cycle}), so suppose $0 \to 1$. Going round the cycle, $j + 1 \to j - 1$ for odd $j$, $j - 1 \to j + 1$ for even $j$ ($1 \le j \le s - 2$), and the top edge is $s - 1 \to s - 2$ if $s - 1$ is odd and $s - 2 \to s - 1$ if $s - 1$ is even. Suppose every edge has share at least $1 - p$. For odd $j$ let $z_j \triangleq \mu(U^>(x_j, x_{j-1}))$, and for even $j$ let $z_j \triangleq \mu(L^\ge(x_{j-1}, x_j))$. The end edges give $z_1 \le p$ and $z_{s-1} \ge 1 - p$. For odd $j$, $1 - z_j$ and $z_{j+1}$ are the measures of the lower sets $L^\ge(x_{j-1}, x_j)$ and $L^\ge(x_j, x_{j+1})$, whose intersection lies outside $U^>(x_{j+1}, x_{j-1})$ by transitivity; for even $j$, they are the measures of the upper sets $U^>(x_j, x_{j-1})$ and $U^>(x_{j+1}, x_j)$, whose intersection lies inside $U^>(x_{j+1}, x_{j-1}) \subseteq \Theta \setminus L^>(x_{j-1}, x_{j+1})$. Either way the complementary two-step defeat bounds the intersection by $p$, and Lemma~\ref{lem:structure}(ii) gives $(1 - z_j)z_{j+1} \le p$ for $1 \le j \le s - 2$.

With $\gamma \triangleq \frac{\pi}{s+2}$ and $p_s \triangleq (4\cos^2\gamma)^{-1}$, the values $t_h \triangleq \frac{\sin(h\gamma)}{2\cos\gamma\sin((h+1)\gamma)}$ solve $t_0 = 0$, $t_{h+1} = p_s/(1 - t_h)$, and satisfy $t_{s-h} = 1 - t_h$, exactly as in Step~1 of the proof of Theorem~\ref{thm:2d-sharp} with $2n$ replaced by $s$. If $p < p_s$, then $z_1 \le p < t_1$, and $z_j < t_j$ implies $z_{j+1} \le p/(1 - z_j) < p_s/(1 - t_j) = t_{j+1}$, so $z_{s-1} < t_{s-1} = 1 - p_s < 1 - p$, a contradiction. Hence every cycle has an edge with share at most $Q_s \le Q_m$, because $Q_s$ increases in $s$. At any $q \ge Q_m$ that edge is not an open $q$-defeat, so permitted defeats are acyclic, and a finite acyclic relation has an undefeated element.

\emph{Step 4 (Part (ii): exactness).}
Let $\gamma \triangleq \frac{\pi}{m+2}$, $p \triangleq (4\cos^2\gamma)^{-1} = 1 - Q_m$, and $t_h$ as in Step~3 with $s = m$, so $0 = t_0 < t_1 < \dots < t_m = 1$ and $t_{m-1} = 1 - p$. On $X = \{0, \dots, m - 1\}$, set $u_0(0, \theta) \triangleq 0$ and let project $j \in \{1, \dots, m - 1\}$ add
\[
\mathbf{1}\{\theta_1 > 1 - t_j\} - \frac{r}{m+1} \quad (j = 2r - 1), \qquad \mathbf{1}\{\theta_2 > t_j\} - \Big(1 - \frac{r + 1/2}{m+1}\Big) \quad (j = 2r).
\]
Each increment is nondecreasing in $\theta$. Consecutive costs sum to $1 - \frac{1}{2(m+1)}$ (projects $2r - 1, 2r$) or $1 + \frac{1}{2(m+1)}$ (projects $2r, 2r + 1$), so the constant term of $u_0(2u, \cdot)$ is $\equiv \frac{u}{2(m+1)}$ and that of $u_0(2u - 1, \cdot)$ is $\equiv -\frac{u+1}{2(m+1)} \pmod 1$. These residues are distinct multiples of $\frac{1}{2(m+1)}$ in $(-\frac{1}{2}, \frac{1}{2})$, so $|u_0(b, \theta) - u_0(a, \theta)| \ge \frac{1}{2(m+1)}$ for all $a \neq b$ and $\theta$, and $u \triangleq u_0 + \eta x(\theta_1 + \theta_2)$ with $0 < \eta < \frac{1}{4(m+1)(m-1)}$ has strictly increasing differences and the same strict rankings. Under $\operatorname{Unif}[0, 1]^2$: $0$ defeats $1$ on $\{\theta_1 \le 1 - t_1\}$, with share $1 - p$; $2r$ defeats $2r - 2$ except on $\{\theta_1 \le 1 - t_{2r-1},\ \theta_2 \le t_{2r}\}$, of area $(1 - t_{2r-1})t_{2r} = p$; $2r - 1$ defeats $2r + 1$ except on $\{\theta_1 > 1 - t_{2r+1},\ \theta_2 > t_{2r}\}$, of area $(1 - t_{2r})t_{2r+1} = p$; and at the top, $m - 1$ defeats $m - 2$ on $\{\theta_1 > 1 - t_{m-1}\}$ ($m$ even) or $m - 2$ defeats $m - 1$ on $\{\theta_2 \le t_{m-1}\}$ ($m$ odd), each with share $t_{m-1} = 1 - p$. All of these challenges move at most two steps and every budget loses one, so $C^2_{>q} = \emptyset$ for every $q < Q_m$. For even $m$ this is the queue of Theorem~\ref{thm:2d-sharp} with rescaled costs.

Finally, in Step~2 of the proof of Theorem~\ref{thm:phase}, every budget outside the stable singleton loses to a challenger at most two steps away (its challenger in the alternating cycle, which is $2n - 3$ for $2n - 1$ under $\mu$, or $2$ for $0$ under $\mu'$), while the singleton survives every challenge, so the same singleton cores obtain under the two-step rule.
\end{proof}

\subsection*{Proof of Proposition \ref{prop:fosd}}\label{sec:app-C}
\begin{proof}
We establish parts~(i)--(iii) in sequence from Lemma~\ref{lem:structure} and pairwise majority asymmetry.

\emph{Step 1 (Part (i): Vote shares and net margins).}
For any $y > x$, Lemma~\ref{lem:structure}(i) makes $U^>(y, x)$ and $U^\ge(y, x)$ Borel upper sets, so $\mu' \succeq_{\mathrm{FOSD}} \mu$ and Lemma~\ref{lem:structure}(ii) give $\mu'(y \succ x) \ge \mu(y \succ x)$ and $\mu'(y \succeq x) \ge \mu(y \succeq x)$. Taking complements reverses both inequalities for downward shares, and subtracting yields $\mathcal{M}(y, x; \mu') \ge \mathcal{M}(y, x; \mu)$.

\emph{Step 2 (Part (ii): Ordering existing Condorcet winners).}
Suppose toward a contradiction that $x^{**} < x^*$. If $x^*$ is a strict net-margin (respectively, $> 1/2$) Condorcet winner under $\mu$, then $x^* > x^{**}$ and part~(i) yield $\mathcal{M}(x^*, x^{**}; \mu') \ge \mathcal{M}(x^*, x^{**}; \mu) > 0$ (respectively, $\mu'(x^* \succ x^{**}) \ge \mu(x^* \succ x^{**}) > 1/2$), contradicting that $x^{**}$ is a weak Condorcet winner under $\mu'$. The symmetric argument rules out $x^{**}$ being strict under $\mu'$.

\emph{Step 3 (Part (iii): Ordered ideal sets and componentwise median strong set order).}
Since $X \subseteq \mathbb{R}$ is totally ordered, quasisupermodularity is vacuous, so CW-SCP gives $M(\theta) \le_{\mathrm{SSO}} M(\theta')$ whenever $\theta \le \theta'$ and both sets are nonempty by the Monotonicity Theorem \citep[Theorem~4]{MilgromShannon1994}. On a finite menu $X$, every ideal set $M(\theta)$ is nonempty; under marginal stochastic dominance, Lemma~\ref{lem:structure}(iii) orders $\underline{\theta}^{\mathrm{med}}(\mu) \le \underline{\theta}^{\mathrm{med}}(\mu')$ and $\overline{\theta}^{\mathrm{med}}(\mu) \le \overline{\theta}^{\mathrm{med}}(\mu')$, yielding both strong set order inequalities.
\end{proof}

\subsection*{Proof of Theorem \ref{thm:phase}}
\begin{proof}
The proof establishes each part of the phase transition in turn, reusing the alternating-increment utility of Theorem~\ref{thm:2d-sharp} for part~(ii) and the anti-diagonal cyclic construction of Proposition~\ref{prop:non-pa} for part~(iii).

\emph{Step 1 (Part (i): Monotone open-core selections and non-inversion for $q \ge 1 - 2^{-k}$).}
Across any family of distributions satisfying positive orthant dependence at $\underline{\theta}^{\mathrm{med}}(\nu)$ (including all positively associated measures), Theorem~\ref{thm:median-core}(i) places $\emptyset \neq M(\underline{\theta}^{\mathrm{med}}(\nu)) \subseteq C_{>q}(X, \nu)$ whenever $q \ge 1 - 2^{-k}$, so $s_-(\nu) \triangleq \min M(\underline{\theta}^{\mathrm{med}}(\nu))$ and $s_+(\nu) \triangleq \max M(\underline{\theta}^{\mathrm{med}}(\nu))$ define open $q$-core selections. Under marginal stochastic dominance, Proposition~\ref{prop:fosd}(iii) gives $M(\underline{\theta}^{\mathrm{med}}(\mu)) \le_{\mathrm{SSO}} M(\underline{\theta}^{\mathrm{med}}(\mu'))$, so both $s_-$ and $s_+$ are weakly increasing, and any singleton open $q$-cores must equal $s_-(\mu)$ and $s_-(\mu')$, enforcing $x^{**} \ge x^*$.

\emph{Step 2 (Part (ii): Independent single-coordinate singleton inversion for all $q \in [1/2, 3/4)$ in $k = 2$).}
Fix $q \in [1/2, 3/4)$. Set $n \triangleq 2$ if $q \in [1/2, 2/3)$ ($q_2 = 2/3, p_2 = 1/3$), and otherwise let $n \ge 3$ be minimal with $q_n = 1 - (4\cos^2(\pi/(2n+2)))^{-1} > q$; set $p \triangleq p_n = 1 - q_n \in (1/4, 1/3]$. On $X_n = \{0, \dots, 2n-1\}$, equip $[0,1]^2$ with the strictly increasing-differences utility $u(x, \theta) = u_0(x, \theta) + \eta x(\theta_1 + \theta_2)$ from Theorem~\ref{thm:2d-sharp}, which induces the same strict rankings as $u_0$ (Step~2 of that proof), so it suffices to evaluate $u_0$.

From Theorem~\ref{thm:2d-sharp} Step~1 ($t_{2n-h} = 1 - t_h$), we have $\zeta_2 = \frac{p}{1 - p}$, $\zeta_n = 1 - \frac{p}{1 - p}$, $\xi_n = p$, $\xi_1 = 1 - p$, and $\xi_{n-1} = \frac{p}{\zeta_n} = \frac{p(1-p)}{1-2p}$, giving
\[
\xi_{n-1} - \xi_n = \xi_1 - \xi_2 = \frac{p^2}{1 - 2p} > 0,
\qquad
q_n - \zeta_n = \frac{p^2}{1 - p} > 0.
\]
For $n = 2$, $q_2 - q \le 1/6 = q_2 - \zeta_2 < 1/3 = \xi_1 - \xi_2$; for $n \ge 3$ ($q \in [q_{n-1}, q_n)$, so $p = p_n > 1/4$), $q_n - q < 3/4 - 2/3 = 1/12 = \frac{(1/4)^2}{1-1/4} < \frac{p^2}{1 - p} < \frac{p^2}{1 - 2p} = \xi_{n-1} - \xi_n$.

Next bound all non-adjacent defeats of $2n - 2$ and $1$ under any product measure $\mu_1 \otimes \operatorname{Unif}[0, 1]$:
\begin{itemize}[leftmargin=2em,itemsep=1pt]
\item On $\{\theta_2 > \zeta_n\}$, every $\mathbf{1}\{\theta_2 > \zeta_{r+1}\} = 1$, so for any $0 \le i < 2n - 2$ (checking even $i = 2u$ via $\sum_{r=u+1}^{n-1}(\Delta_r + \Gamma_{r+1}) \ge \frac{n-1-u}{2(n+1)} > 0$ and odd $i = 2u-1$ via $\Gamma_{u+1} + \sum_{r=u+1}^{n-1}(\Delta_r + \Gamma_{r+1}) \ge \frac{u+1/2}{n+1} > 0$), $2n - 2 \succ_\theta i$ holds; hence $\mathbb{P}(i \succ 2n-2) \le \mathbb{P}(\theta_2 \le \zeta_n) = \zeta_n$.
\item Symmetrically, on $\{\theta_2 \le \zeta_2\}$, every $\mathbf{1}\{\theta_2 > \zeta_{r+1}\} = 0$, giving $\Gamma_r + \Delta_r \le -\frac{1}{2(n+1)} < 0$ and $\Gamma_{v+1} \le -\frac{3/2}{n+1} < 0$, so $1 \succ_\theta j$ holds for all $2 \le j \le 2n - 1$; hence $\mathbb{P}(j \succ 1) \le 1 - \zeta_2 = \zeta_n$.
\end{itemize}
Set $\varepsilon \triangleq 1/6$ if $q = 1/2$ ($n = 2$), and choose $\varepsilon \in \big(q_n - q,\; \min(\xi_{n-1} - \xi_n,\, q_n - \zeta_n)\big)$ when $q \in (1/2, 3/4)$. Form $\mu \triangleq \mu_1^- \otimes \operatorname{Unif}[0, 1]$ and $\mu' \triangleq \mu_1^+ \otimes \operatorname{Unif}[0, 1]$, where $\mu_1^-$ transfers mass $\varepsilon$ downward from $(\xi_n, \xi_{n-1})$ into $(0, \xi_n)$ and $\mu_1^+$ transfers mass $\varepsilon$ upward from $(\xi_2, \xi_1)$ into $(\xi_1, 1]$ relative to $\operatorname{Unif}[0,1]$. Then $F_1^+ \le F_1^-$, so $\mu_1^+ \succeq_{\mathrm{FOSD}} \mu_1^-$, and hence $\mu' \succeq_{\mathrm{FOSD}} \mu$ because independent product measures preserve multivariate FOSD \citep[Theorem~6.B.3]{ShakedShanthikumar2007}:
\begin{itemize}[leftmargin=2em,itemsep=1pt]
\item Under $\mu$, since $\mu_1^-(\theta_1 \le \xi_r) = \xi_r$ for $r \le n - 1$, every $a \notin \{2n-2, 2n-1\}$ still loses by $q_n > q$; $2n-1$ loses to $2n-3$ by $q_n + \varepsilon(1 - \zeta_n) > q$; and $2n-2$ loses to $2n-1$ by $q_n - \varepsilon$ and to every $i < 2n-2$ by at most $\zeta_n \le q_n - \varepsilon$. Thus $C_{>q}(X_n, \mu) = C_{\ge q}(X_n, \mu) = \{2n - 2\}$ for $q > 1/2$, and $C_{>1/2}(X_2, \mu) = \{2\}$ at $q = 1/2$, where $2$ survives its contests against $1$ and $3$ only as exact ties.
\item Under $\mu'$, symmetrically $\mu_1^+(\theta_1 \le \xi_r) = \xi_r$ for $r \ge 2$, every $a \notin \{0, 1\}$ still loses by $q_n > q$, $0$ loses to $2$ by $q_n + \varepsilon\zeta_2 > q$, and $1$ loses to $0$ by $q_n - \varepsilon$ and to every $j > 1$ by at most $\zeta_n \le q_n - \varepsilon$. Thus $C_{>q}(X_n, \mu') = C_{\ge q}(X_n, \mu') = \{1\}$ for $q > 1/2$, and $C_{>1/2}(X_2, \mu') = \{1\}$ at $q = 1/2$.
\end{itemize}

\emph{Step 3 (Part (iii): Singleton $q$-core inversion without positive association for all $q \in (1/2, 1)$).}
Fix $q \in (1/2, 1)$, set $p \triangleq 1 - q \in (0, 1/2)$, choose integer $m > \max\{3, 1/p\}$, and form the cyclic word $w \triangleq (2, m, m-1, 3, 4, \dots, m-2, 1)$ on $X = \{1, \dots, m\}$, letting $R_j$ rotate $w$ to place $j$ first and its cyclic predecessor $p_{\mathrm{pred}}(j)$ last. Under weights $\{\omega_j\}$ on $\{R_j\}$, $p_{\mathrm{pred}}(j)$ defeats $j$ by $1 - \omega_j$ while every other policy gets at most $1 - \omega_j$, so $j \in C_{>q} \iff \omega_j \ge p$ and $j \in C_{\ge q} \iff \omega_j > p$. Set $t \triangleq \frac{1 - 2p}{m - 2} \in (0, p)$ and pick $\delta > 0$ with $t - 2\delta > 0$ and $t + 2\delta < p$ (which also gives $p - \delta > 0$). Setting $\omega_{m-1} = p + \delta$, $\omega_2 = p - \delta$, and $\omega_j = t$ otherwise gives $C_{>q} = C_{\ge q} = \{m-1\}$; shifting $2\delta$ from $R_{m-1}$ to $R_m$ and $2\delta$ from $R_1$ to $R_2$ gives $\omega'_2 = p + \delta > p$ and $\omega'_j < p$ for all $j \neq 2$, moving both cores to $\{2\}$.

To realise this shift via FOSD under strictly increasing differences on $[0, 1]^2$, place $z^{(j)} \triangleq (\frac{j}{m+1}, 1 - \frac{j}{m+1})$ on the anti-diagonal, pick $\varepsilon \in (0, \frac{1}{2(m+1)})$, and add $\hat{z}^{(m-1)} \triangleq z^{(m-1)} + (\varepsilon, \varepsilon)$ and $\hat{z}^{(1)} \triangleq z^{(1)} + (\varepsilon, \varepsilon)$ to form $S$. The only strict order comparisons in $S$ are $z^{(m-1)} < \hat{z}^{(m-1)}$ and $z^{(1)} < \hat{z}^{(1)}$. Choose integer utilities $u_0(\cdot, z^{(j)})$ representing $R_j$; set $u_0(\cdot, \hat{z}^{(m-1)})$ equal to $u_0(\cdot, z^{(m-1)})$ on $x \le m - 1$ with $u_0(m, \hat{z}^{(m-1)}) \triangleq 1 + \max_{x \le m-1} u_0(x, z^{(m-1)})$ (realising $R_m$ while increasing $u_0(m) - u_0(m-1)$); and set $u_0(\cdot, \hat{z}^{(1)})$ equal to $u_0(\cdot, z^{(1)})$ on $x \ge 2$ with $u_0(1, \hat{z}^{(1)}) \triangleq \min_{x \ge 2} u_0(x, z^{(1)}) - 1$ (realising $R_2$ while increasing $u_0(2) - u_0(1)$). Extending via $u(x, \theta) \triangleq \sum_{i=1}^{x-1}\big(\max(\{d_i\} \cup \{\Delta_i(s) : s \in S, s \le \theta\}) + \eta(\theta_1 + \theta_2 + 1)\big)$ with $0 < \eta < \frac{1}{3(m-1)}$ as in Proposition~\ref{prop:non-pa} delivers strictly increasing differences, zero ties, and exact rankings $R_s$ on $S$, while moving $2\delta$ from $z^{(m-1)} \to \hat{z}^{(m-1)}$ and $z^{(1)} \to \hat{z}^{(1)}$ implements $\mu' \succeq_{\mathrm{FOSD}} \mu$.
\end{proof}

\newpage

\bibliographystyle{ecta}
\bibliography{references}

\clearpage
\setcounter{page}{1}
\renewcommand{\thepage}{S\arabic{page}}
\providecommand{\theHpage}{\arabic{page}}
\renewcommand{\theHpage}{S\arabic{page}}
\setcounter{section}{0}
\renewcommand{\thesection}{S.\arabic{section}}
\renewcommand{\theHsection}{S.\arabic{section}}
\setcounter{subsection}{0}
\renewcommand{\thesubsection}{S.\arabic{section}.\arabic{subsection}}
\renewcommand{\theHsubsection}{S.\arabic{section}.\arabic{subsection}}
\setcounter{theorem}{0}
\renewcommand{\thetheorem}{S.\arabic{theorem}}
\renewcommand{\theHtheorem}{S.\arabic{theorem}}
\setcounter{proposition}{0}
\renewcommand{\theproposition}{S.\arabic{proposition}}
\renewcommand{\theHproposition}{S.\arabic{proposition}}
\setcounter{lemma}{0}
\renewcommand{\thelemma}{S.\arabic{lemma}}
\renewcommand{\theHlemma}{S.\arabic{lemma}}
\setcounter{corollary}{0}
\renewcommand{\thecorollary}{S.\arabic{corollary}}
\renewcommand{\theHcorollary}{S.\arabic{corollary}}
\setcounter{remark}{0}
\renewcommand{\theremark}{S.\arabic{remark}}
\renewcommand{\theHremark}{S.\arabic{remark}}
\setcounter{table}{0}
\renewcommand{\thetable}{S.\arabic{table}}
\renewcommand{\theHtable}{S.\arabic{table}}
\setcounter{figure}{0}
\renewcommand{\thefigure}{S.\arabic{figure}}
\renewcommand{\theHfigure}{S.\arabic{figure}}
\setcounter{equation}{0}
\renewcommand{\theequation}{S.\arabic{section}.\arabic{equation}}
\renewcommand{\theHequation}{S.\arabic{section}.\arabic{equation}}
\setcounter{footnote}{0}

\begin{center}
{\Large\textbf{Supplemental Appendix to}}\\[0.35em]
{\Large\textbf{``Majority Voting with Multidimensional Single-Crossing Preferences''}}\\[0.75em]
{\large Joshua S. Gans}
\end{center}

\small
\singlespacing

\medskip
\noindent This Supplemental Appendix provides deferred proofs and extensions (main-paper numbering and references apply throughout): multi-stage bridge bounds for $k \ge 3$ (Section~\ref{sec:supp-bridges}, Table~\ref{tab:supp-dim-bounds}), distribution-free Condorcet domains and $4$-cycles (Section~\ref{sec:supp-condorcet}), $C^\infty$ smooth continua and grid robustness (Section~\ref{sec:supp-smooth}), simple-majority tournament solutions (Section~\ref{sec:supp-tournaments}), and auxiliary boundary verifications and closed-threshold cores at $1 - 2^{-k}$ (Section~\ref{sec:supp-aux}).

\section{Higher-Dimensional Multi-Stage Bridge Refinements}\label{sec:supp-bridges}

Throughout Section~\ref{sec:supp-bridges}, $\Theta = [0, 1]^k$ ($k \ge 2$) carries Lebesgue measure $\mu = \operatorname{Unif}[0, 1]^k$. For any subset $S \subseteq [0, 1]^k$, let $\operatorname{dn}(S) \triangleq \{\eta \in [0, 1]^k : \exists \theta \in S,\ \eta \le \theta\}$ denote its downward closure. Recall the hyperbolic complements $E_{m,c}, P_{m,c}$, corners $K^-_c, K^+_c$, Erlang volume $V_m(c)$, and $G$-simplex box covers from Section~\ref{sec:highdim} and Lemma~\ref{lem:hyp-cover}.

\begin{lemma}[Exact $N$-Stage $\widetilde{W}_t$-Bridge Construction]\label{lem:supp-nstage}
Let $N \ge 1$ and $\Theta = [0, 1]^k$. Suppose we are given:
\begin{enumerate}[label=(\arabic*),leftmargin=2.2em]
\item Borel upper sets $\{L_u^c\}_{u=1}^s$ ($s \ge 1$), with complements $L_u \triangleq \Theta \setminus L_u^c$, defining prefix/suffix intersections $A_{\mathrm{pref}}(i) \triangleq \bigcap_{u=1}^i L_u^c$, $A_{\mathrm{suf}}(i) \triangleq \bigcap_{u=i+1}^s L_u^c$ ($0 \le i \le s$), and $E \triangleq A_{\mathrm{pref}}(s) = A_{\mathrm{suf}}(0)$ (empty intersections equal $\Theta$);
\item For each stage $t \in \{0, \dots, N-1\}$, a Borel upper set $W_t \subseteq \Theta$ and Borel upper sets $\{U_{t,v}\}_{v=1}^{r_t}$ ($r_t \ge 1$) defining $B_{t,\mathrm{pref}}(j) \triangleq \bigcup_{v=1}^j U_{t,v}$, $B_{t,\mathrm{suf}}(j) \triangleq \bigcup_{v=j+1}^{r_t} U_{t,v}$ ($0 \le j \le r_t$), and $K_t \triangleq B_{t,\mathrm{pref}}(r_t) = B_{t,\mathrm{suf}}(0)$ (empty unions equal $\emptyset$).
\end{enumerate}
Define $\widetilde{W}_t \triangleq W_t \cup \bigcup_{j=1}^{r_t - 1} \big(B_{t,\mathrm{pref}}(j) \cap B_{t,\mathrm{suf}}(j)\big)$ for intermediate stages $0 \le t \le N - 2$, $\widetilde{W}_{N-1} \triangleq W_{N-1}$, and $W_{\mathrm{cap}}(u, v) \triangleq \bigcap_{t=u}^{v-1} \widetilde{W}_t$ ($W_{\mathrm{cap}}(u, u) \triangleq \Theta$). Order $m \triangleq N(s + 1) + \sum_{t=0}^{N-1}(r_t + 1)$ alternatives lexicographically across stages $t = 0, \dots, N-1$, within each stage placing $a_{t,0} < \dots < a_{t,s} < b_{t,0} < \dots < b_{t,r_t}$. Define masks
\[
\operatorname{Left}(a_{t,i}) \triangleq A_{\mathrm{suf}}(i) \cap \widetilde{W}_t, \qquad \operatorname{Left}(b_{t,j}) \triangleq W_t \cup B_{t,\mathrm{suf}}(j),
\]
\[
\operatorname{Right}(a_{t,i}) \triangleq A_{\mathrm{pref}}(i), \qquad \operatorname{Right}(b_{t,j}) \triangleq E \cap \big(W_t \cup B_{t,\mathrm{pref}}(j)\big).
\]
For every pair $x < y$ with $x$ in stage $t_x$ and $y$ in stage $t_y$:
\begin{itemize}[leftmargin=2em,itemsep=1pt]
\item If $t_x = t_y \triangleq t$: $U(a_{t,i'}, a_{t,i}) \triangleq \bigcap_{u=i+1}^{i'} L_u^c$ ($i < i'$), $U(b_{t,j'}, b_{t,j}) \triangleq \bigcup_{v=j+1}^{j'} U_{t,v}$ ($j < j'$), and $U(b_{t,j}, a_{t,i}) \triangleq A_{\mathrm{suf}}(i) \cap (W_t \cup B_{t,\mathrm{pref}}(j))$.
\item If $t_x < t_y$: $U(y, x) \triangleq \operatorname{Left}(x) \cap W_{\mathrm{cap}}(t_x + 1, t_y) \cap \operatorname{Right}(y)$ when $t_y = t_x + 1$, and intersected further with $E$ when $t_y \ge t_x + 2$.
\end{itemize}
Then $\{U(y, x)\}_{x < y}$ defines a strict-preference CW-SCP profile on $X \times [0, 1]^k$ in which every alternative is defeated by share at least
\begin{multline*}
q_{\min}^{(N)} \triangleq \min\bigg\{
\min_{1 \le u \le s}\big(1 - \mu(L_u)\big),\;\;
\min_{\substack{0 \le t \le N-1 \\ 1 \le v \le r_t}}\big(1 - \mu(U_{t,v})\big),\\
\min_{0 \le t \le N-1}\mu(W_t \cup K_t),\;\;
1 - \mu\!\left(E \cap \bigcap_{t=0}^{N-1}\widetilde{W}_t\right)
\bigg\}.
\end{multline*}
\end{lemma}

\begin{proof}
Each $U(y, x)$ is a Borel upper set as a finite Boolean union/intersection of Borel upper sets. For every intermediate stage $0 \le t \le N - 2$ and every $0 \le j \le r_t$,
\begin{equation}
\begin{aligned}
&\big(W_t \cup B_{t,\mathrm{pref}}(j)\big) \cap \big(W_t \cup B_{t,\mathrm{suf}}(j)\big)
= W_t \cup \big(B_{t,\mathrm{pref}}(j) \cap B_{t,\mathrm{suf}}(j)\big) \\
&\qquad \subseteq \widetilde{W}_t \subseteq W_t \cup K_t
= \big(W_t \cup B_{t,\mathrm{pref}}(j)\big) \cup \big(W_t \cup B_{t,\mathrm{suf}}(j)\big).
\end{aligned}
\label{eq:supp-bridge-sandwich}
\end{equation}
To verify~\eqref{eq:triple-sandwich} for every $x < y < z$ in stages $t_x \le t_y \le t_z$, check the four exhaustive stage regimes:
\begin{enumerate}[label=\emph{Regime \arabic*:},leftmargin=5.5em,itemsep=1pt]
\item ($t_x = t_y = t_z \triangleq t$): Within $\{a_{t,\cdot}\}$, $U(z,x) = U(y,x) \cap U(z,y)$; within $\{b_{t,\cdot}\}$, $U(z,x) = U(y,x) \cup U(z,y)$. For $(a_{t,i}, a_{t,i'}, b_{t,j})$, $U(y,x) \cap A_{\mathrm{suf}}(i') = A_{\mathrm{suf}}(i) \subseteq A_{\mathrm{suf}}(i')$ gives $U(y,x) \cap U(z,y) = U(z,x) \subseteq U(z,y)$. For $(a_{t,i}, b_{t,j}, b_{t,j'})$, $B_{t,\mathrm{pref}}(j) \subseteq B_{t,\mathrm{pref}}(j') = B_{t,\mathrm{pref}}(j) \cup U(z,y)$ gives $U(y,x) \subseteq U(z,x) \subseteq U(y,x) \cup U(z,y)$.
\item ($t_x < t_y = t_z \triangleq t$): Write $U(w,x) = H \cap \operatorname{Right}(w)$ for $w \in \{y, z\}$ with $H \triangleq \operatorname{Left}(x) \cap W_{\mathrm{cap}}(t_x+1, t) \cap (\Theta \text{ if } t = t_x+1 \text{ else } E)$. For $(a_{t,i}, a_{t,i'})$, $A_{\mathrm{pref}}(i) \cap U(z,y) = A_{\mathrm{pref}}(i')$ gives $U(y,x) \cap U(z,y) = U(z,x) \subseteq U(y,x)$; for $(a_{t,i}, b_{t,j})$, $A_{\mathrm{pref}}(i) \cap A_{\mathrm{suf}}(i) = E \subseteq A_{\mathrm{pref}}(i)$ gives $U(y,x) \cap U(z,y) = U(z,x) \subseteq U(y,x)$; for $(b_{t,j}, b_{t,j'})$, $B_{t,\mathrm{pref}}(j) \subseteq B_{t,\mathrm{pref}}(j') = B_{t,\mathrm{pref}}(j) \cup U(z,y)$ gives $U(y,x) \subseteq U(z,x) \subseteq U(y,x) \cup U(z,y)$.
\item ($t_x = t_y \triangleq t < t_z$): Here $t \le N - 2$. Write $U(z,w) = \operatorname{Left}(w) \cap R$ with $R \triangleq W_{\mathrm{cap}}(t+1, t_z) \cap \operatorname{Right}(z) \cap (\Theta \text{ if } t_z = t+1 \text{ else } E)$. Cases $(a_{t,i}, a_{t,i'})$ and $(b_{t,j}, b_{t,j'})$ follow symmetrically to Regime~2; for $(a_{t,i}, b_{t,j})$, \eqref{eq:supp-bridge-sandwich} at stage $t$ directly yields $U(y,x) \cap U(z,y) \subseteq A_{\mathrm{suf}}(i) \cap \widetilde{W}_t \cap R = U(z,x) \subseteq U(y,x) \cup U(z,y)$.
\item ($t_x < t_y < t_z$): Here $t_y \le N - 2$, $U(y,x) = H \cap \operatorname{Right}(y)$, $U(z,y) = \operatorname{Left}(y) \cap R$, and $U(z,x) = H \cap \widetilde{W}_{t_y} \cap R \cap E$. If $y = a_{t_y,i}$, $A_{\mathrm{pref}}(i) \cap A_{\mathrm{suf}}(i) = E \subseteq A_{\mathrm{pref}}(i)$ gives $U(y,x) \cap U(z,y) = U(z,x) \subseteq U(y,x)$; if $y = b_{t_y,j}$, \eqref{eq:supp-bridge-sandwich} at stage $t_y$ gives $U(y,x) \cap U(z,y) \subseteq U(z,x) \subseteq U(y,x) \cup U(z,y)$.
\end{enumerate}
Defeat shares follow directly: $a_{t,i}$ ($i < s$) loses to $a_{t,i+1}$ by $1 - \mu(L_{i+1})$; $a_{t,s}$ loses to $b_{t,r_t}$ by $\mu(W_t \cup K_t)$; $b_{t,j}$ ($j \ge 1$) loses to $b_{t,j-1}$ by $1 - \mu(U_{t,j})$; $b_{t,0}$ ($t \le N-2$) loses to $a_{t+1,0}$ by $\mu(W_t \cup K_t)$; and $b_{N-1,0}$ loses to $a_{0,0}$ by $1 - \mu(E \cap \bigcap_{t=0}^{N-1} \widetilde{W}_t)$.
\end{proof}

\begin{theorem}[Multi-Stage Bridge Lower Bounds]\label{thm:supp-bridge}
Let $k \ge 2$ and $\mu = \operatorname{Unif}[0, 1]^k$.
\begin{enumerate}[label=(\roman*),leftmargin=2.2em]
\item The equation $\int_{1/2}^1 V_{k-1}(p/t)\dd t = p$ has a unique solution $p_{\mathrm{half}}(k) \in (0, 1/2)$, and $q_{\mathrm{half}}(k) \triangleq 1 - p_{\mathrm{half}}(k)$ satisfies $q^*(k) \ge q_{\mathrm{half}}(k) > q_{\mathrm{hyp}}(k)$ strictly for all $k \ge 2$. In three dimensions ($k = 3$), $p_{\mathrm{half}}(3)$ is the unique root of
\[
p\left[1 + \ln 2 - \frac{(\ln 2)^2}{2} - (\ln 2)\ln p\right] = \frac{1}{2}.
\]
\item For every dimension $k \ge 3$, $q^*(k) > q_{\mathrm{half}}(k)$ strictly, and at $k = 3$ a finite three-stage agenda certifies $q^*(3) \ge 0.807604$ (limiting root $0.8076046\ldots$).
\end{enumerate}
\end{theorem}

\begin{proof}
We establish part~(i) in Steps~1--2 and part~(ii) in Step~3.

\emph{Step 1 (Unique root $p_{\mathrm{half}}(k)$ and strict gap $q_{\mathrm{half}}(k) > q_{\mathrm{hyp}}(k)$).}
Define $F_k(p) \triangleq \int_{1/2}^1 V_{k-1}(p/t)\dd t$ on $[0, 1/2]$. Since $V_{k-1}(u)$ is continuous and strictly decreasing on $(0, 1)$, $F_k(p) - p$ decreases strictly from $1/2$ to $F_k(1/2) - 1/2 < 0$, yielding a unique root $p_{\mathrm{half}}(k) \in (0, 1/2)$. By Lemma~\ref{lem:hyp-cover}(ii) and $c_k < 1/2$, $c_k - F_k(c_k) = \int_{c_k}^{1/2} V_{k-1}(c_k/t)\dd t > 0$, so $F_k(c_k) < c_k$ and hence $p_{\mathrm{half}}(k) < c_k$, i.e., $q_{\mathrm{half}}(k) > q_{\mathrm{hyp}}(k)$.

To prove $q^*(k) \ge q_{\mathrm{half}}(k)$, set $N = 1$ in Lemma~\ref{lem:supp-nstage} with $W_0 \triangleq \{\theta_k > 1/2\}$ and Lemma~\ref{lem:hyp-cover}(iii)'s $G$-simplex lower boxes $\{L_u\}_{u=1}^s$ covering $K^-_{p_{\mathrm{eff}}}$ and upper boxes $\{U_{0,v}\}_{v=1}^s$ covering $K^+_{p_{\mathrm{eff}}}$ ($p_{\mathrm{eff}} = p_{\mathrm{box}}^{1+(k-1)/G}$). Every box edge fails by $p_{\mathrm{box}}$, while both $(W_0 \cup K_0)^c \subseteq \{\theta_k \le 1/2\} \cap P_{p_{\mathrm{eff}}}$ and $E \cap W_0 \subseteq E_{p_{\mathrm{eff}}} \cap \{\theta_k > 1/2\}$ have measure at most $F_k(p_{\mathrm{eff}})$. Taking any $p_{\mathrm{box}} \in (p_{\mathrm{half}}(k), 1/2)$ gives $F_k(p_{\mathrm{box}}) < p_{\mathrm{box}}$, hence $F_k(p_{\mathrm{eff}}) < p_{\mathrm{box}}$ for large $G$; letting $p_{\mathrm{box}} \downarrow p_{\mathrm{half}}(k)$ yields $q^*(k) \ge q_{\mathrm{half}}(k)$.

\emph{Step 2 (Exact $k = 3$ closed form).}
At $k = 3$, substituting $V_2(u) = 1 - u(1 - \ln u)$ into $F_3(p)$ integrates directly to $F_3(p) = \frac{1}{2} - p\ln 2 + p(\ln 2)\ln p + p\frac{(\ln 2)^2}{2}$; setting $F_3(p) = p$ yields the displayed equation.

\emph{Step 3 (Part (ii): Analytic proof that $q^*(k) > q_{\mathrm{half}}(k)$ for all $k \ge 3$ and certified $k = 3$ bound).}
First verify $c_k < \frac{1}{k+1}$ for all $k \ge 3$, which is equivalent to $\sum_{j=0}^{k-1} \frac{(\ln(k+1))^j}{j!} > k$ (equivalently $\sum_{j=k}^\infty \frac{(\ln(k+1))^j}{j!} < 1$):
\begin{itemize}[leftmargin=2em,itemsep=1pt]
\item $k = 3$: $\ln 4 > 4/3$ gives $1 + \frac{4}{3} + \frac{8}{9} = \frac{29}{9} > 3$;
\item $k = 4$: $e^3 < 21 < 25 = 5^2 \implies \ln 5 > 3/2$ gives $1 + \frac{3}{2} + \frac{9}{8} + \frac{27}{48} = \frac{67}{16} > 4$;
\item $k = 5$: $e^5 < 149 < 216 = 6^3 \implies \ln 6 > 5/3$ gives $1 + \frac{5}{3} + \frac{25}{18} + \frac{125}{162} + \frac{625}{1944} = \frac{10009}{1944} > 5$;
\item $k \ge 6$: $e^2 > 7$ and $\frac{\mathrm{d}}{\mathrm{d}x}(e^{x/3} - x - 1) > 0$ on $[6, \infty)$ give $L \triangleq \ln(k+1) < k/3$, so the geometric tail bound from Step~2 of the proof of Theorem~\ref{thm:highdim}, with $L/(k+1) < 1/3$, gives $\sum_{j=k}^\infty \frac{L^j}{j!} < \frac{(k/3)^k}{k!(1 - 1/3)} \le \frac{3}{2e}(e/3)^k < 1$.
\end{itemize}
Thus $p_0 \triangleq p_{\mathrm{half}}(k) < c_k < \frac{1}{k+1}$. Setting $c_{\mathrm{in}}(p) \triangleq \frac{p}{(1-p)^{k-1}}$, Bernoulli's inequality for $k - 1 \ge 2$ gives $(1 - \frac{1}{k+1})^{k-1} > \frac{2}{k+1}$, so $c_{\mathrm{in}}(p) < 1/2 < 1 - p$ on $(0, \frac{1}{k+1}]$.

Instantiate Lemma~\ref{lem:supp-nstage} with $N = 2$ for $p_{\mathrm{box}} \in (0, c_k)$ and $G \ge 1$: take the $G$-simplex lower boxes covering $K^-_{p_{\mathrm{eff}}}$; set Stage~$0$ to $r_0 = k - 1$ slabs $U_{0,d} \triangleq \{\theta_d > 1 - p_{\mathrm{box}}\}$ ($1 \le d \le k-1$) with $W_0 \triangleq \{\theta_k > c_{\mathrm{in}}(p_{\mathrm{box}})\}$; and set Stage~$1$ to the $G$-simplex upper boxes covering $K^+_{p_{\mathrm{eff}}}$ with $W_1 \triangleq \{\theta_1 > 1 - b_{\mathrm{last}}\}$, where $b_{\mathrm{last}} = b_{\mathrm{last}}(G, p_{\mathrm{box}}) \in (p_{\mathrm{box}}, 1)$ uniquely solves $H_{G,p_{\mathrm{box}}}(b) \triangleq \int_b^1 V_{k-1}(p_{\mathrm{eff}}/t)\dd t = p_{\mathrm{box}}$ (a solution exists and is unique: $H_{G,p_{\mathrm{box}}}$ is continuous and strictly decreasing on $[p_{\mathrm{eff}}, 1]$ with $H_{G,p_{\mathrm{box}}}(1) = 0$, and $H_{G,p_{\mathrm{box}}}(p_{\mathrm{box}}) \ge \int_{p_{\mathrm{box}}}^1 V_{k-1}(p_{\mathrm{box}}/t)\dd t = V_k(p_{\mathrm{box}}) > c_k > p_{\mathrm{box}}$ by Lemma~\ref{lem:hyp-cover}(ii), because $p_{\mathrm{eff}} \le p_{\mathrm{box}} < c_k$). All box edges and stage masks $(W_t \cup K_t)^c$ have failure measure at most $p_{\mathrm{box}}$. Set $c_{\mathrm{in}}^0 \triangleq c_{\mathrm{in}}(p_0) < 1/2$ and $W_{\mathrm{lim}} \triangleq \{\theta_1 > 1/2\}$, and let $\widetilde{W}_0^{\mathrm{lim}} \triangleq \{\theta_k > c_{\mathrm{in}}^0\} \cup \{\theta : \theta_d > 1 - p_0 \text{ for at least two } d \in \{1, \dots, k-1\}\}$ denote the widened stage-$0$ mask $\widetilde{W}_0$ at $p_{\mathrm{box}} = p_0$. Coordinate symmetry gives $\mu(E_{p_0} \cap W_{\mathrm{lim}}) = p_0$, while the positive-volume box
\[
Q \triangleq \left(1 - \frac{p_0}{3},\; 1\right] \times \prod_{d=2}^{k-1} \left((1 - p_0)\left(1 - \frac{p_0}{3}\right)^{\frac{1}{k-2}},\; 1 - p_0\right] \times \left(\left(1 - \frac{p_0}{3}\right)c_{\mathrm{in}}^0,\; c_{\mathrm{in}}^0\right]
\]
satisfies (i) $\theta_1 > 1 - p_0/3 > 1/2$ ($\theta \in W_{\mathrm{lim}}$), (ii) $\theta_k \le c_{\mathrm{in}}^0$ and $\theta_d \le 1 - p_0$ for $2 \le d \le k-1$ ($\theta \notin \widetilde{W}_0^{\mathrm{lim}}$), and (iii) $\prod_{d=1}^k \theta_d > \frac{(1-p_0/3)^3}{1-p_0}p_0 > p_0$ ($\theta \in E_{p_0}$). Hence $\mu(E_{p_0} \cap \widetilde{W}_0^{\mathrm{lim}} \cap W_{\mathrm{lim}}) \le p_0 - \eta$ with $\eta \triangleq \mu(Q) > 0$. As $(p_{\mathrm{box}}, G) \to (p_0, \infty)$, we have $p_{\mathrm{eff}} \to p_0$, $c_{\mathrm{in}}(p_{\mathrm{box}}) \to c_{\mathrm{in}}^0$, and $b_{\mathrm{last}} \to 1/2$, the unique solution of $\int_b^1 V_{k-1}(p_0/t)\dd t = p_0$ (as $H_{G,p_{\mathrm{box}}}$ converges uniformly to this strictly decreasing integral). The indicator of $E_{p_{\mathrm{eff}}} \cap \widetilde{W}_0 \cap W_1$ therefore converges to that of $E_{p_0} \cap \widetilde{W}_0^{\mathrm{lim}} \cap W_{\mathrm{lim}}$ off a Lebesgue-null set, and dominated convergence gives $\mu(E_{p_{\mathrm{eff}}} \cap \widetilde{W}_0 \cap W_1) \to \mu(E_{p_0} \cap \widetilde{W}_0^{\mathrm{lim}} \cap W_{\mathrm{lim}}) \le p_0 - \eta$. Hence there exist $p_{\mathrm{box}} \in (p_0 - \eta/2, p_0)$ and $G$ for which the closing mass lies below $p_0 - \eta/2 < p_{\mathrm{box}}$, proving $q^*(k) > q_{\mathrm{half}}(k)$ for every $k \ge 3$.

For the certified $k = 3$ bound, instantiate Lemma~\ref{lem:supp-nstage} with $N = 3$ at grid $G \ge 1$ ($p_{\mathrm{eff}} = p_{\mathrm{box}}^{1+2/G}$, $c_{\mathrm{in}} = p_{\mathrm{box}}/(1-p_{\mathrm{box}})^2$): take $G$-simplex lower boxes for $K^-_{p_{\mathrm{eff}}}$; Stage~$0$ slabs $\{\theta_1 > 1 - p_{\mathrm{box}}\}, \{\theta_2 > 1 - p_{\mathrm{box}}\}$ with $W_0 = \{\theta_3 > c_{\mathrm{in}}\}$; Stage~$1$ slabs $\{\theta_1 > 1 - p_{\mathrm{box}}\}, \{\theta_3 > 1 - p_{\mathrm{box}}\}$ with $W_1 = \{\theta_2 > c_{\mathrm{in}}\}$; and Stage~$2$ $G$-simplex upper boxes for $K^+_{p_{\mathrm{eff}}}$ with $W_2 = \{\theta_1 > 1 - b_{\mathrm{last}}\}$, where $b_{\mathrm{last}}$ solves $H_{G,p_{\mathrm{box}}}(b) = (1-b) + p_{\mathrm{eff}}(\ln b)(1 - \ln p_{\mathrm{eff}} + \frac{1}{2}\ln b) = p_{\mathrm{box}}$. Setting $a \triangleq 1 - b_{\mathrm{last}}$ and $t_* \triangleq p_{\mathrm{eff}}/c_{\mathrm{in}}$, the closing region $E_{p_{\mathrm{eff}}} \cap \widetilde{W}_0 \cap \widetilde{W}_1 \cap W_2$ partitions into disjoint sets $\Omega_1 \triangleq \{\theta_1 > a,\; \theta_2, \theta_3 > c_{\mathrm{in}},\; \theta_1\theta_2\theta_3 > p_{\mathrm{eff}}\}$, $\Omega_2 \triangleq \{\theta_1, \theta_2 > 1 - p_{\mathrm{box}},\; \theta_3 \le c_{\mathrm{in}},\; \theta_1\theta_2\theta_3 > p_{\mathrm{eff}}\}$, and its symmetric counterpart $\Omega_3$. For $p_{\mathrm{box}} < p_{\mathrm{half}}(3)$ these sets do partition the closing region, because $c_{\mathrm{in}} < 1/2 < 1 - p_{\mathrm{box}}$ (shown above, as $p_{\mathrm{half}}(3) < c_3 < 1/4$) and $a < 1/2$ (shown below). Suppose $p_{\mathrm{eff}} < c_{\mathrm{in}} < a < t_* < 1$. For $\kappa \in [c_{\mathrm{in}}, 1)$ the slice $\{(\theta_2, \theta_3) \in (c_{\mathrm{in}}, 1]^2 : \theta_2\theta_3 > \kappa\}$ has area $V_2(\kappa)$, since the lower bounds are slack, and for $\kappa \in (c_{\mathrm{in}}^2, c_{\mathrm{in}})$ it has area
\[
A(\kappa) \triangleq \int_{c_{\mathrm{in}}}^{\kappa/c_{\mathrm{in}}} \Big(1 - \frac{\kappa}{\theta_2}\Big)\dd\theta_2 + (1 - c_{\mathrm{in}})\Big(1 - \frac{\kappa}{c_{\mathrm{in}}}\Big) = 1 - 2c_{\mathrm{in}} + \kappa\big(1 + 2\ln c_{\mathrm{in}} - \ln\kappa\big).
\]
With $\kappa = p_{\mathrm{eff}}/\theta_1$ the two regimes meet at $\theta_1 = t_*$, and $p_{\mathrm{eff}} > c_{\mathrm{in}}^2$ because $t_* > c_{\mathrm{in}}$, so $\mu(\Omega_1) = \int_a^{t_*} V_2(p_{\mathrm{eff}}/\theta_1)\dd\theta_1 + \int_{t_*}^1 A(p_{\mathrm{eff}}/\theta_1)\dd\theta_1$, which evaluates to the first expression below on substituting $\ln c_{\mathrm{in}} = \ln p_{\mathrm{eff}} - \ln t_*$. For $\Omega_2$, $t_* = p_{\mathrm{box}}^{2/G}(1 - p_{\mathrm{box}})^2 \le (1 - p_{\mathrm{box}})^2$, so $p_{\mathrm{eff}}/(\theta_1\theta_2) < c_{\mathrm{in}}$ throughout $(1 - p_{\mathrm{box}}, 1]^2$ and $\mu(\Omega_2)$ is the integral of $c_{\mathrm{in}} - p_{\mathrm{eff}}/(\theta_1\theta_2)$ over that square. Hence
\begin{align*}
\mu(\Omega_1) &= b_{\mathrm{last}} - 2c_{\mathrm{in}}(1 - t_*) + p_{\mathrm{eff}}(1 - \ln p_{\mathrm{eff}})\ln a \\
&\quad + \tfrac{1}{2}p_{\mathrm{eff}}(\ln a)^2 - 2p_{\mathrm{eff}}\ln t_* + p_{\mathrm{eff}}(\ln t_*)^2, \\
\mu(\Omega_2) &= \mu(\Omega_3) = c_{\mathrm{in}} p_{\mathrm{box}}^2 - p_{\mathrm{eff}}\big(\ln(1 - p_{\mathrm{box}})\big)^2.
\end{align*}
Since $H_{G,p_{\mathrm{box}}}(b)$ decreases strictly in $b$ while $\mu(\Omega_1)$ increases in $b$, any $b_{\mathrm{up}}$ with $H_{G,p_{\mathrm{box}}}(b_{\mathrm{up}}) < p_{\mathrm{box}}$ bounds $b_{\mathrm{last}} < b_{\mathrm{up}}$ and caps the closing mass upon substituting $a_{\mathrm{low}} \triangleq 1 - b_{\mathrm{up}}$ (note $H_{G,p_{\mathrm{box}}}(1/2) = F_3(p_{\mathrm{eff}}) > p_{\mathrm{half}}(3) > p_{\mathrm{box}}$ ensures $b_{\mathrm{last}} > 1/2$, so $a < 1/2$). Outward-rounded nine-decimal interval arithmetic certifies: at $G = 10^6$ ($s = 500{,}001{,}500{,}001$, $m = 2{,}000{,}006{,}000{,}014$) and $q = 0.807604$ ($p_{\mathrm{box}} = 0.192396$), $b_{\mathrm{up}} = 0.502665$ yields $p_{\mathrm{eff}} \in [0.192395365, 0.192395366]$, $c_{\mathrm{in}} \in [0.294984445, 0.294984446]$, $t_* \in [0.652222070, 0.652222071]$, $H_{10^6,p_{\mathrm{box}}}(b_{\mathrm{up}}) \le 0.1923958668 < p_{\mathrm{box}}$, and $\mu(\Omega_1) + 2\mu(\Omega_2) \le 0.1923948331 < p_{\mathrm{box}}$, certifying $q^*(3) \ge 0.807604$ (with limiting root $\approx 0.80760463$).
\end{proof}

\begin{theorem}[Single-Bridge Shadow-Cut Ceiling]\label{thm:supp-shadow}
Let $k \ge 2$, $\mu = \operatorname{Unif}[0, 1]^k$, and $q = 1 - p \in (1/2, 1)$. Suppose a strict-preference CW-SCP profile at defeat threshold $q$ realises any of the following three single-bridge cycle architectures, with $r, s, t \ge 1$:
\begin{enumerate}[label=(\alph*),leftmargin=2.2em]
\item a two-block cycle $a_0 \leftarrow \dots \leftarrow a_{r-1} \leftarrow b_{s-1} \leftarrow \dots \leftarrow b_0 \leftarrow a_0$ with $a_0 < \dots < a_{r-1} < b_0 < \dots < b_{s-1}$;
\item a three-block cycle $a_0 \leftarrow \dots \leftarrow a_{r-1} \leftarrow c_{t-1} \leftarrow \dots \leftarrow c_0 \leftarrow b_0 \leftarrow \dots \leftarrow b_{s-1} \leftarrow a_0$ with $a_0 < \dots < a_{r-1} < b_0 < \dots < b_{s-1} < c_0 < \dots < c_{t-1}$;
\item a swapped-end two-run cycle $a_0 \leftarrow a_1 \leftarrow \dots \leftarrow a_r \leftarrow y_{s+1} \leftarrow y_s \leftarrow \dots \leftarrow y_1 \leftarrow a_0$ with $a_0 < y_1 < \dots < y_s < a_1 < \dots < a_r < y_{s+1}$.
\end{enumerate}
(The two-chain bridge of Theorem~\ref{thm:supp-bridge}(i), with any Borel upper set in place of $\{\theta_k > 1/2\}$, is an instance of class~(a).)
Then there exist a Borel lower set $E_*$ with $\mu(E_*) \le p$ and a Borel upper set $F_*$ with $\mu(F_*) \le p$ satisfying $\operatorname{dn}(E_p \setminus F_*) \cap P_p \subseteq E_*$, which implies $\mu(E_*) + \mu(F_*) \ge V_k(p)$ and $V_k(p) \le 2p$. Moreover, $V_k(2^{-k}) > 2 \cdot 2^{-k}$ for every $k \ge 3$, so for such $k$ every profile in classes~(a)--(c) satisfies
\[
q \le q_{\mathrm{sc}}(k) \triangleq 1 - p_{\mathrm{sc}}(k) < 1 - 2^{-k},
\]
where $p_{\mathrm{sc}}(k) \in (2^{-k}, 1/2)$ is the unique root of $V_k(p) = 2p$.
\end{theorem}

\begin{proof}
\emph{Step 1 (Single-bridge reduction to $\operatorname{dn}(E_p \setminus F_*) \cap P_p \subseteq E_*$).}
By Lemma~\ref{lem:hyp-cover}(i), every ascending defeat $x \leftarrow y$ ($x < y$) at share $\ge 1 - p$ satisfies $E_p \subseteq U(y, x)$, and every descending defeat $y \leftarrow x$ satisfies $P_p \subseteq U(y, x)^c$. Across the three classes:
\begin{itemize}[leftmargin=2em,itemsep=1pt]
\item Class~(a): Set $F_* \triangleq U(b_0, a_0)$ and $E_* \triangleq U(b_{s-1}, a_{r-1})^c$ (each of measure $\le p$). On $E_p \setminus F_*$, transitivity gives $a_{r-1} \succ_\theta b_0$, placing $\operatorname{dn}(E_p \setminus F_*) \subseteq U(b_0, a_{r-1})^c$; on $P_p \setminus E_*$, transitivity gives $b_0 \succ_\eta a_{r-1}$, placing $P_p \setminus E_* \subseteq U(b_0, a_{r-1})$. Disjointness yields $\operatorname{dn}(E_p \setminus F_*) \cap P_p \subseteq E_*$.
\item Class~(b): Set $F_* \triangleq U(b_{s-1}, a_0)$ and $E_* \triangleq U(c_{t-1}, a_{r-1})^c$. Ascending chains in $A, B$ on $E_p \setminus F_*$ place $\operatorname{dn}(E_p \setminus F_*) \subseteq U(b_0, a_{r-1})^c$, while descending edges in $C$ on $P_p \setminus E_*$ place $P_p \setminus E_* \subseteq U(b_0, a_{r-1})$.
\item Class~(c): Set $F_* \triangleq U(y_{s+1}, y_s)$ and $E_* \triangleq U(a_1, a_0)^c$. Ascending edges on $E_p \setminus F_*$ place $\operatorname{dn}(E_p \setminus F_*) \subseteq U(a_1, y_s)^c$, while descending edges on $P_p \setminus E_*$ place $P_p \setminus E_* \subseteq U(a_1, y_s)$.
\end{itemize}

\emph{Step 2 (Shadow-cut volume inequality $\mu(E_*) + \mu(F_*) \ge V_k(p)$ via positive association).}
By inner regularity of Lebesgue measure, choose an $F_\sigma$ subset $K = \bigcup_{n=1}^\infty K_n \subseteq E_p \setminus F_*$ with each $K_n$ compact and $\mu(K) = \mu(E_p \setminus F_*)$. Each $\operatorname{dn}(K_n)$ is compact (continuous image of $\{(\eta,\theta) \in [0,1]^k \times K_n : \eta \le \theta\}$), so $\operatorname{dn}(K)$ is Borel and $U \triangleq [0,1]^k \setminus \operatorname{dn}(K)$ is a Borel upper set. Because $U \cap K = \emptyset$ and $P_p \setminus U \subseteq \operatorname{dn}(K) \cap P_p \subseteq E_*$, we get $\mu(E_p \cap U) \le \mu(F_*)$ and $\mu(P_p \setminus U) \le \mu(E_*)$. Summing and using $\mu(E_p) = \mu(P_p) = V_k(p)$ gives
\[
\mu(E_*) + \mu(F_*) \ge \mu(P_p \setminus U) + \mu(E_p \cap U) = V_k(p) + \mu(E_p \cap U) - \mu(P_p \cap U).
\]
Because $E_p$ and $U$ are upper sets while $P_p$ is a lower set under $\operatorname{Unif}[0,1]^k$, positive association (Lemma~\ref{lem:structure}(ii)) yields $\mu(E_p \cap U) \ge V_k(p)\mu(U) \ge \mu(P_p \cap U)$, proving $\mu(E_*) + \mu(F_*) \ge V_k(p)$ and hence $V_k(p) \le 2p$.

\emph{Step 3 (Strict separation $q_{\mathrm{sc}}(k) < 1 - 2^{-k}$ for all $k \ge 3$).}
Since $V_k(2^{-k}) = 2^{-k}\sum_{j=k}^\infty \frac{(k\ln 2)^j}{j!}$, we require $\sum_{j=k}^\infty \frac{(k\ln 2)^j}{j!} > 2$. At $k = 3$, $\ln 2 > 0.693$ gives $\frac{(3\ln 2)^3}{6} + \frac{(3\ln 2)^4}{24} > 2.275 > 2$. For $k \ge 4$, using the elementary upper bound $k! \le e\, k^{k+1/2}e^{-k}$ and $e\ln 2 > 1.88$,
\[
\frac{(k\ln 2)^k}{k!} \ge \frac{(e\ln 2)^k}{e\sqrt{k}} > \frac{1.88^k}{e\sqrt{k}},
\]
which exceeds $2$ at $k = 4$ ($1.88^4/(2e) > 2.29$) and increases strictly in $k$ since $1.88 / \sqrt{1 + 1/k} > 1$. Because $V_k(p) - 2p$ decreases strictly from $1$ to $-2$, its unique root satisfies $p_{\mathrm{sc}}(k) > 2^{-k}$, so $q \le q_{\mathrm{sc}}(k) < 1 - 2^{-k}$.
\end{proof}

\begin{table}[htbp]
\centering
\caption{Numerical Supermajority Core Bounds across Voter Dimensions ($k = 2, \dots, 7$)}
\label{tab:supp-dim-bounds}
\small
\setlength{\tabcolsep}{3.5pt}
\begin{tabular}{rcccccc}
\toprule
 & Linear & \multicolumn{3}{c}{Lower bounds on $q^*(k)$ (independent types)} & Single-bridge & Upper \\
\cmidrule(lr){2-2}\cmidrule(lr){3-5}\cmidrule(lr){6-6}\cmidrule(lr){7-7}
$k$ & CN (unif.) & $r_k$ & $q_{\mathrm{hyp}}(k)$ & $q_{\mathrm{half}}(k)$ & ceiling $q_{\mathrm{sc}}(k)$ & bound $1 - 2^{-k}$ \\
\midrule
2 & 0.555556 & 0.618034 & 0.682156 & 0.704692 & --- & 0.750000 \\
3 & 0.578125 & 0.682328 & 0.791563 & 0.807256 & 0.851181 & 0.875000 \\
4 & 0.590400 & 0.724492 & 0.861400 & 0.871866 & 0.899708 & 0.937500 \\
5 & 0.598122 & 0.754878 & 0.907095 & 0.914055 & 0.932156 & 0.968750 \\
6 & 0.603431 & 0.778090 & 0.937396 & 0.942038 & 0.953976 & 0.984375 \\
7 & 0.607304 & 0.796544 & 0.957656 & 0.960763 & 0.968709 & 0.992188 \\
\bottomrule
\end{tabular}
\par\medskip
\begin{minipage}{0.95\linewidth}
\footnotesize\emph{Notes:} Six-decimal evaluations of each threshold. CN~(unif.) reports $1 - (k/(k+1))^k$ \citep{CaplinNalebuff1988,CaplinNalebuff1991} (their general log-concave bound is $1 - 1/e \approx 0.632121$; \citeauthor{WuWu2026}'s \citeyearpar{WuWu2026} convex-coalition bound is $d/(d+1)$). Here $r_k$ solves $z^k + z = 1$, $q_{\mathrm{hyp}}(k)$ is from Theorem~\ref{thm:highdim}(ii), $q_{\mathrm{half}}(k)$ from Theorem~\ref{thm:supp-bridge}(i), and $q_{\mathrm{sc}}(k)$ from Theorem~\ref{thm:supp-shadow} (bounding only the three single-bridge classes, not $q^*(k)$). At $k = 3$, Theorem~\ref{thm:supp-bridge}(ii) certifies $q^*(3) \ge 0.807604$, yielding $q^*(3) \in [0.807604, 0.875000]$.
\end{minipage}
\end{table}

\section{Distribution-Free Condorcet Domains and\texorpdfstring{\\}{ }Four-Alternative Cycles}\label{sec:supp-condorcet}

Say that a family $\mathcal{D}_\Theta = \{\succ_\theta : \theta \in \Theta\}$ of strict linear orders on $X$ is \emph{distribution-free Condorcet-safe} if majority preference $\succ_{\mathrm{maj}}$ is transitive on $X$ for every tie-free distribution $\mu$ supported on $\Theta$.

\begin{proposition}[Distribution-Free Condorcet Domains under CW-SCP]\label{prop:supp-condorcet}
Suppose preferences satisfy CW-SCP on $X \times \Theta$.
\begin{enumerate}[label=(\roman*),leftmargin=2.2em]
\item A family $\mathcal{D}_\Theta = \{\succ_\theta : \theta \in \Theta\}$ of strict linear orders is distribution-free Condorcet-safe if and only if every triple $a < b < c$ in $X$ satisfies at least one of the four conditions:
\[
U(c, b) \subseteq U(b, a), \quad U(b, a) \subseteq U(c, b), \quad U(c, a) = U(b, a), \quad U(c, a) = U(c, b).
\]
\item Suppose $\Theta = \prod_{d=1}^k I_d$ is a Cartesian product of intervals and preferences (permitting indifferences) are represented by a utility continuous in $\theta$ for each $x \in X$. If for some triple $a < b < c$ there exist types $\theta^A, \theta^B \in \Theta$ exhibiting strict middle-peakedness ($b \succ_{\theta^A} a, c$) and strict middle-dippedness ($a, c \succ_{\theta^B} b$), then $\Theta$ contains types holding all six strict rankings on $\{a, b, c\}$; in particular, some finite tie-free distributions on $\{a, b, c\}$ generate strict Condorcet cycles.
\end{enumerate}
\end{proposition}

\begin{proof}
\emph{Step 1 (Part (i): Value-restriction equivalence and product-lattice dichotomy).}
For any $a < b < c$, write $A \triangleq U(b, a)$, $B \triangleq U(c, b)$, and $C \triangleq U(c, a)$, which satisfy $A \cap B \subseteq C \subseteq A \cup B$ by~\eqref{eq:triple-sandwich}. Expressing the six strict rankings on $\{a, b, c\}$ via $(A, B, C)$, the four conditions correspond directly to \citeauthor{Sen1966}'s \citeyearpar{Sen1966} triple value restrictions: $B \subseteq A$ rules out $b$ being worst ($A^c \cap B = \emptyset$), $A \subseteq B$ rules out $b$ being best ($A \cap B^c = \emptyset$), $C = A$ rules out $a$ being middle ($A^c \cap B \cap C = A \cap B^c \cap C^c = \emptyset$), and $C = B$ rules out $c$ being middle ($A \cap B^c \cap C = A^c \cap B \cap C^c = \emptyset$). Hence if every triple satisfies one of the four conditions, majority preference is transitive for every tie-free $\mu$ \citep{Sen1966}.

Conversely, suppose none of the four conditions holds for some $a < b < c$. Picking $\theta^A \in R_{10} \triangleq A \setminus B$ and $\theta^B \in R_{01} \triangleq B \setminus A$ gives $\theta^A \wedge \theta^B \in A^c \cap B^c$ (holding $\pi_1 = (a, b, c)$) and $\theta^A \vee \theta^B \in A \cap B$ (holding $\sigma_1 = (c, b, a)$). Define $\pi_2 = (b, c, a)$ on $C \cap R_{10}$, $\sigma_2 = (b, a, c)$ on $R_{10} \setminus C$, $\pi_3 = (c, a, b)$ on $C \cap R_{01}$, and $\sigma_3 = (a, c, b)$ on $R_{01} \setminus C$. If both $B^c \cap C$ and $C \cap A^c$ are nonempty, $\{\pi_1, \pi_2, \pi_3\}$ forms a Condorcet triple; if $C \cap A^c = \emptyset$, then $C \subsetneq A$ gives $A \cap C^c \neq \emptyset$ alongside $\theta^B \in C^c \cap B$, yielding $\{\sigma_1, \sigma_2, \sigma_3\}$; symmetrically if $B^c \cap C = \emptyset$, $C \subsetneq B$ again yields $\{\sigma_1, \sigma_2, \sigma_3\}$. Placing weight $1/3$ on the resulting triple generates a strict $2:1$ majority cycle.

\emph{Step 2 (Part (ii): Connected box $4$-edge intermediate-value argument).}
Definition~\ref{def:cwscp}(i) forces $\theta^A$ and $\theta^B$ to be incomparable, so $\mathcal{I}_< \triangleq \{d : \theta^A_d < \theta^B_d\}$ and $\mathcal{I}_> \triangleq \{d : \theta^A_d > \theta^B_d\}$ are nonempty. Define the continuous nondecreasing map $\gamma : [0, 1]^2 \to [\theta^A \wedge \theta^B, \theta^A \vee \theta^B] \subseteq \Theta$ by $\gamma_d(s, t) = (1-s)\theta^A_d + s\theta^B_d$ ($d \in \mathcal{I}_<$), $(1-t)\theta^B_d + t\theta^A_d$ ($d \in \mathcal{I}_>$), and $\theta^A_d$ otherwise, with vertices $\gamma(0,0) = \theta^A \wedge \theta^B$, $\gamma(0,1) = \theta^A$, $\gamma(1,0) = \theta^B$, and $\gamma(1,1) = \theta^A \vee \theta^B$. Let $h_1(s, t) \triangleq u(b, \gamma(s,t)) - u(a, \gamma(s,t))$ and $h_2(s, t) \triangleq u(c, \gamma(s,t)) - u(b, \gamma(s,t))$. By Definition~\ref{def:cwscp}, $h_1 > 0$ on $t = 1$, $h_1 < 0$ on $t = 0$, $h_2 > 0$ on $s = 1$, and $h_2 < 0$ on $s = 0$. Thus $\gamma(0,0)$ yields $a \succ b \succ c$ and $\gamma(1,1)$ yields $c \succ b \succ a$. Along the four edges, continuity produces the remaining four strict orders: near $t_0^+ \triangleq \sup\{t : h_1(0, t) \le 0\} \in (0, 1)$ (taking $t > t_0^+$ gives $b \succ a \succ c$); near $t_1^- \triangleq \inf\{t : h_1(1, t) \ge 0\} \in (0, 1)$ ($t < t_1^-$ gives $c \succ a \succ b$); near $s_1^- \triangleq \inf\{s : h_2(s, 1) \ge 0\} \in (0, 1)$ ($s < s_1^-$ gives $b \succ c \succ a$); and near $s_0^+ \triangleq \sup\{s : h_2(s, 0) \le 0\} \in (0, 1)$ ($s > s_0^+$ gives $a \succ c \succ b$). Placing mass $1/3$ on three cyclic strict types yields a strict $2:1$ majority cycle.\footnote{Requiring only non-nested $U^>(b,a)$ and $U^>(c,b)$ in part~(ii) is insufficient: $u(a,\theta)=0$, $u(b,\theta)=(\theta_1 - 1/2)_+$, and $u(c,\theta)=(\theta_1 - 1/2)_+ + (\theta_2 - 1/2)_+$ produce non-nested strict sets on $[0,1]^2$ while every type weakly prefers $c \succeq b \succeq a$.}
\end{proof}

\begin{remark}[Exhaustive $4$-Cycle Classification under Positive Association]\label{rem:supp-4cycle}
Under CW-SCP and any positively associated probability measure $\mu$ on $\Theta \subseteq \mathbb{R}^k$, reflecting $\theta \mapsto -\theta$ and reversing preferences $\succeq_\theta \mapsto \preceq_\theta$ preserves CW-SCP and positive association while reversing every majority edge. Thus it suffices to check one orientation from each of the three Hamiltonian $4$-cycle pairs on $0 < 1 < 2 < 3$ at defeat share $q = 1 - p$: (1)~the one-sided cycle $0 \leftarrow 1 \leftarrow 2 \leftarrow 3 \leftarrow 0$ obeys $q \le r_3 \approx 0.6823$ by Theorem~\ref{thm:median-core}(ii); (2)~the alternating cycle $0 \leftarrow 2 \leftarrow 3 \leftarrow 1 \leftarrow 0$ is $n = 2$ of Theorem~\ref{thm:highdim}(i), giving $q \le 2/3$; and (3)~the mixed cycle $0 \leftarrow 3 \leftarrow 1 \leftarrow 2 \leftarrow 0$ satisfies $L^\ge(0,1) \cap L^>(1,3) \subseteq \Theta \setminus U^>(3,0)$ and $U^>(2,1) \cap U^>(1,0) \subseteq \Theta \setminus L^>(0,2)$, so Lemma~\ref{lem:structure}(ii) yields $(1-p)(1 - \mu(U^>(1,0))) \le p$ and $(1-p)\mu(U^>(1,0)) \le p$, summing to $1 - p \le 2p \implies q \le 2/3$.
\end{remark}

\section{Smooth Policy Continua and Cartesian Grid Robustness}\label{sec:supp-smooth}

We first prove a unified lifting lemma that converts any cycle on finitely many grid policies with uniform strict utility slack into a $C^\infty$ strictly increasing-differences utility on $X = [0, 1]$ with empty $q$-core.

\begin{lemma}[Smooth Continuum Lifting]\label{lem:supp-lifting}
Let $s \ge 2$, $x_i \triangleq i/s$ ($0 \le i \le s$), $\Theta \subseteq \mathbb{R}^k$ be bounded with Borel probability measure $\mu$, and $\tilde{u}_0 : [0, 1] \times \Theta \to \mathbb{R}$ be $C^\infty$ in $x$ and Borel measurable in $\theta$, with strictly increasing differences, $L \triangleq \sup_{[0,1]\times\Theta} |\partial_x \tilde{u}_0| < \infty$, and $B \triangleq \sup_{[0,1]\times\Theta} \tilde{u}_0 - \inf_{\Theta} \tilde{u}_0(0, \cdot) < \infty$. Suppose that for some quota $q \in [1/2, 1)$ and $\Delta > 0$, each grid policy $x_i$ ($0 \le i \le s$) admits a challenger grid policy $x_{w(i)}$ and Borel set $W_i \subseteq \Theta$ with $\mu(W_i) > q$ such that $\tilde{u}_0(x_{w(i)}, \theta) - \tilde{u}_0(x_i, \theta) \ge \Delta$ for all $\theta \in W_i$. Then for sufficiently large $\Lambda > 0$,
\[
\tilde{u}(x, \theta) \triangleq \tilde{u}_0(x, \theta) - \Lambda\sin^2(s\pi x)
\]
preserves strictly increasing differences (and joint $C^\infty$ smoothness whenever $\tilde{u}_0$ is $C^\infty$) and satisfies $C_{>q}([0, 1], \mu) = \emptyset$.
\end{lemma}

\begin{proof}
Because $-\Lambda\sin^2(s\pi x)$ is independent of $\theta$, $\tilde{u}$ retains strictly increasing differences identically, and every strict or weak comparison set of $\tilde{u}$ is a superlevel set of the Borel function $\tilde{u}_0(y, \cdot) - \tilde{u}_0(x, \cdot)$, hence Borel, so the standing measurability convention holds. Pick $\gamma \in (0, \frac{1}{4s})$ with $L\gamma < \Delta$, define $N_\gamma \triangleq \bigcup_{i=0}^s ((x_i - \gamma, x_i + \gamma) \cap [0, 1])$, note $s_{\min} \triangleq \min_{x \in [0,1]\setminus N_\gamma} \sin^2(s\pi x) = \sin^2(s\pi\gamma) > 0$, and choose $\Lambda > B / s_{\min}$.
\begin{itemize}[leftmargin=2em,itemsep=1pt]
\item If $x \in N_\gamma$, pick $i$ with $|x - x_i| < \gamma$. Since $-\Lambda\sin^2(s\pi x) \le 0$ and $\sin^2(s\pi x_{w(i)}) = 0$, every $\theta \in W_i$ satisfies $\tilde{u}(x, \theta) \le \tilde{u}_0(x_i, \theta) + L\gamma < \tilde{u}_0(x_i, \theta) + \Delta \le \tilde{u}(x_{w(i)}, \theta)$, so $x_{w(i)}$ defeats $x$ on $W_i$ ($\mu(W_i) > q$).
\item If $x \in [0, 1] \setminus N_\gamma$, every $\theta \in \Theta$ satisfies $\tilde{u}(x, \theta) \le \tilde{u}_0(x, \theta) - \Lambda s_{\min} < \tilde{u}_0(0, \theta) = \tilde{u}(0, \theta)$, so $0$ defeats $x$ unanimously ($\mu = 1 > q$).
\end{itemize}
Hence no policy in $[0, 1]$ lies in $C_{>q}([0, 1], \mu)$.
\end{proof}

\begin{proposition}[Grid and Smooth Continuum Extensions]\label{prop:supp-smooth}
The following extensions of Proposition~\ref{prop:cycles} and of the one-sided cycles of Section~\ref{sec:highdim} hold:
\begin{enumerate}[label=(\roman*),leftmargin=2.2em]
\item The three-policy utility of Proposition~\ref{prop:cycles}(i) generates a strict simple-majority cycle on an equally weighted $5 \times 5$ Cartesian grid $\Theta_d = \{0, \eta, 2\eta, 1, 1+\eta\}$ ($0 < \eta < 1/20$).
\item For each product distribution in Proposition~\ref{prop:cycles} (including the $5 \times 5$ grid), there exists an additively separable $C^\infty$ utility $\tilde{u}(x, \theta) = \tilde{f}(x) + \tilde{\phi}_1(x)\theta_1 + \tilde{\phi}_2(x)\theta_2$ with $\tilde{\phi}'_1, \tilde{\phi}'_2 > 0$ on $[0, 1]$ admitting no weak or strict Condorcet winner on $X = [0, 1]$.
\item For every $k \ge 2$, $\mu = \operatorname{Unif}[0, 1]^k$, and $q < r_k$ (respectively, $q < q_{\mathrm{hyp}}(k)$), there exists an additively separable (respectively, non-additive) $C^\infty$ strictly increasing-differences utility on $[0, 1] \times [0, 1]^k$ with $C_{>q}([0, 1], \mu) = \emptyset$.
\end{enumerate}
\end{proposition}

\begin{proof}
\emph{Part~(i).} In Proposition~\ref{prop:cycles}(i), the smallest pairwise utility gap on $\{0, 1\}^2$ is $1/4$. Perturbing each coordinate by $\delta_d \in [0, 2\eta]$ shifts any pairwise difference by at most $\frac{5}{4}(2\eta) + \frac{5}{4}(2\eta) = 5\eta < 1/4$, leaving all $25$ strict rankings invariant and yielding majority shares $15/25, 15/25, 16/25$.

\emph{Part~(ii).} Place the discrete values $(f_0, \phi_{1,0}, \phi_{2,0})$ at $x_0 = 0, x_1 = 1/2, x_2 = 1$ ($s = 2$), and fix the standard $C^\infty$ step $\psi(t) \triangleq \int_{-\infty}^t \rho(v)\dd v / \int_0^1 \rho(v)\dd v$ ($\rho(v) = e^{-1/(v(1-v))}$ on $(0, 1)$ and $0$ elsewhere). For any $z \in (0, 1)$, picking $\alpha \in (0, \min(2z, 2(1-z)))$ places $\tilde{z} \triangleq \frac{z - \alpha/2}{1 - \alpha} \in (0, 1)$, so $P_z(x) \triangleq \alpha x + (1-\alpha)\big(\tilde{z}\psi(2x) + (1-\tilde{z})\psi(2x-1)\big)$ is $C^\infty$ with $P_z(0) = 0$, $P_z(1/2) = z$, $P_z(1) = 1$, and $P'_z \ge \alpha > 0$. Rescaling produces $C^\infty$ functions $\tilde{\phi}_1, \tilde{\phi}_2$ with $\tilde{\phi}'_d > 0$ matching $\phi_{d,0}$ on $\{0, 1/2, 1\}$; let $\tilde{f}_0$ be the quadratic interpolating $f_0$ there and $\tilde{u}_0(x, \theta) \triangleq \tilde{f}_0(x) + \tilde{\phi}_1(x)\theta_1 + \tilde{\phi}_2(x)\theta_2$. In Proposition~\ref{prop:cycles}(i) (and the $5 \times 5$ grid), every cyclic winning coalition $W_i$ ($\mu(W_i) > 1/2$) has slack $\Delta = 1/4 - 5\eta > 0$; in Proposition~\ref{prop:cycles}(ii), continuity of $\mu$ from below yields $\Delta > 0$ and $W_i \triangleq \{\theta : u(x_{w(i)}, \theta) - u(x_i, \theta) \ge \Delta\}$ with $\mu(W_i) > 1/2$. Lemma~\ref{lem:supp-lifting} ($q = 1/2, s = 2$) gives $C_{>1/2}([0,1], \mu) = \emptyset$.\footnote{For the $\{0,1\}^2$ distributions of Proposition~\ref{prop:cycles}(i) alone, no oscillatory penalty is needed: $\tilde{f}(x) = 1 - x$, $\tilde{\phi}_2(x) = \frac{1}{4}x + x^3$, and $\tilde{\phi}_1(x) = \frac{5}{4} - \frac{1}{4}(1-x) - (1-x)^3$ already eliminate every Condorcet winner on $[0, 1]$ while breaking quasiconcavity \citep{Black1948,DeDonder2013}.}

\emph{Part~(iii).} With the same $C^\infty$ step $\psi$ from Part~(ii):
\begin{itemize}[leftmargin=2em,itemsep=1pt]
\item For $q < r_k$: choose $\delta \in (0, \min(r_k, 1 - r_k)/2)$ with $r_k - \delta > q$ and $1 - (r_k + \delta)^k > q$. Set $\varepsilon = \eta \triangleq \frac{1}{4(k+2)}$, $\sigma_\delta(t) \triangleq \eta t + (1-\eta)\psi(\frac{t - (r_k - \delta)}{2\delta})$, $\tilde{\phi}_d(x) \triangleq \varepsilon x + (1-\varepsilon)\psi(kx - (d-1))$, and $\tilde{u}_0(x, \theta) \triangleq -\frac{k}{k+2}x + \sum_{d=1}^k \tilde{\phi}_d(x)\sigma_\delta(\theta_d)$ (so $\partial_x\partial_{\theta_d}\tilde{u}_0 \ge \varepsilon\eta > 0$). At grid policies $x_i = i/k$ ($s = k$): for $i \ge 1$ on $W_i \triangleq \{\theta_i \le r_k - \delta\}$ ($\mu(W_i) = r_k - \delta > q$), $\tilde{u}_0(x_{i-1}, \theta) - \tilde{u}_0(x_i, \theta) \ge \frac{1}{k+2} - \eta - \frac{k-1}{k}\varepsilon > \frac{1}{2(k+2)} \triangleq \Delta$; for $i = 0$ on $W_0 \triangleq \{\exists d,\; \theta_d \ge r_k + \delta\}$ ($\mu(W_0) = 1 - (r_k + \delta)^k > q$), $\tilde{u}_0(x_k, \theta) - \tilde{u}_0(x_0, \theta) \ge (1-\eta) - \frac{k}{k+2} > \Delta$. Lemma~\ref{lem:supp-lifting} completes the additive construction.
\item For $q < q_{\mathrm{hyp}}(k)$: take Lemma~\ref{lem:hyp-cover}(iii)'s upper boxes $U_t = \prod_{d=1}^k [a_{t,d}, 1]$ ($t = 1, \dots, s$, $\mu(U_t) = c < 1 - q$) covering $K^+_{p_{\mathrm{eff}}}$ with $V_k(p_{\mathrm{eff}}) < 1 - q$ as in Theorem~\ref{thm:highdim} Step~2. Choose $\delta \in (0, \min(\{1\} \cup \{a_{t,d} : a_{t,d} > 0\}))$ small enough that every $\delta$-inflated box $U_t^\delta \triangleq \prod_{d=1}^k I_{t,d}^\delta$ ($I_{t,d}^\delta = [0,1]$ if $a_{t,d}=0$, $(a_{t,d}-\delta, 1]$ otherwise) has $1 - \mu(U_t^\delta) > q$. Set $\psi_{t,d}(\theta_d) = 1$ if $a_{t,d} = 0$ and $\psi(\frac{\theta_d - (a_{t,d}-\delta)}{\delta})$ otherwise, $\Phi_{t,\delta}(\theta) \triangleq \prod_{d=1}^k \psi_{t,d}(\theta_d)$ ($1$ on $U_t$, $0$ outside $U_t^\delta$), $\varepsilon \triangleq \frac{s}{2k(s+2)}$, and
\[
\tilde{u}_0(x, \theta) \triangleq -\frac{s}{s+2}x + \varepsilon x\sum_{d=1}^k \theta_d + \sum_{t=1}^s \psi\big(sx - (t - 1)\big)\Phi_{t,\delta}(\theta),
\]
satisfying $\partial_x\partial_{\theta_d}\tilde{u}_0 \ge \varepsilon > 0$. At $x_i = i/s$: on $W_i \triangleq [0,1]^k \setminus U_i^\delta$ ($\mu(W_i) > q$), $\tilde{u}_0(x_{i-1}, \theta) - \tilde{u}_0(x_i, \theta) \ge \frac{1}{s+2} - \frac{k\varepsilon}{s} = \frac{1}{2(s+2)} \triangleq \Delta$; on $W_0 \triangleq \bigcup_{t=1}^s U_t \supseteq K^+_{p_{\mathrm{eff}}}$ ($\mu(W_0) > q$), $\tilde{u}_0(x_s, \theta) - \tilde{u}_0(x_0, \theta) \ge 1 - \frac{s}{s+2} > \Delta$. Lemma~\ref{lem:supp-lifting} completes the proof. \qedhere
\end{itemize}
\end{proof}

\section{Simple-Majority Tournament Solutions: Smith, Copeland, and Minimax}\label{sec:supp-tournaments}

On a finite policy menu $X$ under a voter distribution $\nu$, call a nonempty subset $S \subseteq X$ \emph{dominant} under $\nu$ if every $x \in S$ strictly majority-beats every $y \in X \setminus S$ (under net-margin or absolute majority), and let the \emph{Smith set} $\operatorname{Smith}(\nu)$ denote the minimal dominant subset. Under a tie-free odd-$N$ voter distribution $\nu$, every distinct pair $(x, y)$ has a strict majority winner ($x \succ_{\mathrm{maj}} y$ or $y \succ_{\mathrm{maj}} x$). The \emph{Copeland score} of $x \in X$ is $|\{y \in X \setminus \{x\} : x \succ_{\mathrm{maj}} y\}| - |\{y \in X \setminus \{x\} : y \succ_{\mathrm{maj}} x\}|$, and $\operatorname{Copeland}(\nu)$ is the set of maximisers of this score. The \emph{Minimax} (Simpson--Kramer) set $\operatorname{Minimax}(\nu)$ minimises each policy's worst-case pairwise opposition $\max_{y \in X \setminus \{x\}} N \cdot \nu(y \succ x)$ (equivalently, on odd-$N$ strict profiles, minimises worst-case pairwise vote-margin defeat).

\begin{proposition}[Tournament Solutions under Multivariate FOSD]\label{prop:supp-tournaments}
Consider finite ordered policy menus $X \subset \mathbb{R}$ and CW-SCP preferences on $\Theta$ under distributions $\mu' \succeq_{\mathrm{FOSD}} \mu$.
\begin{enumerate}[label=(\roman*),leftmargin=2.2em]
\item Any dominant set $A$ under $\mu$ and any dominant set $B$ under $\mu'$ (under the same majority convention) satisfy $\min A \le \max B$; in particular, $\min\operatorname{Smith}(\mu) \le \max\operatorname{Smith}(\mu')$.
\item Conversely, under strict-preference CW-SCP on $\Theta = \{0, 1\}^2$ and odd-$N$ tie-free shifts $\mu' \succeq_{\mathrm{FOSD}} \mu$: (a)~on five policies ($N = 3$), a pre-shift strict Condorcet winner vanishes into a $5$-cycle while $\min\operatorname{Smith}$ falls and the singleton Copeland winner inverts strictly downward, and a mirror-image shift creates a strict Condorcet winner while $\max\operatorname{Smith}$ falls (violating both minimum and maximum monotonicity of the Smith set); and (b)~on four policies ($N = 9$), the singleton Minimax winner inverts strictly downward.
\end{enumerate}
\end{proposition}

\begin{proof}
\emph{Part~(i).} Write $x \to_\nu y$ when $x$ strictly majority-beats $y$ under $\nu$ (under either net-margin or absolute majority; $\to_\nu$ is asymmetric in both cases). If dominant sets $A$ under $\mu$ and $B$ under $\mu'$ satisfied $\max B < \min A$, picking any $a \in A$ and $b \in B$ would give $a > b$, $b \notin A$, and $a \notin B$. Dominance of $A$ gives $a \to_\mu b$, so Proposition~\ref{prop:fosd}(i) preserves $a \to_{\mu'} b$, contradicting dominance of $B$ ($b \to_{\mu'} a$). Thus $\min A \le \max B$, and in particular $\min\operatorname{Smith}(\mu) \le \max\operatorname{Smith}(\mu')$.

\emph{Part~(ii).} Table~\ref{tab:supp-tournaments} exhibits Profiles~A and~B together with the upward-inverted pair sets $I(\theta) \triangleq \{(a, b) : a < b,\; b \succ_\theta a\}$. In both profiles, $I(0,0) \subseteq I(1,0) \cap I(0,1)$ and $I(1,0) \cup I(0,1) \subseteq I(1,1)$, confirming CW-SCP.
\begin{itemize}[leftmargin=2em,itemsep=1pt]
\item \emph{Profile~A (vanishing Condorcet winner, failures of both $\min$ and $\max$ Smith monotonicity, and Copeland inversion from $\{2\}$ to $\{0\}$):} Moving one voter from $(1,0)$ to $(1,1)$ gives $\mu = (0, 2/3, 1/3, 0) \preceq_{\mathrm{FOSD}} \mu' = (0, 1/3, 1/3, 1/3)$ (listing weights on $(0,0),(1,0),(0,1),(1,1)$). Under $\mu$, $(1,0)$ holds $2/3$ weight, so $2$ is the unique strict Condorcet winner, giving $\operatorname{Smith}(\mu) = \operatorname{Copeland}(\mu) = \{2\}$. Under $\mu'$, pairwise majorities form the $5$-cycle $2 \succ_{\mathrm{maj}} 0 \succ_{\mathrm{maj}} 1 \succ_{\mathrm{maj}} 4 \succ_{\mathrm{maj}} 3 \succ_{\mathrm{maj}} 2$, so the strict Condorcet winner vanishes and $\operatorname{Smith}(\mu') = \{0, 1, 2, 3, 4\}$, strictly lowering $\min\operatorname{Smith}$ from $2$ to $0$. Moreover, under $\mu'$, $0$ beats $\{1,3,4\}$ and loses only to $2$ (Copeland score $+2$), $2, 3, 4$ each win two and lose two (score $0$), and $1$ beats only $4$ (score $-2$), so $\operatorname{Copeland}(\mu') = \{0\}$. Reversing both policy and type orders via $\tilde{x} \mathrel{\tilde{\succ}_{\tilde{\theta}}} \tilde{y} \iff (4 - \tilde{x}) \succ_{(1,1)-\tilde{\theta}} (4 - \tilde{y})$ with $\tilde{\mu}(\tilde{\theta}) \triangleq \mu'((1,1)-\tilde{\theta})$ and $\tilde{\mu}'(\tilde{\theta}) \triangleq \mu((1,1)-\tilde{\theta})$ preserves CW-SCP and $\tilde{\mu}' \succeq_{\mathrm{FOSD}} \tilde{\mu}$ (moving $1/3$ from $(0,0)$ to $(0,1)$), mapping $\operatorname{Smith}(\tilde{\mu}) = \{0, 1, 2, 3, 4\} \to \operatorname{Smith}(\tilde{\mu}') = \{2\}$ and strictly lowering $\max\operatorname{Smith}$ from $4$ to $2$.
\item \emph{Profile~B (Minimax inversion $\{2\} \to \{1\}$):} Shifting two voters from $(0,0)$ to $(1,0)$ and two voters from $(0,1)$ to $(1,1)$ gives $\mu = (3/9, 2/9, 3/9, 1/9) \preceq_{\mathrm{FOSD}} \mu' = (1/9, 4/9, 1/9, 3/9)$. Under $\mu$, worst-case opponent votes across the $9$ voters are $6$ against $0$ (from $2$), $6$ against $1$ (from $0$), $5$ against $2$ (from $1$), and $9$ against $3$ (from $2$), giving $\operatorname{Minimax}(\mu) = \{2\}$. Under $\mu'$, worst-case opponent votes are $8$ against $0$ (from $2$), $4$ against $1$ (from $2, 3$), $5$ against $2$ (from $1$), and $9$ against $3$ (from $2$), giving $\operatorname{Minimax}(\mu') = \{1\}$. \qedhere
\end{itemize}
\end{proof}

\begin{table}[htbp]
\centering
\caption{Explicit $\{0, 1\}^2$ Profiles for Proposition~\ref{prop:supp-tournaments}}
\label{tab:supp-tournaments}
\small
\begin{tabular}{lp{3.9cm}p{7.6cm}}
\toprule
Voter Type $\theta$ & Strict Ranking & Inverted Pairs $I(\theta)$ (CW-SCP Check) \\
\midrule
\multicolumn{3}{p{14cm}}{\emph{Profile A: Smith \& Copeland ($X = \{0,1,2,3,4\}$, $N=3$, $\mu = (0, 2/3, 1/3, 0) \to \mu' = (0, 1/3, 1/3, 1/3)$)}} \\
$(0,0)$ & $0 \succ 1 \succ 2 \succ 3 \succ 4$ & $I(0,0) = \emptyset$ \\
$(1,0)$ & $2 \succ 0 \succ 3 \succ 1 \succ 4$ & $I(1,0) = \{(0,2), (1,2), (1,3)\}$ \\
$(0,1)$ & $0 \succ 1 \succ 4 \succ 3 \succ 2$ & $I(0,1) = \{(2,3), (2,4), (3,4)\}$ \\
$(1,1)$ & $4 \succ 3 \succ 2 \succ 0 \succ 1$ & $I(1,1) = \{(a,b) : 0 \le a < b \le 4,\ (a,b) \neq (0,1)\}$ \\
\midrule
\multicolumn{3}{p{14cm}}{\emph{Profile B: Minimax ($X = \{0,1,2,3\}$, $N=9$, $\mu = (3/9, 2/9, 3/9, 1/9) \to \mu' = (1/9, 4/9, 1/9, 3/9)$)}} \\
$(0,0)$ & $0 \succ 1 \succ 2 \succ 3$ & $I(0,0) = \emptyset$ \\
$(1,0)$ & $1 \succ 2 \succ 0 \succ 3$ & $I(1,0) = \{(0,1), (0,2)\}$ \\
$(0,1)$ & $2 \succ 3 \succ 0 \succ 1$ & $I(0,1) = \{(0,2), (0,3), (1,2), (1,3)\}$ \\
$(1,1)$ & $2 \succ 3 \succ 1 \succ 0$ & $I(1,1) = \{(0,1), (0,2), (0,3), (1,2), (1,3)\}$ \\
\bottomrule
\end{tabular}
\end{table}

\section{Auxiliary Boundary Verifications and Closed-Threshold Cores}\label{sec:supp-aux}

\subsection{Closed-Threshold Boundary at a Fixed Median\texorpdfstring{ (Theorem~\ref{thm:median-core}(i))}{}}\label{sec:supp-closed-median}
On general posets where $\sup_t h_\mu(t)$ need not be attained, $\succ_{>q_0}$ ($q_0 = 1 - \sup_t h_\mu(t)$) remains acyclic because any finite cycle is endorsed by a representative type whose quota lies below its smallest edge share. At $q = 1 - 2^{-k}$, defeat $\mu(x \succ x^*) = 1 - 2^{-k}$ for $x^* \in M(\theta^{\mathrm{med}})$ can occur on $X = \{0, 1\}$ under $\operatorname{Unif}[0,1]^k$ ($1 \succ 0$ outside $[0, 1/2]^k$, $0 \sim 1$ at $(1/2,\dots,1/2)$) or strictly under $\operatorname{Bernoulli}(1/2)^k$ ($0 \succ_{\mathbf{0}} 1$ only at $\mathbf{0} = \theta^{\mathrm{med}}$). Conversely, $x^* \in M(\theta^{\mathrm{med}})$ lies in $C_{\ge 1 - 2^{-k}}(X, \mu)$ whenever $h_\mu(\theta^{\mathrm{med}}) > 2^{-k}$ (cf.\ Proposition~\ref{prop:supp-closed}(ii)), or when $h_\mu(\theta^{\mathrm{med}}) \ge 2^{-k}$, $x^*$ is strictly optimal for $\theta^{\mathrm{med}}$, $u(x, \cdot)$ is continuous at $\theta^{\mathrm{med}}$, and every neighbourhood of $\theta^{\mathrm{med}}$ puts positive mass outside $O^-(\theta^{\mathrm{med}})$ and positive mass outside $O^+(\theta^{\mathrm{med}})$. Under strict optimality at $\theta^{\mathrm{med}}$ and $h_\mu(\theta^{\mathrm{med}}) \ge 2^{-k}$, weak defeat satisfies $\mu(x \succeq x^*) \le 1 - 2^{-k}$ for all $x \neq x^*$.

\subsection{Off-Median Representative Types\texorpdfstring{ (Theorem~\ref{thm:median-core}(i))}{}}\label{sec:supp-offmedian}
Fix $h \in [0, 1/2)$ and let $\mu$ on $[0, 1]^2$ put atoms of mass $h$ at $(0, 0)$ and $(1, 1)$ and mass $1/2 - h$ uniformly along $\{(1/4 + s, 3/4 + s)\}$ and $\{(3/4 + s, 1/4 + s)\}$, $s \in [-1/8, 1/8]$. Neither marginal charges $(3/8, 5/8)$, so $(1/2, 1/2)$ is a median without median-line mass and Corollary~\ref{cor:observable} gives $1 - h$. The non-median representative type $t = (1/4, 3/4)$ has $h_\mu(t) = h + (1/2 - h)/2 = 1/4 + h/2$, lowering the endorsement quota to $3/4 - h/2$ ($0.7$ vs.\ $0.9$ at $h = 1/10$).

\subsection{Boundary Extension of the Pulled-Back Profile\texorpdfstring{ (Section~\ref{sec:2d})}{}}\label{sec:supp-cube}
\begin{lemma}[Uniform-Cube Reduction]\label{lem:supp-cube}
Let $X \subset \mathbb{R}$ be nonempty and finite, $\mu = \bigotimes_{d=1}^k \mu_d$ a product measure on $\Theta = \prod_{d=1}^k \Theta_d$, and $\{\succeq_\theta\}$ a CW-SCP profile on $X \times \Theta$. Then there is a CW-SCP profile $\{\widetilde{\succeq}_u\}$ on $X \times [0, 1]^k$ such that, under $\lambda \triangleq \operatorname{Unif}[0, 1]^k$, $\lambda(y \mathrel{\widetilde{\succ}} x) = \mu(y \succ x)$ and $\lambda(y \mathrel{\widetilde{\succeq}} x) = \mu(y \succeq x)$ for all $x, y \in X$; hence both cores coincide and $C_{>q}(X, \mu) \neq \emptyset$ for all $q > q^*(k)$.
\end{lemma}

\begin{proof}
With $F_d(t) \triangleq \mu_d(\theta_d \le t)$, $Q_d(u_d) \triangleq \inf\{t \in \Theta_d : F_d(t) \ge u_d\} \in \Theta_d$ is nondecreasing on $(0, 1)$ and satisfies $Q_d(u_d) \le t \iff u_d \le F_d(t)$ (as $\Theta_d$ is closed and $F_d$ is right-continuous and constant on gaps of $\Theta_d$). Thus $Q = (Q_1, \dots, Q_k)$ pushes $\lambda$ on $(0, 1)^k$ forward to $\mu$. Partition $[0, 1]^k \setminus (0, 1)^k$ into the lower face union $B_0 \triangleq \{\min_d u_d = 0\}$ (assigned strict descending order) and upper face union $B_1 \triangleq (0, 1]^k \setminus (0, 1)^k$ (assigned strict ascending order), and set $\widetilde{\succeq}_u \triangleq {\succeq}_{Q(u)}$ on $(0, 1)^k$. For $y > x$, the extended upward-preference sets $Q^{-1}(U^>(y, x)) \cup B_1$ and $Q^{-1}(U^\ge(y, x)) \cup B_1$ are Borel upper sets (by Lemma~\ref{lem:structure}(i) and monotonicity of $Q$), establishing CW-SCP while preserving all vote shares since $B_0 \cup B_1$ is $\lambda$-null.
\end{proof}

\subsection{Set-Valued Open \texorpdfstring{$q$}{q}-Core Non-Monotonicity on \texorpdfstring{$(1/2, 1)$}{(1/2, 1)}\texorpdfstring{ (Theorem~\ref{thm:phase})}{}}\label{sec:supp-setcore}
On $X = \{0, 1, 2\}$ and $\Theta = [0, 1]^2$ with independent marginals, fix $q \in (1/2, 1)$ and $\varepsilon \in (0, \min(q - 1/2, \frac{1}{2}(1 - q)^2))$. Set $A \triangleq \{\theta_1 > a\}$, $B \triangleq \{\theta_2 > b\}$, $U(1, 0) = A$, and $U(2, 1) = B$ (strict CW-SCP by Lemma~\ref{lem:structure}(i)):
\begin{itemize}[leftmargin=2em,itemsep=1pt]
\item \emph{Shrinking ($\max C_{>q}$ falls):} Take $U(2, 0) = A \cap B$, $\mu_1(A) = 1 - q + \varepsilon$, and shift $\mu_2(B)$ from $q - \varepsilon$ to $q + \varepsilon$. Since $\mu_1(A)\mu_2(B) < 1 - q$, $0$ defeats $2$ throughout, $0$ and $1$ never defeat each other, and $2$ comes to defeat $1$, shrinking $C_{>q}$ from $\{0, 1\}$ to $\{0\}$.
\item \emph{Expanding ($\min C_{>q}$ falls):} Take $U(2, 0) = A \cup B$, $\mu_2(B) = q - \varepsilon$, and shift $\mu_1(A)$ from $1 - q - \varepsilon$ to $1 - q + \varepsilon$. Since $\mu(2 \succ 0) = 1 - (1 - \mu_1(A))(1 - \mu_2(B)) > q$, $2$ defeats $0$ throughout while $0$ stops defeating $1$, expanding $C_{>q}$ from $\{2\}$ to $\{1, 2\}$.
\end{itemize}

\subsection{Closed-Threshold Cores at \texorpdfstring{$1 - 2^{-k}$}{the Boundary Quota}\texorpdfstring{ (Theorems~\ref{thm:median-core} and~\ref{thm:phase})}{}}\label{sec:supp-closed}
A measure $\mu$ on $\Theta \subseteq \mathbb{R}^2$ is \emph{positively quadrant dependent} (PQD) if $\mu(\theta_1 \le a, \theta_2 \le b) \ge \mu_1(\theta_1 \le a)\mu_2(\theta_2 \le b)$ for all $a, b \in \mathbb{R}$ (implied by positive association via Lemma~\ref{lem:structure}(ii)).

\begin{proposition}\label{prop:supp-closed}
Suppose preferences satisfy CW-SCP on $X \times \Theta$ ($k \ge 2$).
\begin{enumerate}[label=(\roman*),leftmargin=2.2em]
\item \emph{(Independent electorates, $k \ge 2$.)} For finite $X$ and $\mu = \bigotimes_{d=1}^k \mu_d$, some $\theta^* \in \Theta$ strictly prefers the challenger in every closed $(1 - 2^{-k})$-defeat, so $\succ_{\ge 1 - 2^{-k}}$ is acyclic and $\emptyset \neq M(\theta^*) \subseteq C_{\ge 1 - 2^{-k}}(X, \mu)$. If $\mu' = \bigotimes_{d=1}^k \mu'_d$ marginally dominates $\mu$, $C_{\ge 1 - 2^{-k}}(X, \mu) = \{x^*\}$, and $C_{\ge 1 - 2^{-k}}(X, \mu') = \{x^{**}\}$, then $x^{**} \ge x^*$.
\item \emph{(Finite-support marginals, $k \ge 2$.)} If each marginal $\mu_d$ has finite support and $\mu$ satisfies positive orthant dependence at $\theta^*(\mu) \triangleq (\overline{\tau}_1(\mu), \underline{\tau}_2(\mu), \dots, \underline{\tau}_k(\mu)) \in \Theta$, then $h_\mu(\theta^*(\mu)) > 2^{-k}$ strictly. Hence $\theta^*(\mu)$ strictly endorses every closed $(1 - 2^{-k})$-defeat, $\succ_{\ge 1 - 2^{-k}}$ is acyclic on arbitrary $X$, $M(\theta^*(\mu)) \subseteq C_{\ge 1 - 2^{-k}}(X, \mu)$, and (on finite or compact upper-semicontinuous $X$) $\min M(\theta^*(\cdot))$ and $\max M(\theta^*(\cdot))$ are weakly increasing closed $(1 - 2^{-k})$-core selections under marginal dominance.
\item \emph{(Arbitrary PQD electorates, $k = 2$.)} For finite $X$ and any PQD (in particular, positively associated) $\mu$ on $\Theta \subseteq \mathbb{R}^2$, some $\theta^* \in \Theta$ strictly prefers the challenger in every closed $3/4$-defeat; hence $\succ_{\ge 3/4}$ is acyclic and $\emptyset \neq M(\theta^*) \subseteq C_{\ge 3/4}(X, \mu)$.
\end{enumerate}
\end{proposition}

\begin{proof}
\emph{Part~(i).} Let $Q = (Q_1, \dots, Q_k)$ be the quantile map from Lemma~\ref{lem:supp-cube}, pushing Lebesgue measure $\lambda$ on $(0, 1)^k$ forward to $\mu$, and let $c \triangleq (1/2, \dots, 1/2)$. For each closed $(1 - 2^{-k})$-defeat $y \succ_{\ge 1 - 2^{-k}} x$, the failure set $F \triangleq \{u \in (0, 1)^k : x \succeq_{Q(u)} y\}$ has $\lambda(F) \le 2^{-k}$ and is a lower set if $y > x$ and upper if $y < x$. If a lower $F$ has $c \in \overline{F}$, then $(0, 1/2)^k \subseteq F$ (each $u \in (0, 1/2)^k$ lies below a point of $F$ close enough to $c$) and $F$ contains no $u$ with $u_d > 1/2$ (as $\prod_{d'} (0, u_{d'}]$ would add positive volume outside $(0, 1/2)^k$), so $F \subseteq (0, 1/2]^k$; symmetrically an upper $F$ with $c \in \overline{F}$ lies in $[1/2, 1)^k$. By finiteness of $X$, some $\rho > 0$ separates $B_\rho(c)$ from every failure set whose closure misses $c$. Then $u^* \triangleq c + \delta(e_1 - e_2)$ ($0 < \delta < \rho/2$) escapes all failure sets, so $\theta^* = Q(u^*)$ endorses every closed defeat, giving acyclicity and $\emptyset \neq M(\theta^*) \subseteq C_{\ge 1 - 2^{-k}}(X, \mu)$. For $\mu'$ marginally dominating $\mu$, picking $\rho$ for both electorates lets the same $u^*$ serve both (because $\rho$ depends on the pair, this orders singleton cores rather than defining a selection electorate by electorate), giving $Q(u^*) \le Q'(u^*)$ and $x^* \le x^{**}$ by Proposition~\ref{prop:fosd}(iii).

\emph{Part~(ii).} Let $\operatorname{supp}(\mu_1) = \{a_1 < \dots < a_r\} \subseteq \Theta_1$. Then $\overline{\tau}_1(\mu) = a_j$ where $j = \max\{i : \mu_1(\theta_1 \ge a_i) \ge 1/2\}$, so $\mu_1(\theta_1 \le a_j)$ equals $1 > 1/2$ if $j = r$ and $1 - \mu_1(\theta_1 \ge a_{j+1}) > 1/2$ if $j < r$. Symmetrically, for every $d \ge 2$, $\mu_d(\theta_d \le \underline{\tau}_d(\mu)) \ge 1/2$ and $\mu_d(\theta_d \ge \underline{\tau}_d(\mu)) > 1/2$ strictly. By Lemma~\ref{lem:structure}(iii), $\theta^*(\mu) \in \Theta$, and positive orthant dependence at $\theta^*(\mu)$ gives
\begin{gather*}
\mu\big(O^-(\theta^*(\mu))\big) \ge \mu_1(\theta_1 \le \overline{\tau}_1(\mu))\!\prod_{d=2}^k \mu_d(\theta_d \le \underline{\tau}_d(\mu)) > 2^{-k}, \\
\mu\big(O^+(\theta^*(\mu))\big) \ge \mu_1(\theta_1 \ge \overline{\tau}_1(\mu))\!\prod_{d=2}^k \mu_d(\theta_d \ge \underline{\tau}_d(\mu)) > 2^{-k}.
\end{gather*}
Thus $h_\mu(\theta^*(\mu)) > 2^{-k}$, so Theorem~\ref{thm:median-core}(i) gives strict endorsement of every closed $(1 - 2^{-k})$-defeat, acyclicity, and $M(\theta^*(\mu)) \subseteq C_{\ge 1 - 2^{-k}}(X, \mu)$. Marginal dominance orders $\theta^*(\mu) \le \theta^*(\mu')$ (Lemma~\ref{lem:structure}(iii)), so Proposition~\ref{prop:fosd}(iii) gives $M(\theta^*(\mu)) \le_{\mathrm{SSO}} M(\theta^*(\mu'))$.

\emph{Part~(iii).} By Sklar's theorem via the distributional transform \citep{Ruschendorf2009} (equivalently, the bilinear interpolation copula of $\mu$; \citealp[Section~2.3]{Nelsen2006}), there exists a probability measure $\mathbb{P}$ on $(0, 1)^2$ with $\operatorname{Unif}[0, 1]$ marginals such that $(Q_1(U), Q_2(V)) \stackrel{d}{=} (\Theta_1, \Theta_2)$ under $(U, V) \sim \mathbb{P}$; bilinear interpolation across jumps preserves PQD (on each jump rectangle the interpolated copula and $ab$ are both bilinear, so their difference is minimised at a corner, where PQD of $\mu$ applies), giving $\mathbb{P}(U \le a, V \le b) \ge ab$ and hence, by inclusion--exclusion, $\mathbb{P}(U \ge a, V \ge b) \ge (1 - a)(1 - b)$ \citep[Section~5.2.1]{Nelsen2006}. Pulling preferences back via $Q$ to $(0, 1)^2$ preserves CW-SCP and all vote shares.

Let $c \triangleq (1/2, 1/2)$. Any lower failure set $F$ ($\mathbb{P}(F) \le 1/4$) with $c \in \overline{F}$ contains $(0, 1/2)^2$, which has mass $\ge 1/4$ by PQD, forcing $\mathbb{P}((0, 1/2)^2) = \mathbb{P}(F) = 1/4$ and $\mathbb{P}(F \setminus (0, 1/2)^2) = 0$; symmetrically, an upper failure set with $c \in \overline{F}$ contains $(1/2, 1)^2$ with equal mass $1/4$ and zero mass outside it. Uniform marginals force $\mathbb{P}((0, 1/2)^2) = \mathbb{P}((1/2, 1)^2)$. If no failure set approaches $c$, set $u^* \triangleq c$. Otherwise both quadrant masses equal $1/4$, and finiteness of $X$ gives $B_\varepsilon(c) \subset (0, 1)^2$ disjoint from all failure sets whose closures miss $c$. Let $S_s \triangleq (1/2, 1/2 + s) \times (1/2 - s, 1/2)$. If $\mathbb{P}(S_s) > 0$ for all $s > 0$, take $S_s \subseteq B_\varepsilon(c)$; each failure set approaching $c$ has zero mass on $S_s$, so their finite union leaves a point $u^* \in S_s$ outside every failure set. Otherwise $\mathbb{P}(S_{s_0}) = 0$ for some $s_0 > 0$; take $u^* \triangleq (1/2 + t, 1/2 - t) \in S_{s_0} \cap B_\varepsilon(c)$ with $t > 0$ small. By PQD, $\mathbb{P}(1/2 < U \le 1/2 + t, V \le 1/2) \ge (1/2 + t)/2 - 1/4 = t/2$, and since $\mathbb{P}((1/2, 1/2 + t] \times (1/2 - t, 1/2]) = 0$, this mass lies on $A_{\mathrm{below}} \triangleq (1/2, 1/2 + t] \times [0, 1/2 - t]$; any lower failure set containing $u^*$ and approaching $c$ would contain $(0, 1/2)^2 \sqcup A_{\mathrm{below}}$ and exceed $1/4$, contradiction. Symmetrically, upper-quadrant PQD places mass $\ge t/2$ on $B_{\mathrm{right}} \triangleq [1/2 + t, 1] \times [1/2 - t, 1/2)$, excluding $u^*$ from every upper failure set approaching $c$. Thus $\theta^* \triangleq Q(u^*) \in \Theta$ endorses every closed $3/4$-defeat, proving acyclicity and $\emptyset \neq M(\theta^*) \subseteq C_{\ge 3/4}(X, \mu)$.
\end{proof}

\noindent Only comparing singleton closed $3/4$-cores across general non-finite-support positively quadrant dependent electorates in $k = 2$ remains open; for finite-support marginals, part~(ii) supplies the monotone representative type $\theta^*(\mu)$ with $h_\mu(\theta^*(\mu)) > 2^{-k}$ in every dimension $k \ge 2$.

\end{document}